\documentclass[11pt]{article}

\usepackage{epsfig}
\usepackage{geometry}
\usepackage{lscape}
\usepackage{afterpage}
\usepackage{amsmath}
\usepackage[square,comma,numbers,sort&compress]{natbib}
\usepackage{float}
\usepackage{amssymb}
\usepackage{longtable}

\newcommand*{\no}{\noindent}
\newcommand*{\bea}{\begin{eqnarray}}
\newcommand*{\eea}{\end{eqnarray}}
\newcommand*{\be}{\begin{equation}}
\newcommand*{\ee}{\end{equation}}
\newcommand*{\pd}{\partial}

\newcommand*{\pref}[1]{(\ref{#1})}

\newcommand*{\prefr}[2]{(\ref{#1}-\ref{#2})} 
\newcommand*{\nn}{\nonumber}

\newcommand*{\tr}{\mathrm{tr}}

\newcommand{\bma}{\begin{pmatrix}}
\newcommand{\ema}{\end{pmatrix}}

\begin{document}  
 
\vspace*{1mm}

\begin{center}

{\LARGE Finite-density gauge correlation functions in QC$_2$D}
\vskip10mm

Tamer Boz$^\dagger$, Ouraman Hajizadeh$^\star$, Axel Maas$^\star$, Jon-Ivar Skullerud$^\dagger$
\vskip8mm
$^\star$ University of Graz, Institute of Physics, NAWI Graz, Universit\"atsplatz 5, 8010 Graz, Austria

$\dagger$ Department of Theoretical Physics, National University of
Ireland Maynooth, Maynooth, Co Kildare, Ireland 
\end{center}
\vskip15mm

\begin{abstract}

2-color QCD is the simplest QCD-like theory which is accessible to lattice simulations at finite density. It therefore plays an important role to test qualitative features and to provide benchmarks to other methods and models, which do not suffer from a sign problem. To this end, we determine the minimal-Landau-gauge propagators and 3-point vertices in this theory over a wide range of densities, the vacuum, and at both finite temperature and density. The results show that there is essentially no modification of the gauge sector in the low-temperature, low-density phase. Even outside this phase only mild modifications appear, mostly in the chromoelectric sector.

\end{abstract}

\section{Introduction}

It has been argued for a long time that nuclear matter at high density and (relatively) low temperature would undergo a transition to a phase where quarks are the main degrees of freedom. More precisely, at high densities the overlap of the baryonic wave functions becomes substantial, leading to direct interactions of the quarks and in this sense quarks become bulk degrees of freedom. Due to the attractive strong interaction, it is expected that the Fermi surface will easily be disturbed leading to various pairing patterns of quarks and different phases \cite{Leupold:2011zz,BraunMunzinger:2009zz,Berges:1998rc,Rajagopal:2000wf,Alford:2001dt,Schafer:2003vz,Buballa:2003qv}.

To firmly establish these qualitative features demands a
first-principles calculation of QCD at low temperature and high
densities. Unfortunately, this is the regime where perturbative
methods fail, as high energy excitations would cost too much energy
for the system and therefore the physics will be dominated by the low
energy excitations. Thus, a non-perturbative treatment is
mandatory. Lattice QCD, as the mainstay of non-perturbative methods of
studying QCD, suffers from the infamous sign problem. It arises as a
result of introducing a quark-chemical potential in combination with
the complex color representation of the quarks in QCD, which leads to
a complex action in the path-integral. In turn, this has up to now made lattice Monte-Carlo simulations too inefficient. See \cite{Seiler:2017wvd,Bedaque:2017epw} for summaries of recent progress on this problem. An alternative is to use non-lattice methods, either functional methods, see e.\ g.\ \cite{Nickel:2006kc,Nickel:2006vf,Marhauser:2006hy,Nickel:2008ef,Braun:2008pi,Contant:2017gtz,Pawlowski:2010ht,Braun:2011pp,Muller:2013pya,Fischer:2018sdj}, or models/effective field theories, see e.\ g.\ \cite{Leupold:2011zz,Buballa:2003qv,Tripolt:2017zal,Strodthoff:2011tz}. However, these also require assumptions.

One way to circumvent the sign problem on the lattice is to study
QCD-like theories \cite{Kogut:2000ek,Langfeld:2011rh,vonSmekal:2012vx}
that share important features with real QCD \cite{vonSmekal:2012vx},
but are accessible on the lattice at finite density. Among those are
two-color QCD with an even number of fundamental quarks
\cite{Hands:2006ve,Hands:2010gd,Cotter:2012mb,Boz:2013rca,Braguta:2015owi,Braguta:2016cpw,Wellegehausen:2017gba,Holicki:2017psk},
also known as QC$_2$D, G$_2$-QCD
\cite{Maas:2012ts,Maas:2012wr,Wellegehausen:2017gba,Wellegehausen:2013cya},
and QCD with adjoint quarks \cite{Hands:2000ei,Kogut:2000ek}.  In
fact, such studies can even be pushed to study neutron stars in such
theories \cite{Hajizadeh:2017jsw}, allowing for a macroscopic test of
the implications of gauge interactions. Furthermore, while such
theories will certainly differ quantitatively from QCD, they allow us to test qualitative mechanisms, e.g.\ the aforementioned pairing, and provide reliable benchmarks for the assumptions of models and functional methods. Especially the latter has already been done successfully in the vacuum and at finite temperature \cite{Maas:2011se}.

To provide such benchmarks, we will here study the gauge sector of
QC$_2$D at finite density, i.\ e.\ the minimal Landau-gauge
\cite{Maas:2011se} gluon and ghost propagators as well as their
3-point vertices on the lattice. This extends previous studies of the
gluon propagator alone \cite{Hands:2006ve,Boz:2013rca}. In addition,
as a derived quantity, we will determine the running coupling in the
miniMOM scheme \cite{vonSmekal:2009ae}. For comparison, we study the
same theory in the vacuum and in the interior of the phase diagram, as
well as pure Yang-Mills theory. 

The details of the simulations are laid out in section \ref{s:ltt}. A study of systematic errors is relegated to appendix \ref{s:sys}. Results in the vacuum will be discussed in section \ref{s:vacres} and at finite temperature in section \ref{s:tres}.

The main results at finite density and zero temperature are shown in
section \ref{s:res} and at both finite density and temperature in section \ref{s:pd}. Unexpectedly, we do not observe any substantial
dependence of the correlation functions on the density, even in regions
where quark-related quantities are affected
\cite{Hands:2006ve,Hands:2010gd,Cotter:2012mb,Boz:2013rca,Braguta:2015owi,Braguta:2016cpw,Holicki:2017psk}. In
particular, the running coupling remains strong throughout the whole
density range. Only above a critical, chemical-potential-dependent,
temperature do we observe any change. This change is essentially identical to the one observed at zero chemical potential. This agrees with results from simulations using staggered fermions \cite{Astrakhantsev:2018uzd,Bornyakov:2017txe} that no phase transitions occurs at zero temperature in the chemical-potential range studied here. These findings will be summarized in section \ref{s:con}.

On the one hand, our findings imply that keeping the gauge sector only
slightly modified in continuum calculations at low temperatures, as was done in
\cite{Nickel:2006kc,Nickel:2006vf,Marhauser:2006hy,Nickel:2008ef,Contant:2017gtz,Pawlowski:2010ht,Braun:2011pp,Muller:2013pya,Fischer:2018sdj}, is
well justified. On the other hand, this implies that the physics
observed is driven by the quarks, and that the investigated region in the phase diagram is not dominated by weak coupling physics. This is in line with observations made for the Wilson potential  \cite{Hands:2006ve,Hands:2010gd,Cotter:2012mb,Boz:2013rca}. This result should be contrasted with the observation that at low densities the matter is an essentially free diquark superfluid after the silver blaze point at not too high chemical potentials \cite{Hands:2006ve,Hands:2010gd,Cotter:2012mb,Boz:2013rca,Scior:2015vra,Strodthoff:2011tz}.

Some preliminary results for $\beta\le 1.9$ have been presented in \cite{Hajizadeh:2017ewa}. Note that we find here that some of the results on such coarse lattices appear to be lattice artifacts, and thus the preliminary conclusions of \cite{Hajizadeh:2017ewa} are superseded by the ones presented here.
         
\section{Setup, observables, and technical details}\label{s:ltt}         

\subsection{Configurations}

\begin{figure}
\includegraphics[width=\textwidth]{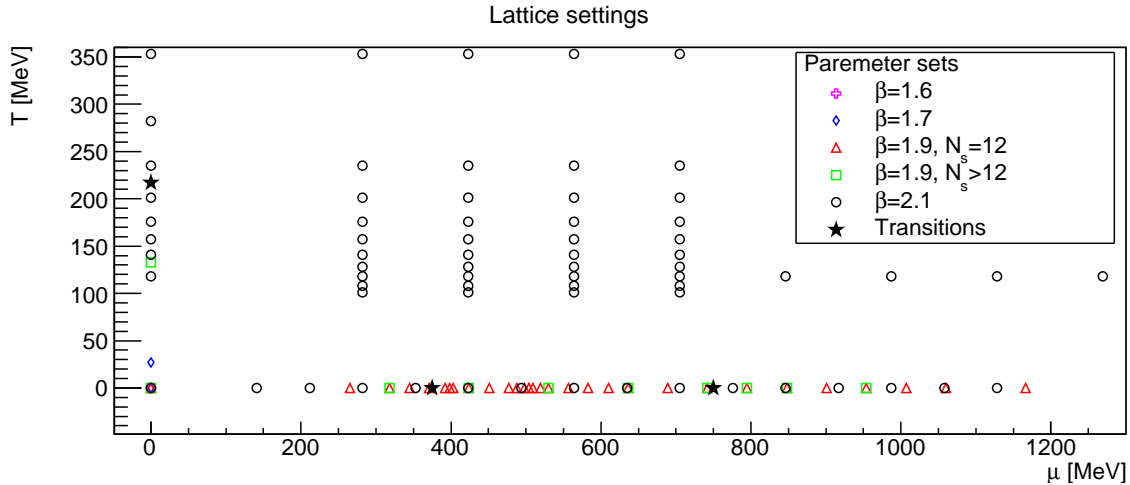}
\caption{\label{fig:pd}Location of the lattice configurations in the phase diagram, see table \ref{conf-sys} for details. The indicated transitions are the silver-blaze transition, the finite-temperature transition and the (likely spurious coarse-lattice) finite-density transition discussed in the text. Note that the temperature is set to zero if $N_t\ge N_s$ in the lattice setup.}
\end{figure}

In the following we use ensembles which have been created using the methods described in \cite{Hands:2006ve,Cotter:2012mb,Boz:2013rca} for the temperatures and densities plotted in figure \ref{fig:pd}. Most of these configurations have also been used in these works. They were created using an unimproved Wilson gauge action with 2 flavors of unimproved Wilson quarks\footnote{Note that for these lattice parameters there are potentially various bulk issues \cite{Wellegehausen:2017gba}. However, the gauge quantities investigated here have not shown any sensitivity to such problems \cite{Maas:2014xma}, and are therefore probably safe.}. The details of the employed lattice parameters and the number of configurations are listed in table \ref{conf-sys} in appendix \ref{s:conf}. The quark mass parameter at finite density was fixed to $\kappa=0.1680$ and $\kappa=0.1577$ at $\beta=1.9$ and $\beta=2.1$, respectively. This corresponds to rather heavy pions with mass $m_\pi=717(25)$ MeV. In comparison, at $\beta=1.7$ and $\kappa=0.178$ it is 668(6) MeV \cite{Cotter:2012mb}.

For the finer lattices at $\beta=1.7$, $\beta=1.9$, and $\beta=2.1$ lattice spacings have been determined using hadronic observables in \cite{Hands:2006ve,Cotter:2012mb,Boz:2013rca}, corresponding to $a=0.229$ fm, $a=0.186$ fm, and $a=0.138$ fm, respectively. Using various observables to extrapolate, most notably the running coupling to be discussed below, we estimate the lattice spacing at $\beta=1.6$ to be $a=0.266$ fm, which we will be using throughout.

Temperature is introduced by using asymmetric lattices $N_t\times N_s$, where $N_t$ is the temporal extent and $N_s$ is the spatial extent. The temperature is then given by $1/(aN_t)$ for $N_t<N_s$. If $N_t\ge N_s$, we set the temperature to zero in the main text. This ignores a 'residual lattice temperature' due to the finite lattice extent. This systematic error is investigated in appendix \ref{a:ar}, and no severe implications for the main text are found. The chemical potential is added explicitly to the action \cite{Gattringer:2010zz,Hands:2006ve,Cotter:2012mb,Boz:2013rca}. Because of the pseudo-reality of SU(2) the quark determinant remains real \cite{Kogut:2000ek}. With two degenerate quark flavours, the square of the determinant enters, and the action is therefore real and positive. Thus, the sign problem is avoided.

At finite density a diquark condensation is expected to take place in 2-color QCD \cite{vonSmekal:2012vx}. As this is a spontaneous symmetry breaking, this requires a limiting process of explicit breaking on a lattice \cite{Frohlich:1976it,Neuberger:1987kt}. To this end, a diquark source $j$ is introduced \cite{Boz:2013rca}, and varied over a range given in table \ref{conf-sys}. In principle, an extrapolation to zero $j$ is then necessary. However, as discussed in appendix \ref{a:sysj}, essentially no statistically significant dependence on $j$ is found for the quantities investigated here.

The configurations have afterwards been fixed to minimal Landau gauge
using an adaptive stochastic overrelaxation algorithm
\cite{Cucchieri:2006tf}. This minimizes the quantity $-\sum_{x,\mu}\tr U_\mu(x)$, where $U_\mu(x)$ are the link variables, which is equivalent to $\pd_\mu A_\mu^a=0$ in the continuum. Concerning Gribov copies, we used the first Gribov copy found, which corresponds to a flat average over all Gribov copies within the first Gribov horizon, i.\ e.\ the Gribov copies with positive semi-definite Faddeev-Popov operator \cite{Maas:2011se}. Details of stochastic overrelaxation are given in \cite{Cucchieri:1995pn}. The algorithm adapts the tuning parameter of \cite{Cucchieri:1995pn} by changing it such that during configuration creation the number of iteration steps is reduced, based on information from already gauge-fixed configurations.

We will occasionally compare to results from
pure Yang-Mills theory. For this purpose, we will use results from
\cite{Maas:2014xma,Maas:2011ez,Maas:unpublished,Fister:2014bpa}, using
as far as possible the same physical volumes and lattice
spacings. This will allow us to estimate the unquenching effects as well as the influence of the finite-density environment. At finite temperature, we will compare to results at roughly the same ratio $T/T_c$, where $T_c=217(23)$ MeV in the QC$_2$D case \cite{Boz:2013rca}.
         
\subsection{Observables}\label{s:obs}

Our primary interest here is the gauge sector. To this end, we determined the longitudinal ((chromo)electric) and transverse ((chromo)magnetic) dressing functions of the gluon propagator \cite{Cucchieri:2007ta}
\bea
D_T(p_0,\vec p^2) &=& \frac{1}{(d-2)N_g} \left\langle\sum_{\mu=1}^3 \, A_\mu^a(p) A_\mu^a(-p)-\frac{p_0^2}{{\vec p\,}^2} \, A_0^a(p)A_0^a(-p)\right\rangle\,,\label{eq:corrt}  \\
D_L(p_0,\vec p^2) &=& \frac{1}{N_g} \left(1+\frac{p_0^2}{{\vec p\,}^2}\right) \langle A_0^a(p)A_0^a(-p)\rangle \label{eq:corrl}\,,\\
A_\mu^a(x)&=&\frac{1}{2i}\tr\left(\tau^a U_\mu(x)\right)\nn
\eea
\no with respect to the heat bath, and both soft modes ($p_0=0$) and hard modes ($p_0=n\pi T$). Here, $d=4$ is the dimensionality and $N_g=3$ is the number of gluons. In the vacuum, both coincide, $D_T=D_L=D$. We define corresponding screening masses
\be
m_{T/L}=\frac{1}{\sqrt{D_{T/L}(0)}}\label{screenmass},
\ee
\no which are also called curvature or constituent masses. We also determine the corresponding dimensionless susceptibility $\chi_\mu=\pd m/\pd\mu$ of \pref{screenmass} by numerical derivation of $m(\mu)$ with respect to the chemical potential in the zero temperature case. The screening mass can be quite different from a, possibly not even existing, pole mass. The latter can potentially be obtained from the corresponding effective mass $m(t)$ \cite{Maas:2011se}
\bea
m_{T/L}(t)&=&-\ln\frac{\Delta_{T/L}(t+a)}{\Delta_{T/L}(t)}\nn\\
\Delta_{T/L}(t)&=&\frac{1}{a\pi}\frac{1}{N_t}\sum_{P_0=0}^{N_t-1}\cos\left(\frac{2\pi tP_0}{N_t}\right)D_{T/L}\left(\frac{2}{a}\sin\left(\frac{2\pi tP_0}{N_t}\right),0\right)\nn,
\eea
\no where $\Delta$ is the Schwinger function, and which we also have investigated. If $m(t\to\infty)$ is positive and time-independent, this defines a pole mass. In the vacuum, this effective mass is not compatible with an ordinary pole \cite{Maas:2011se}. Here, statistical noise precludes any conclusion either from the long-time behavior or from any fit. Thus we concentrate on the screening mass \pref{screenmass}.

We also investigated the scalar ghost propagator, given by
\bea
D_G(p_0,\vec p^2)&=&\frac{1}{V}\langle (M^{-1})^{aa}(p) \rangle\,,\label{eq:DofG}\\
M(y,x)^{ab}\omega_b(x)&=&c\left(\sum_x\big(G^{ab}(x)\omega_b(x)+\sum_\mu A_\mu^{ab}(x)\omega_b(x+e_\mu)+B_\mu^{ab}(x)\omega_b(x-e_\mu)\big)\right)\nn\\
G^{ab}(x)&=&\sum_\mu \tr(\{\tau^a,\tau^b\}(U_\mu(x)+U_\mu(x-e_\mu)))\nn\\
A_\mu^{ab}(x)&=&-2\tr(\tau^a \tau^bU_\mu(x))\nn\\
B_\mu^{ab}(x)&=&-2\tr(\tau^a \tau^bU_\mu^\dagger(x-e_\mu))\nn,
\eea
\no where $M$ is the Faddeev-Popov operator \cite{Cucchieri:2007ta,Zwanziger:1993dh}, again for both hard modes and soft modes. We usually plots its dressing function $G(p_0,\vec p^2)=(p_0^2+\vec p^2)D_G(p_0,\vec p^2)$. The necessary inversion of the Faddeev-Popov operator has been done using a standard conjugate gradient algorithm on a point source for the propagator and on a plane-wave source for the vertex below \cite{Cucchieri:2006tf}.

From these propagators the running coupling in the miniMOM scheme \cite{vonSmekal:1997vx,vonSmekal:2009ae} can be derived. In the vacuum, it is given by
\bea
\alpha(p^2)&=&\alpha(a^{-2})p^6D_G^2(p^2)D(p^2)\,,\label{alpha}\\
\alpha(a^{-2})&=&\frac{1}{\pi\beta}\,.\nn
\eea
\no We now define longitudinal and transverse couplings in a thermodynamic environment as
\bea
\alpha_T(p_0,\vec p^2)&=&\alpha(a^{-2})(p_0^2+\vec p^2)^3D_G^2(p_0,\vec p^2)D_T(p_0,\vec p^2)\label{alphat}\,,\\
\alpha_L(p_0,\vec p^2)&=&\alpha(a^{-2})(p_0^2+\vec p^2)^3D_G^2(p_0,\vec p^2)D_L(p_0,\vec p^2)\label{alphal}\,,
\eea
\no describing the strength of coupling of the longitudinal and the
transverse degrees of freedom, respectively.

Finally, we study the two three-point vertices, the ghost-gluon vertex
and the three-gluon vertex. In the vacuum, we follow the procedures in
\cite{Cucchieri:2006tf}. Accordingly, we determine the dressing-function of the ghost-gluon vertex and the dressing-function of the tree-level tensor of the three-gluon vertex as
\bea
G^X=\frac{\Gamma V^X}{\Gamma D_1 D_2 D_3\Gamma}\label{vertex}\,.
\eea
\no Here $\Gamma$ is the (lattice-improved) \cite{Rothe:2005nw}
tree-level vertex, and $V$ are the three-point vertices
\bea
V^{c\bar{c}A}&=&\langle A_\mu^a M^{bc-1}\rangle\,,\nn\\
V^{A^3}&=&\langle A_\mu^a A_\nu^b A_\rho^c\rangle\,,\nn
\eea
\no for the ghost-gluon and the three-gluon vertex, respectively. The $D_i$ are the corresponding propagators to amputate the lattice vertex \cite{Cucchieri:2004sq}. We use the same momentum configurations, one gluon momentum vanishing, two momenta orthogonal, and all momenta equal, as in \cite{Cucchieri:2006tf}.

In a thermodynamic environment many more tensor structures would arise. We follow here \cite{Fister:2014bpa} and only consider the full transverse zero-modes of the tree-level vertices, which corresponds to evaluating \pref{vertex} with all Matsubara frequencies vanishing and using only the transverse propagators for amputation.
  
\section{Vacuum results}\label{s:vacres}

The effects of unquenching on the gauge sector of QCD has been studied
to some extent on the lattice
\cite{Aouane:2012bk,Bowman:2010zr,Kamleh:2007ud}. In continuum
methods, this is a well-established topic, see
e.\ g.\ \cite{Fischer:2006ub,Binosi:2009qm} for reviews, and
\cite{Williams:2015cvx,Cyrol:2017ewj} for recent determinations. All
these results show no qualitative differences in the gauge sector,
the main effect being a suppression of the gluon propagator at mid-momentum.

\begin{figure}
\includegraphics[width=\textwidth]{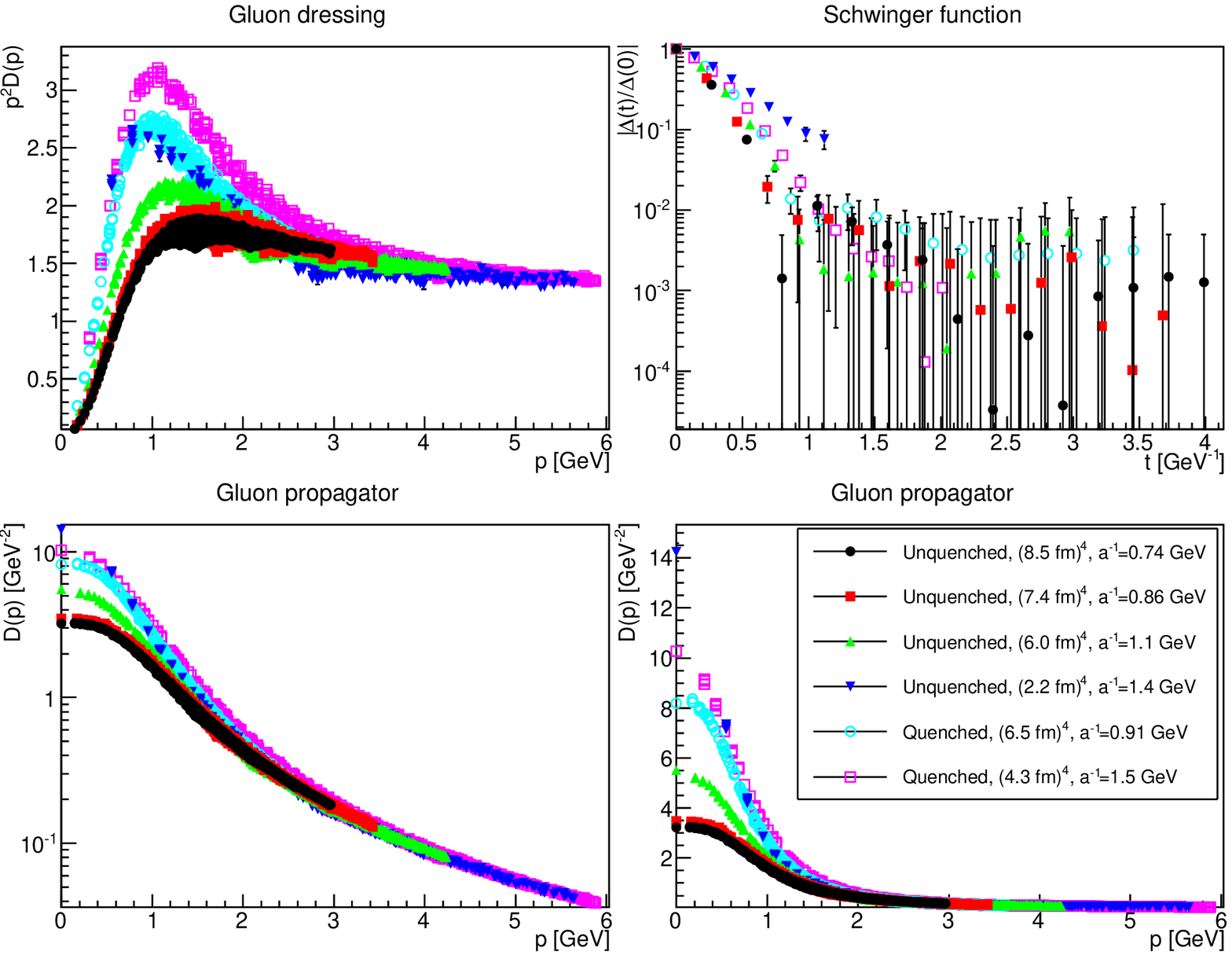}\\
\includegraphics[width=0.5\textwidth]{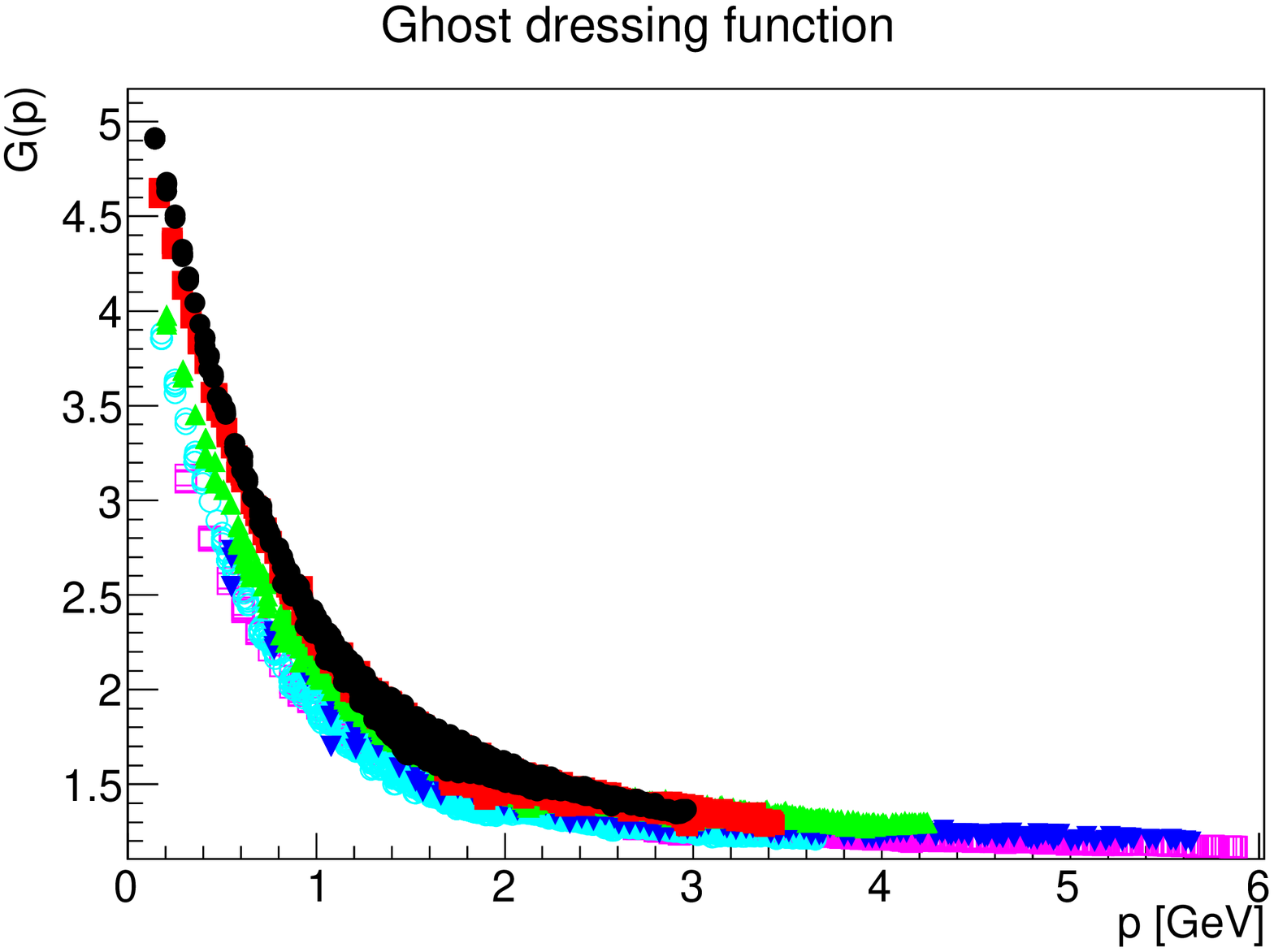}\includegraphics[width=0.5\textwidth]{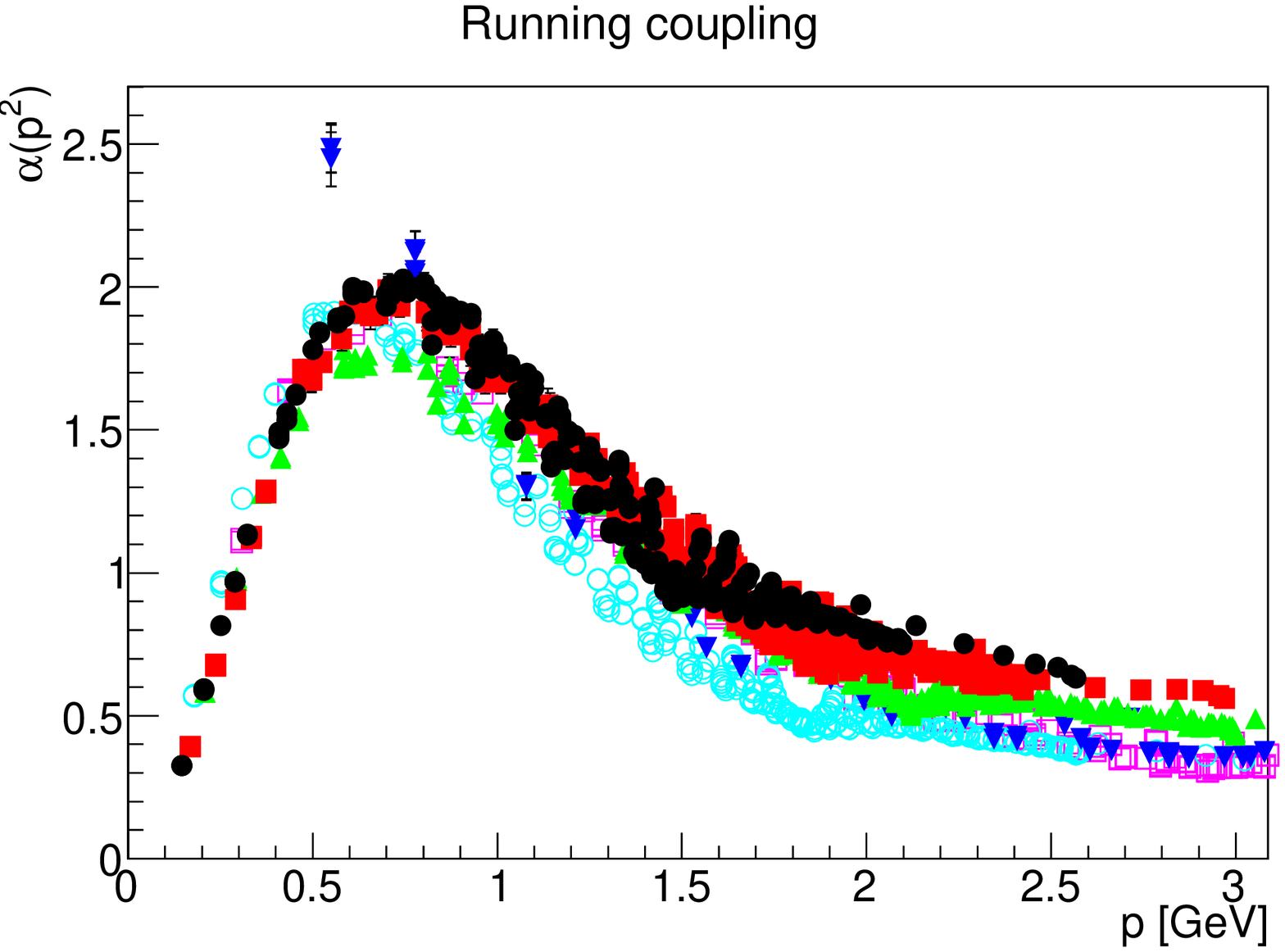}
\caption{\label{fig:vac}The quenched and unquenched gluon dressing function (top-left panel), Schwinger function (top-right panel), gluon propagator (middle panels, logarithmic and linear), ghost dressing function (bottom-left panel) and running coupling (bottom-right panel). Quenched data is from \cite{Maas:2014xma}. Error bars partly smaller than the symbol size. Results have not been renormalized.}
\end{figure}

In the present case of two-color QCD the pattern is similar, as can be
seen in figure \ref{fig:vac}. The gluon dressing function is
substantially suppressed at mid-momentum and in the
infrared. It is also observed that there is a substantial difference,
due to lattice artifacts, between the two coarser lattices on the one
hand and the next finer one. The finest lattice is even more different, but it is on a substantially smaller physical volume.

The ghost sector, which does not have a direct coupling to the matter
sector, is notably less affected. Since the dressing function enters
the running coupling \pref{alpha} quadratically, this also pushes
through to the running coupling. It shows very little difference
between Yang-Mills theory and two-color QCD, except for its
large-momentum running. In particular, its behavior in the momentum range  between a half to about one GeV is almost unchanged. This is of considerable importance, as this region dominates hadron phenomenology \cite{Alkofer:2000wg,Fischer:2006ub,Braun:2011pp,Cyrol:2017ewj,Blank:2010pa}.

\begin{figure}
\includegraphics[width=\textwidth]{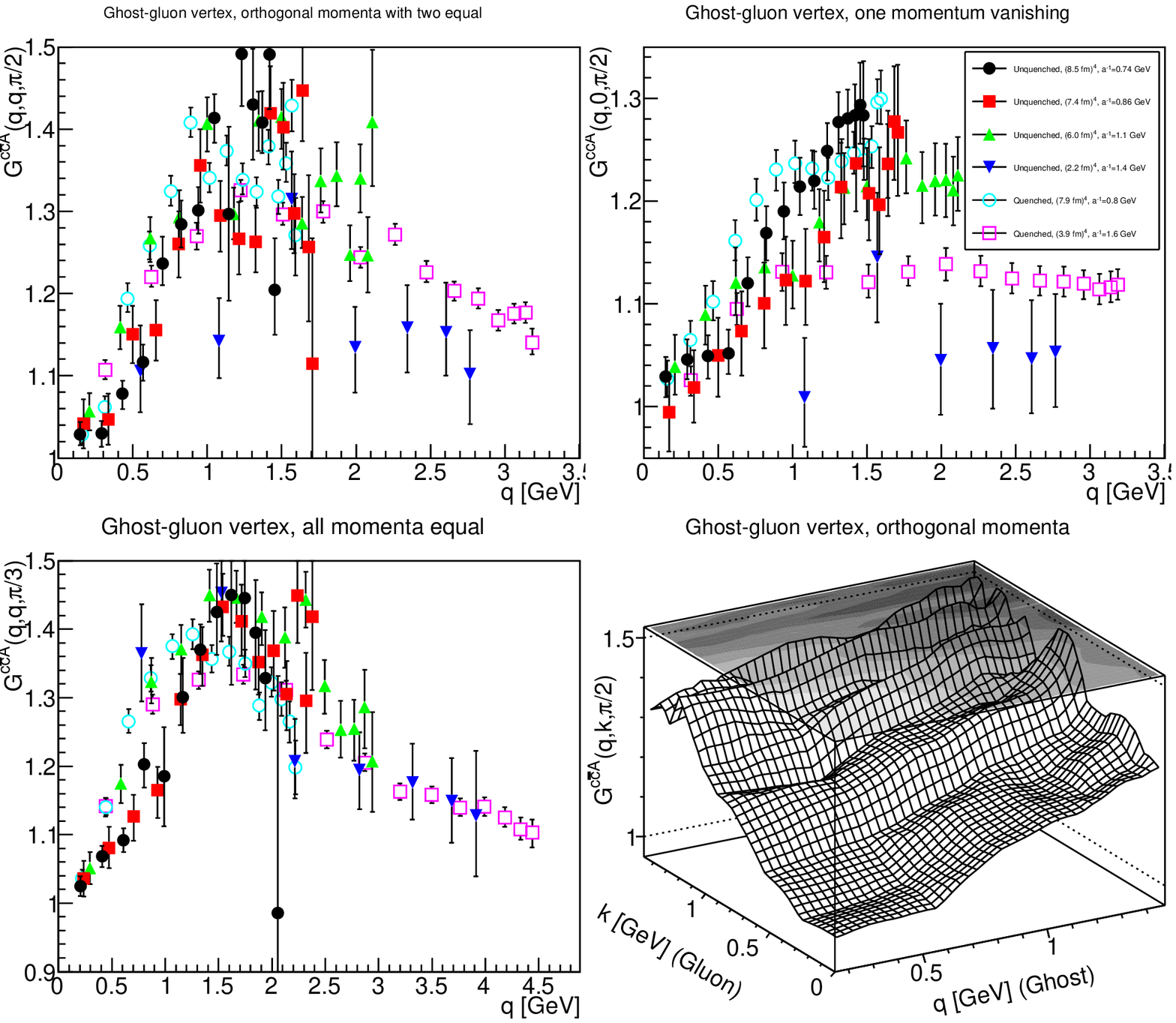}
\caption{\label{fig:vac:ggv}The ghost-gluon vertex dressing for different momentum configurations in comparison to quenched data from \cite{Maas:unpublished}. See text for details. Results have not been renormalized. The lower-right hand-plot is the unquenched set at $a^{-1}=0.74$ GeV.}
\end{figure}

That the ghost sector is quite unaffected by unquenching is also seen for the ghost-gluon vertex in figure \ref{fig:vac:ggv}. Within errors no significant deviations are observed from the quenched case.

\begin{figure}
\includegraphics[width=\textwidth]{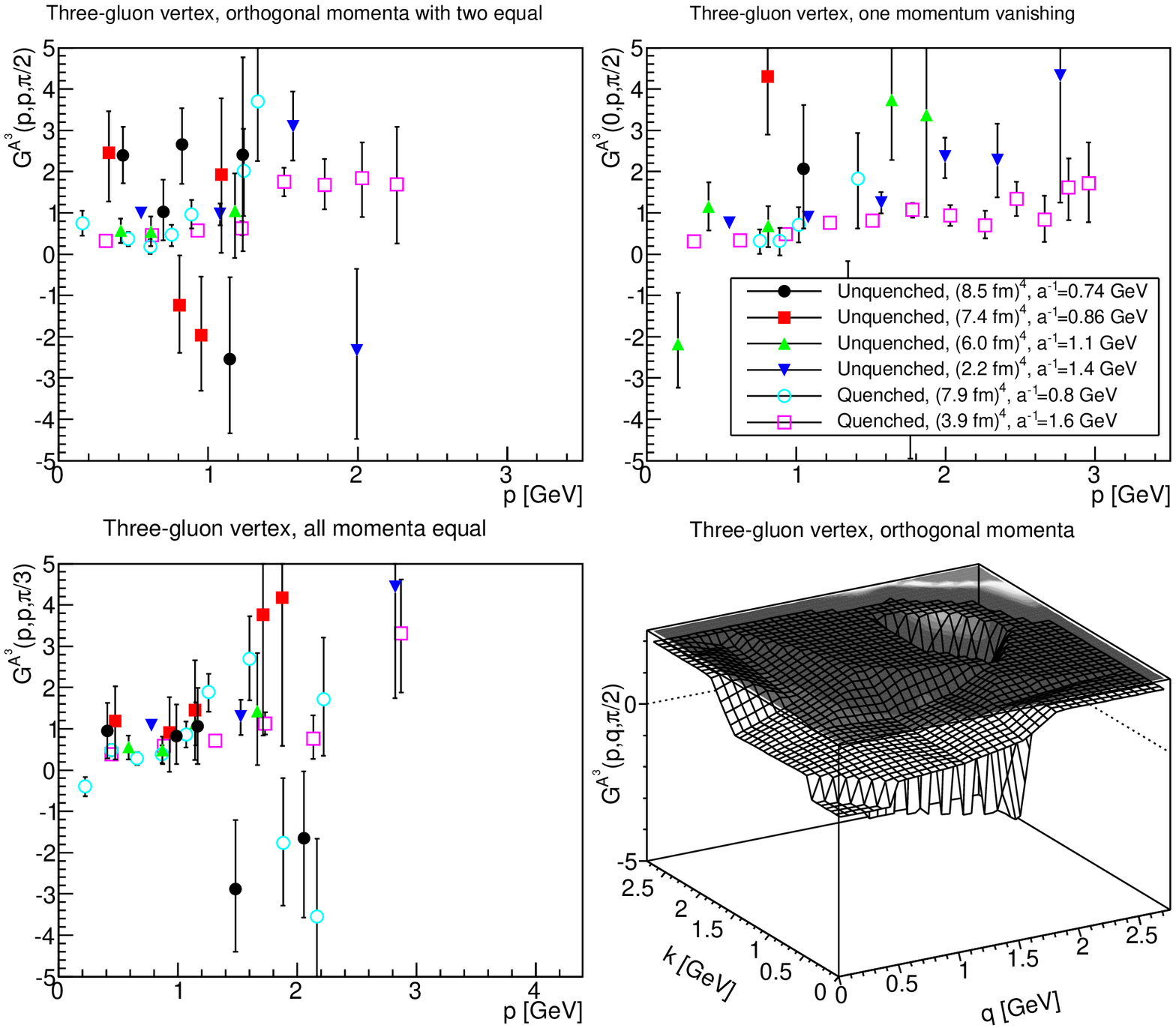}
\caption{\label{fig:vac:g3v}The three-gluon vertex dressing for different momentum configurations in comparison to quenched data from \cite{Maas:unpublished}. Points with a relative error larger than 100\% have been suppressed. See text for details. Results have not been renormalized. The lower-right hand-plot is the unquenched set at $a^{-1}=1.4$ GeV.}
\end{figure}

The three-gluon case is notoriously affected much more strongly by
statistical fluctuations \cite{Cucchieri:2006tf}. Thus, given the
rather low statistics available, the results shown in figure
\ref{fig:vac:g3v} can be taken to be indicative at best. However, within their errors they also do not show a marked difference compared to the quenched case. Also, no exceptional behavior, e.\ g.\ in the statistics dependence, is seen.

\section{Finite-temperature results}\label{s:tres}

At finite temperature the gauge sector of Yang-Mills theory shows a markedly different behavior above and below the critical temperature. In the low-temperature phase both polarizations of the gluon show only small, or possibly even no, dependence on the temperature \cite{Maas:2011ez,Silva:2013maa,Cucchieri:2014nya,Silva:2017feh,Cyrol:2017qkl,Aouane:2011fv}. Above the phase transition especially the longitudinal part shows a marked temperature dependence \cite{Cucchieri:2001tw,Cucchieri:2007ta,Maas:2011ez,Fischer:2010fx,Cyrol:2017qkl,Aouane:2011fv}, with possibly critical behavior around the phase transition \cite{Maas:2011ez}. Due to the crossover nature of the phase change in full QCD there this behavior becomes softened \cite{Aouane:2012bk}. Besides some isolated non-zero temperatures at large volumes, we have for the finest set a range of temperatures at fixed spatial volume available.

\begin{figure}
\includegraphics[width=0.5\textwidth]{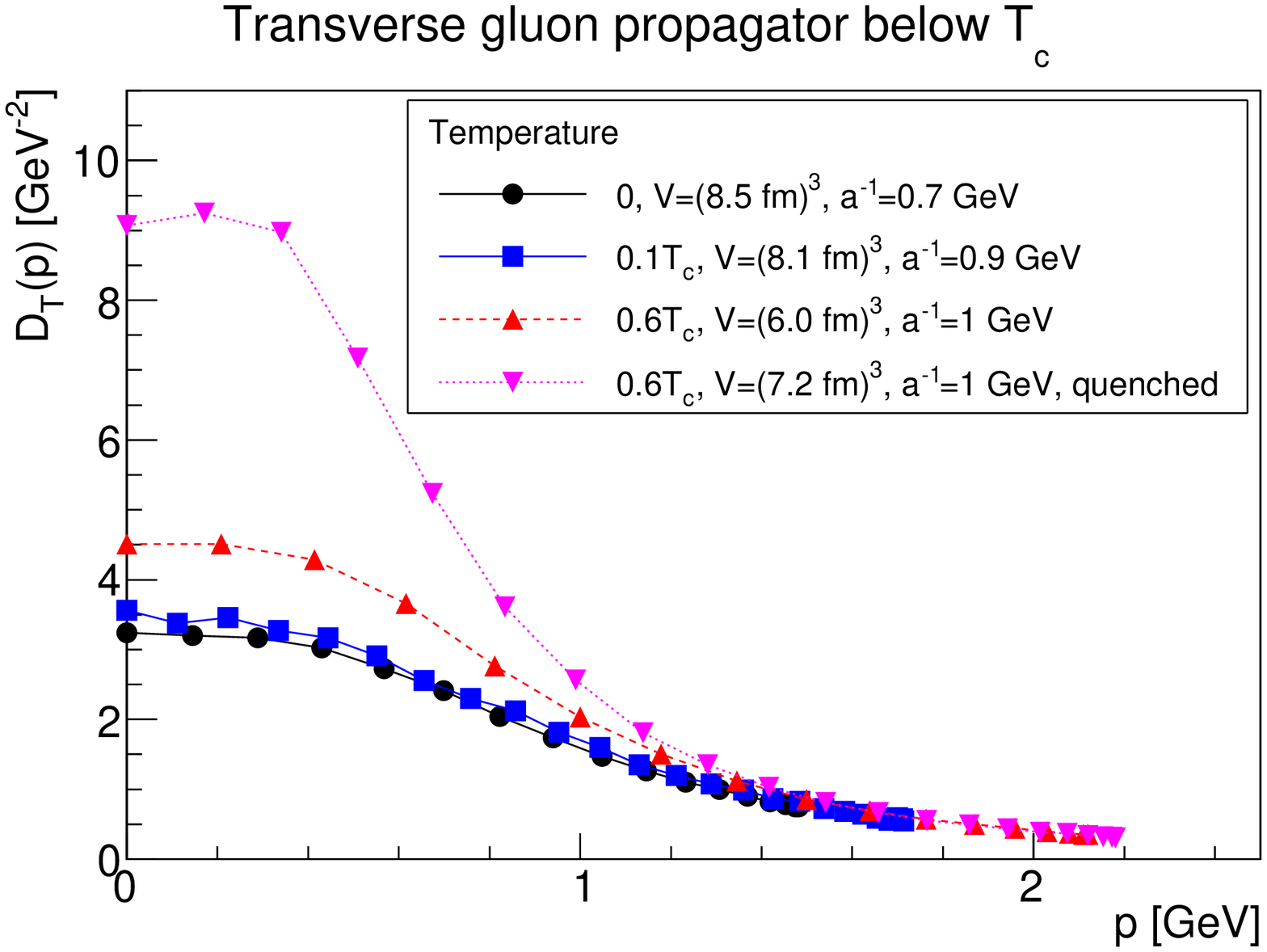}\includegraphics[width=0.5\textwidth]{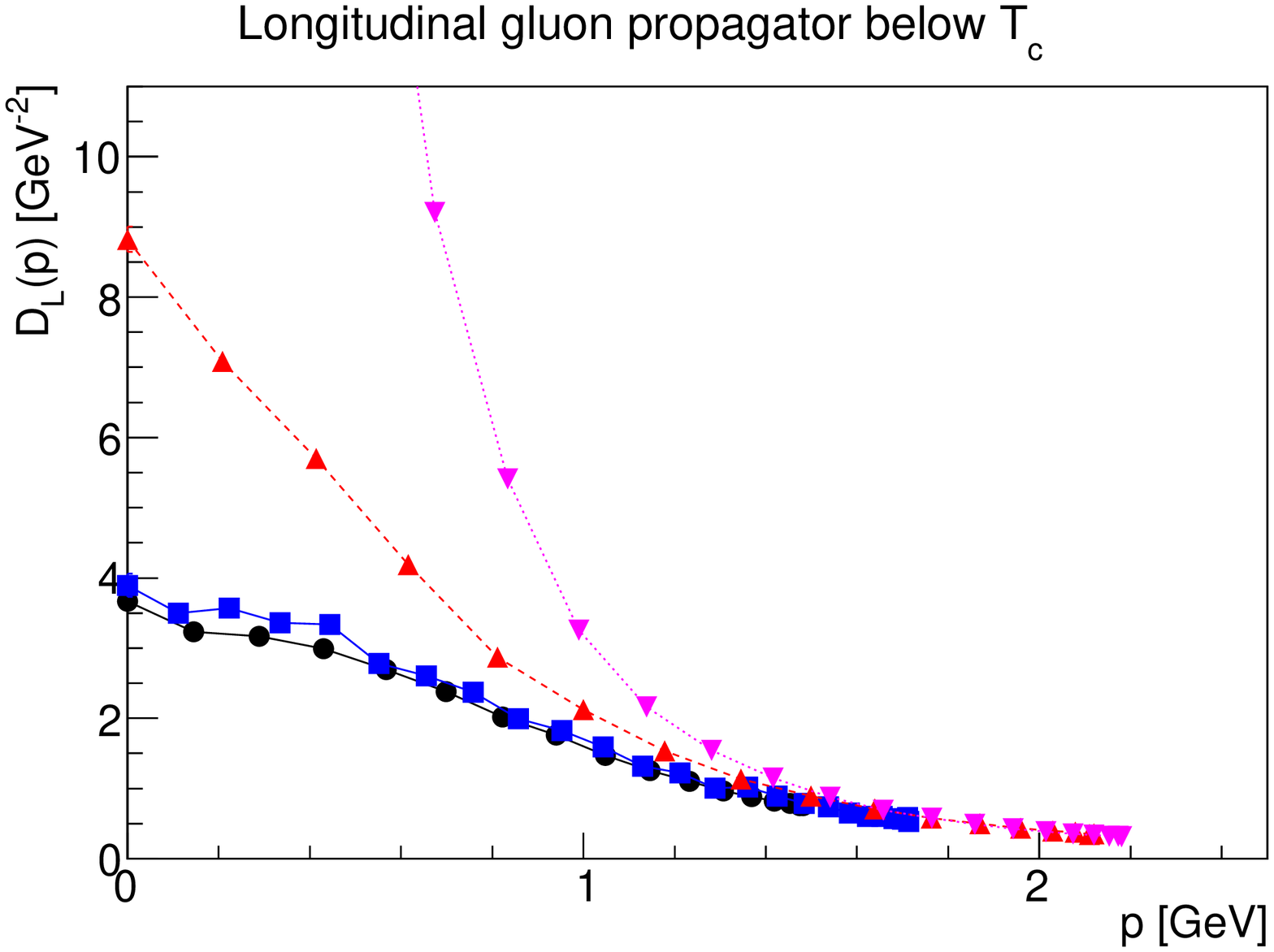}\\
\includegraphics[width=0.5\textwidth]{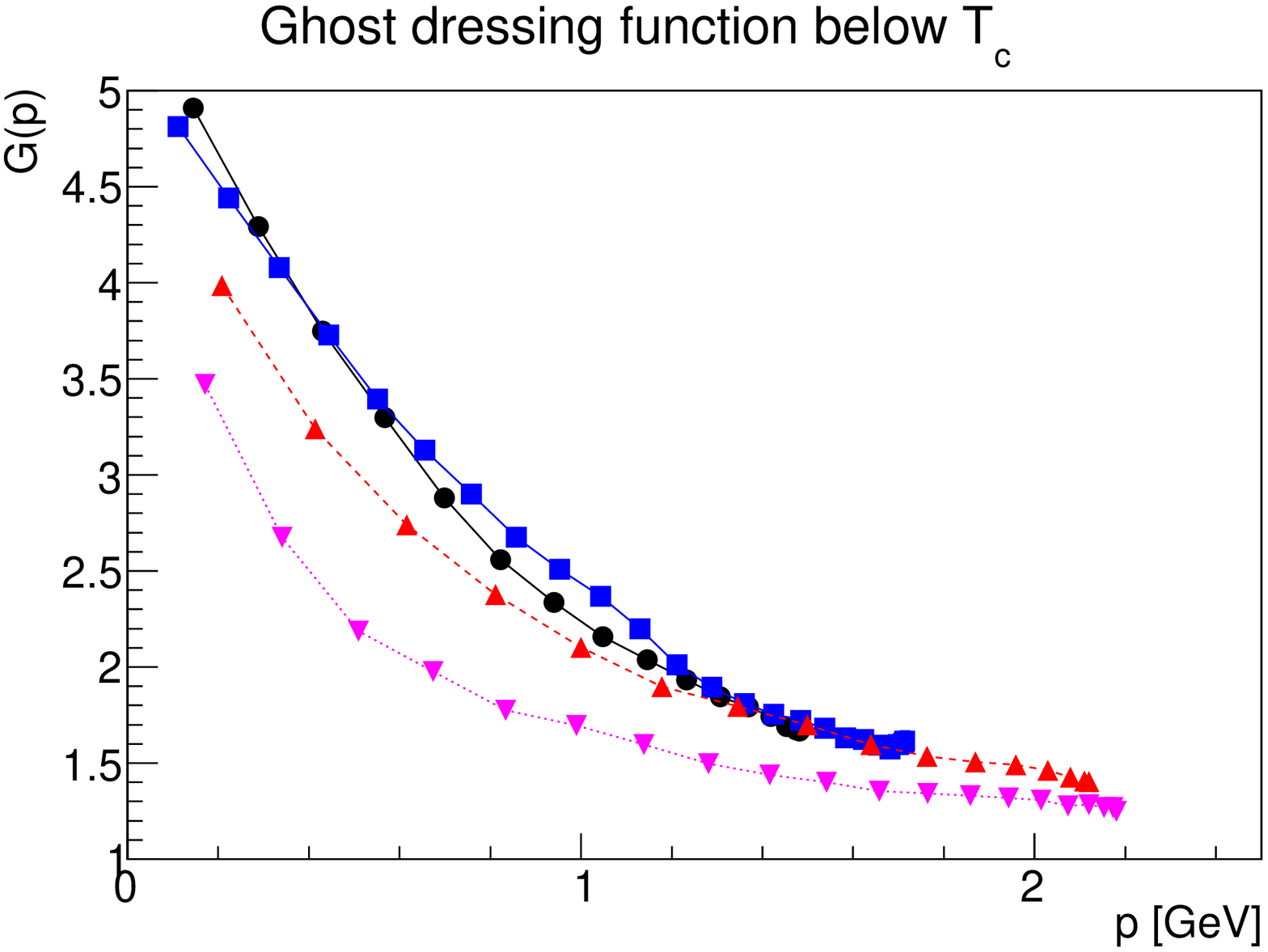}\includegraphics[width=0.5\textwidth]{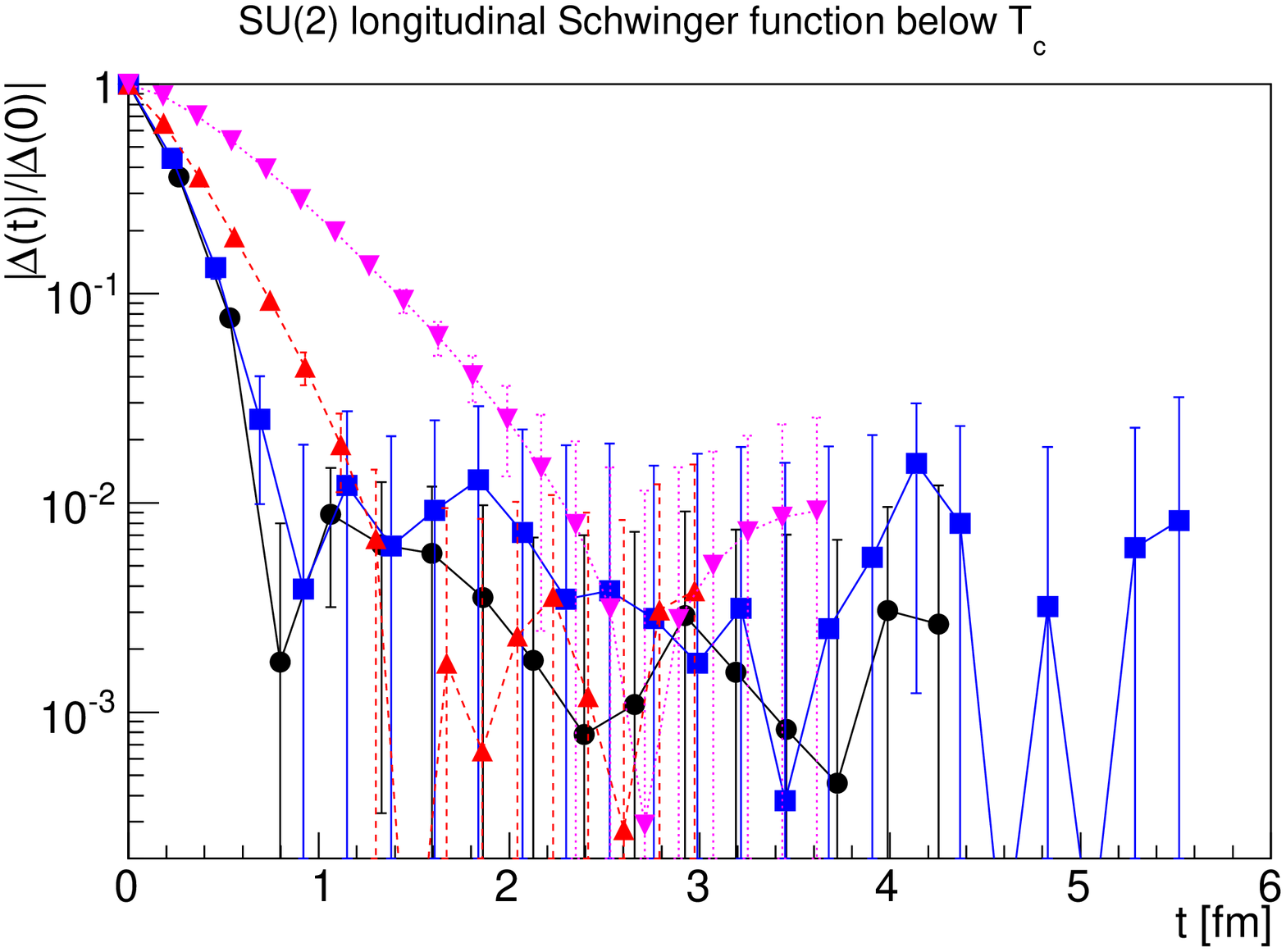}\\
\caption{\label{fig:gpft}The soft mode of the magnetic gluon propagator (top-left panel), the soft mode of the electric gluon propagator (top-right panel), the ghost dressing function (lower-left panel), and the Schwinger function of the electric gluon (lower-right panel) at finite temperature. Quenched results are from \cite{Maas:2011ez}. Results have not been renormalized. Momenta for propagators here and hereafter are along an edge to reduce lattice artifacts at small momenta.}
\end{figure}

\begin{figure}
\includegraphics[width=0.5\textwidth]{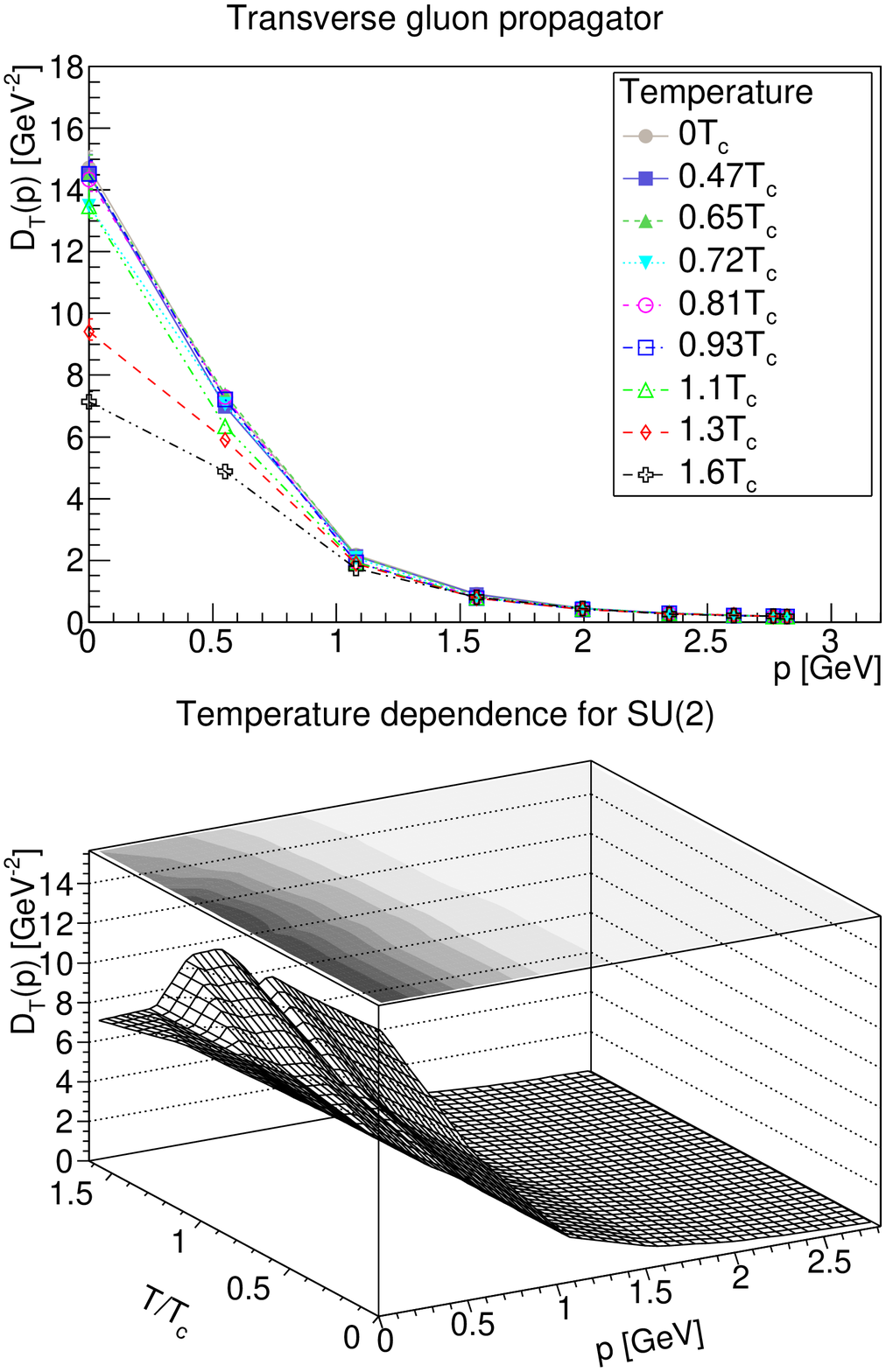}\includegraphics[width=0.5\textwidth]{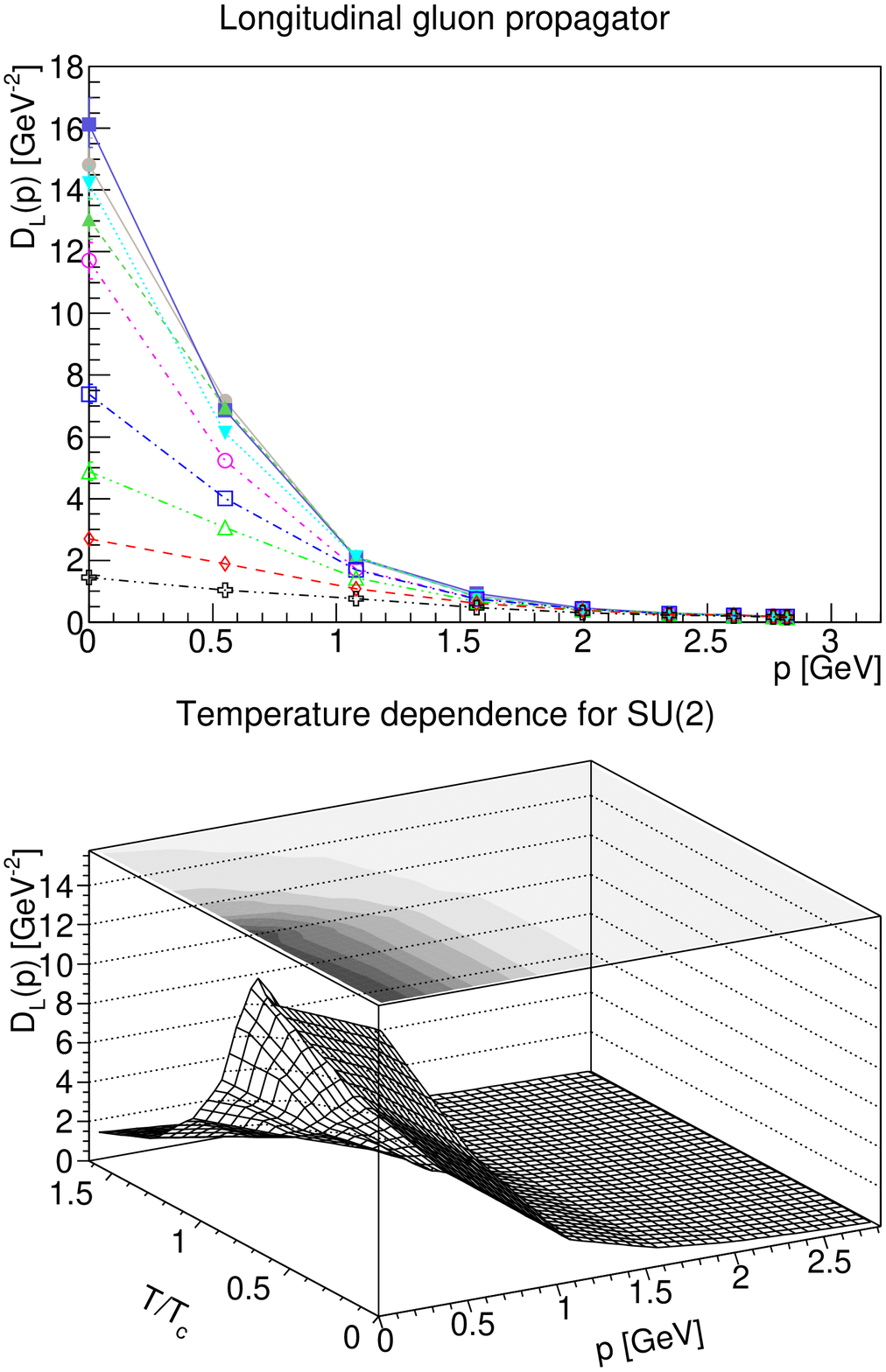}\\
\includegraphics[width=0.5\textwidth]{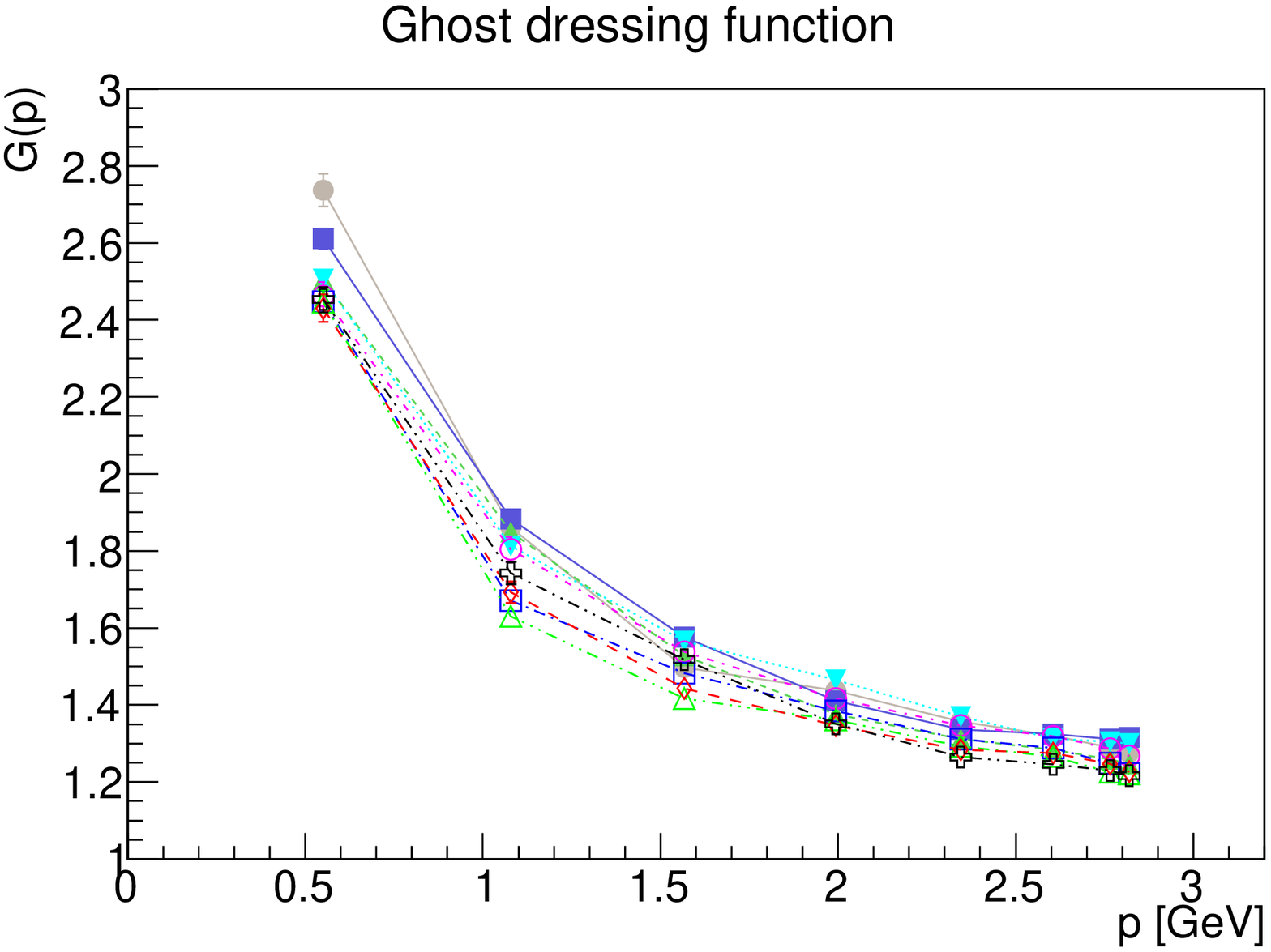}\includegraphics[width=0.5\textwidth]{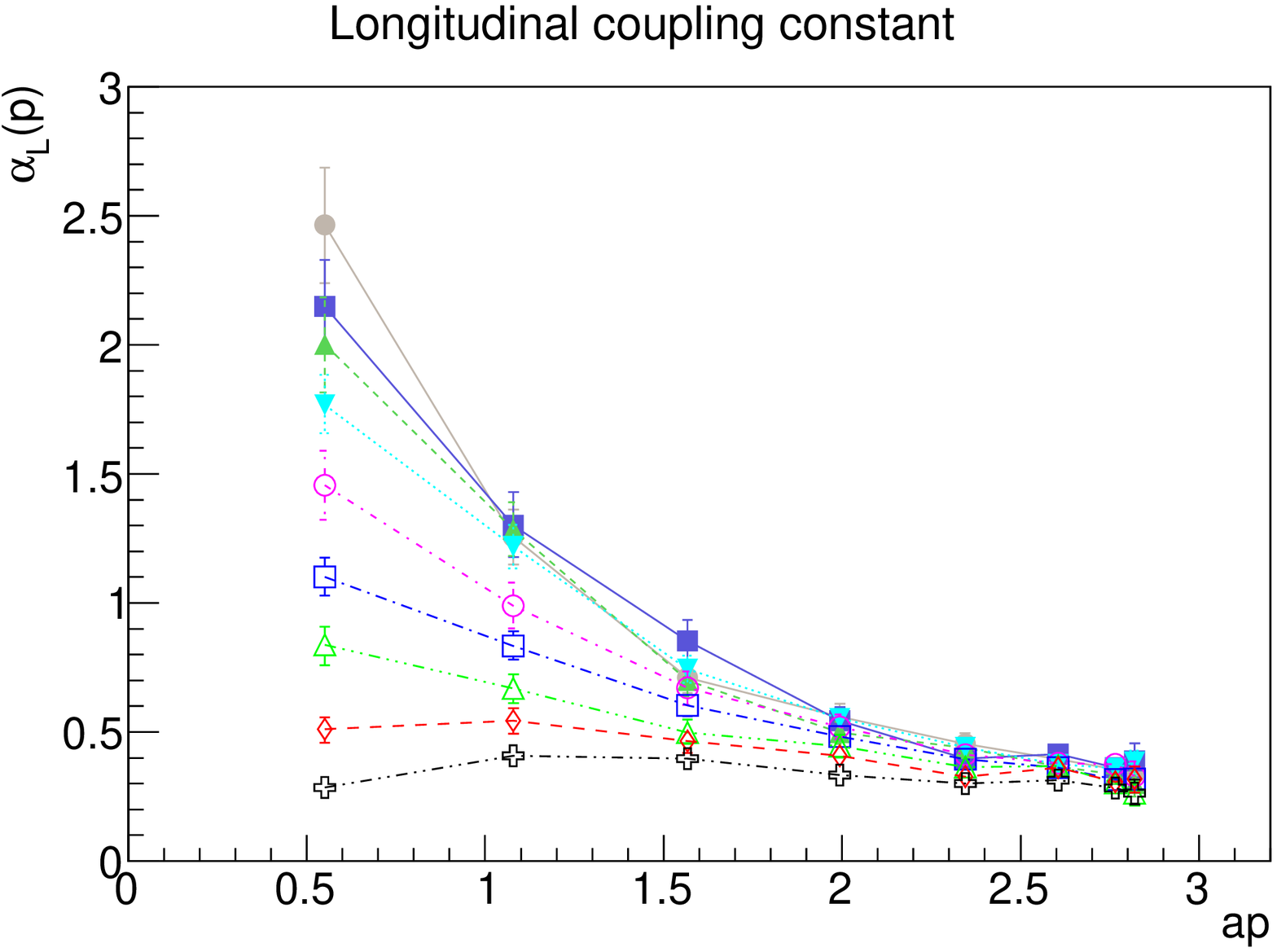}
\caption{\label{fig:gpftr}The soft mode of the magnetic gluon propagator (top-left panels), the soft mode of the electric gluon propagator (top-right panels), the ghost dressing function (lower-left panel), and the running longitudinal coupling (lower-right panel) at finite temperature. Results have not been renormalized.}
\end{figure}

The results for the propagators are shown in figure
\ref{fig:gpft} for the large volumes and in figure \ref{fig:gpftr} for fixed $\beta=2.1$. While the overall scale is much more attenuated in the
unquenched case a very similar behavior is seen for both the quenched
and unquenched case. The magnetic gluon propagator shows very little
influence due to the temperature, except for the usual suppression at very high temperatures. The electric one is quite different. The result in figure \ref{fig:gpft} shows a strong change at low momenta, but this is on different spatial volumes. Keeping the spatial volume fixed, as in figure \ref{fig:gpftr}, this is no longer the case. Then only a quick onset of a strong infrared suppression around the phase transition is observed. Both effects, the enhanced volume dependence and the temperature dependence agree with the situation in the quenched theory \cite{Cucchieri:2014nya,Maas:2011ez,Silva:2013maa,Silva:2017feh,Cyrol:2017qkl,Aouane:2011fv}.

This is in line with observations from 3-color QCD \cite{Aouane:2012bk}. Thus, the same pattern seems to emerge as in ordinary QCD. While not shown
explicitly, the hard modes of Matsubara frequency $n$ behave
essentially as the soft modes evaluated  at $(2\pi n T)^2+\vec p^2$,
as was already observed in the quenched case
\cite{Maas:2011ez,Silva:2017feh}. In the regime up to about $t=1$ fm
the Schwinger function did not show any significant changes with
temperature. At larger times the statistical noise precluded any
statements. The only exception is, as also shown in figure
\ref{fig:gpft}, that in the longitudinal case the Schwinger function
decays more slowly at the highest temperature, just as for the quenched case.

The soft mode ghost dressing function\footnote{The statistically not significant oscillatory behavior at $T=0.1T_c$ is an artifact of using a point-source for inversion, and would vanish with increasing statistics \cite{Cucchieri:2006tf}.}, also shown in figures \ref{fig:gpft} and \ref{fig:gpftr}, shows no qualitative deviation from the quenched case in that it is essentially not responding to temperature, except for some very slight overall suppression with increasing temperature \cite{Maas:2011ez,Cucchieri:2007ta,Fischer:2010fx}. The same also applies again to the not-shown hard modes, which can be described approximately as for the gluonic hard modes.

\begin{figure}
\includegraphics[width=0.5\textwidth]{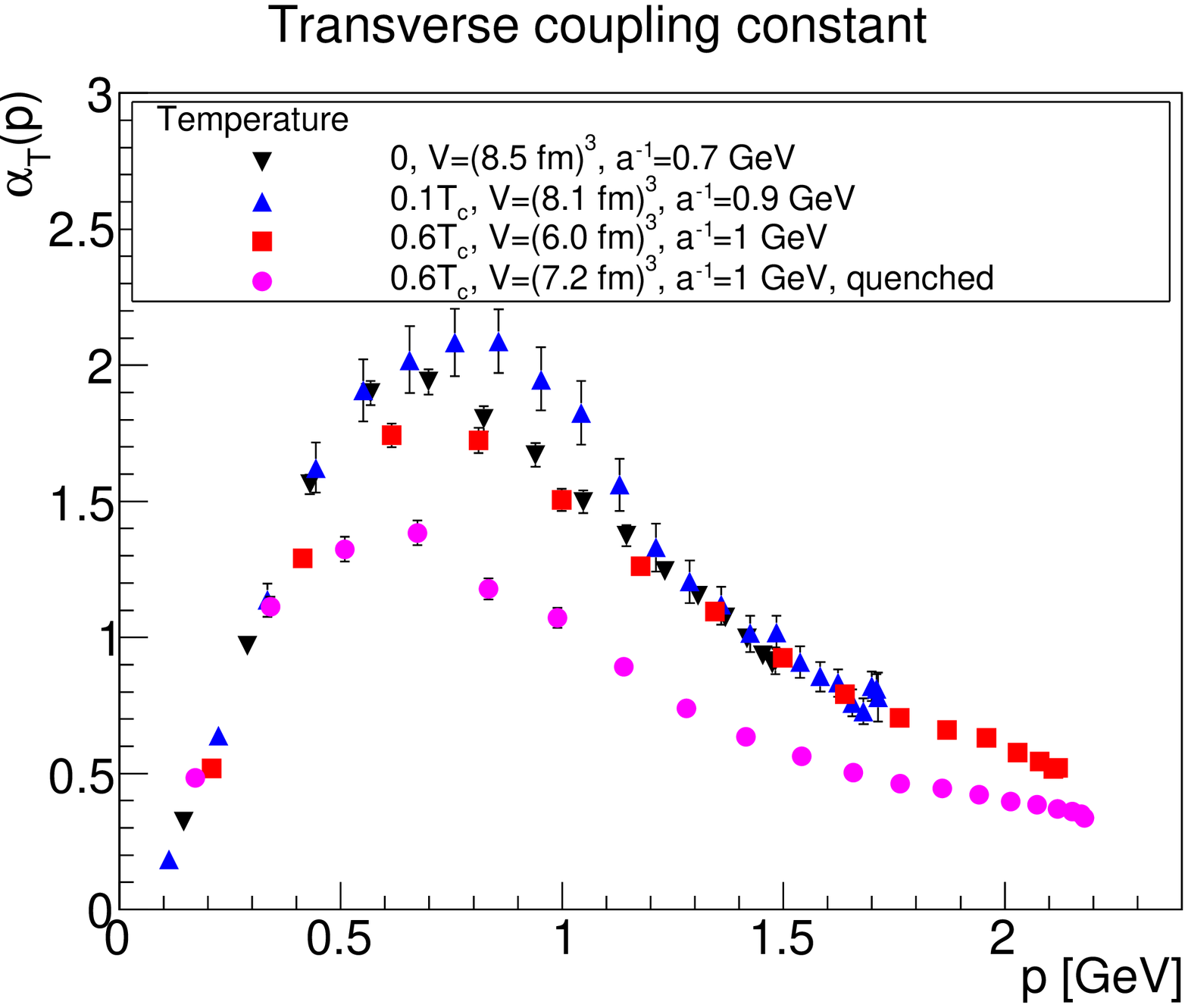}\includegraphics[width=0.5\textwidth]{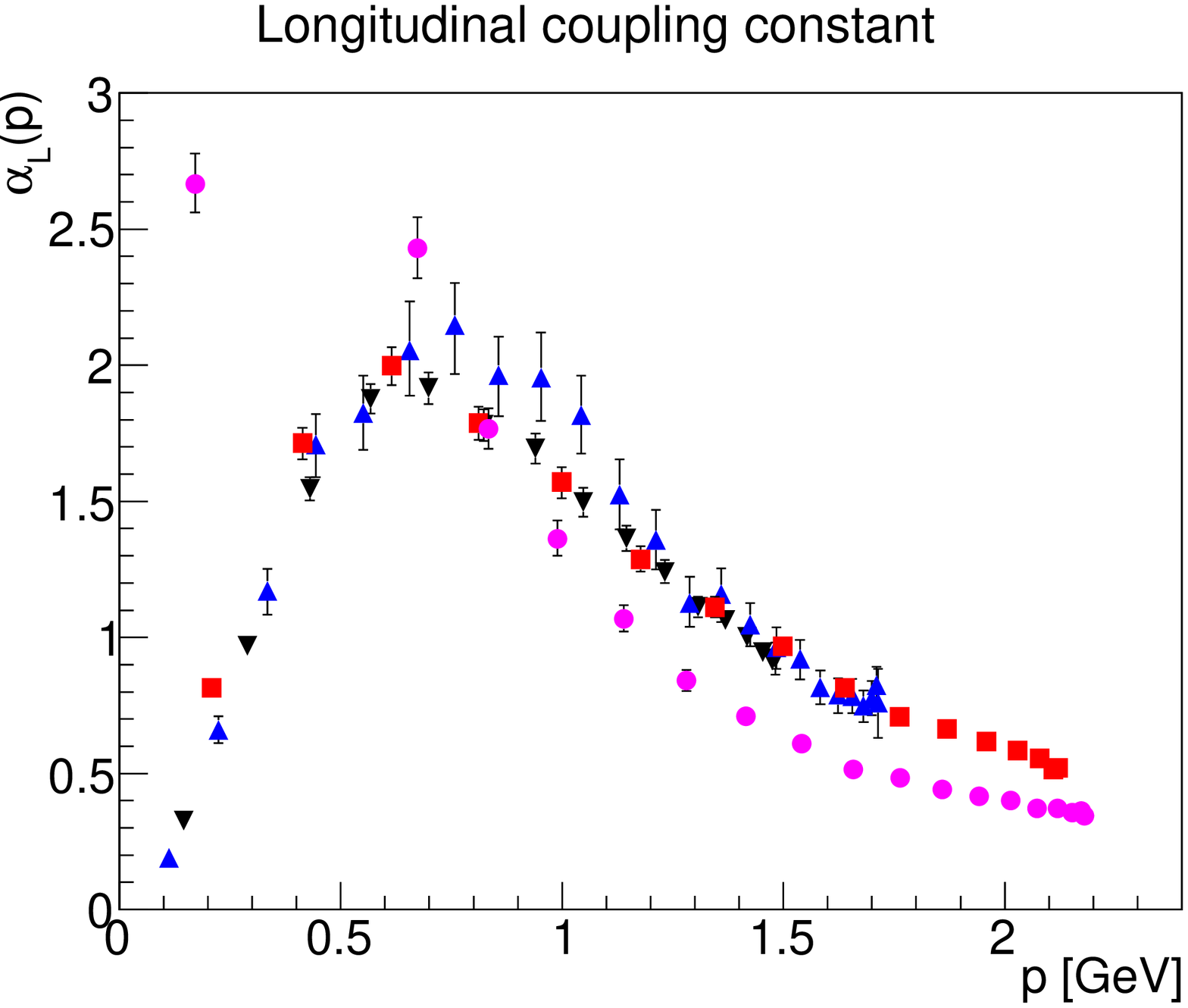}
\caption{\label{fig:alphaft}The transverse (left panel) and longitudinal (right panel) running coupling at finite temperature. Quenched results are from \cite{Maas:2011ez}. Note that the lowest momentum point of the longitudinal coupling in the quenched case is likely strongly affected by finite-volume and discretization effects \cite{Cucchieri:2014nya,Maas:2011ez}.}
\end{figure}

As described in equations \prefr{alphat}{alphal} the running coupling also splits into a magnetic and an electric one. The results are shown in figure \ref{fig:alphaft} and \ref{fig:gpftr}. The results are essentially identical, at low temperatures. At high temperatures the almost unaffected ghost propagator in combination with the strong suppression of the gluon propagator induces a suppression of the running coupling, which is much stronger for the electric one.

\begin{figure}
\includegraphics[width=\textwidth]{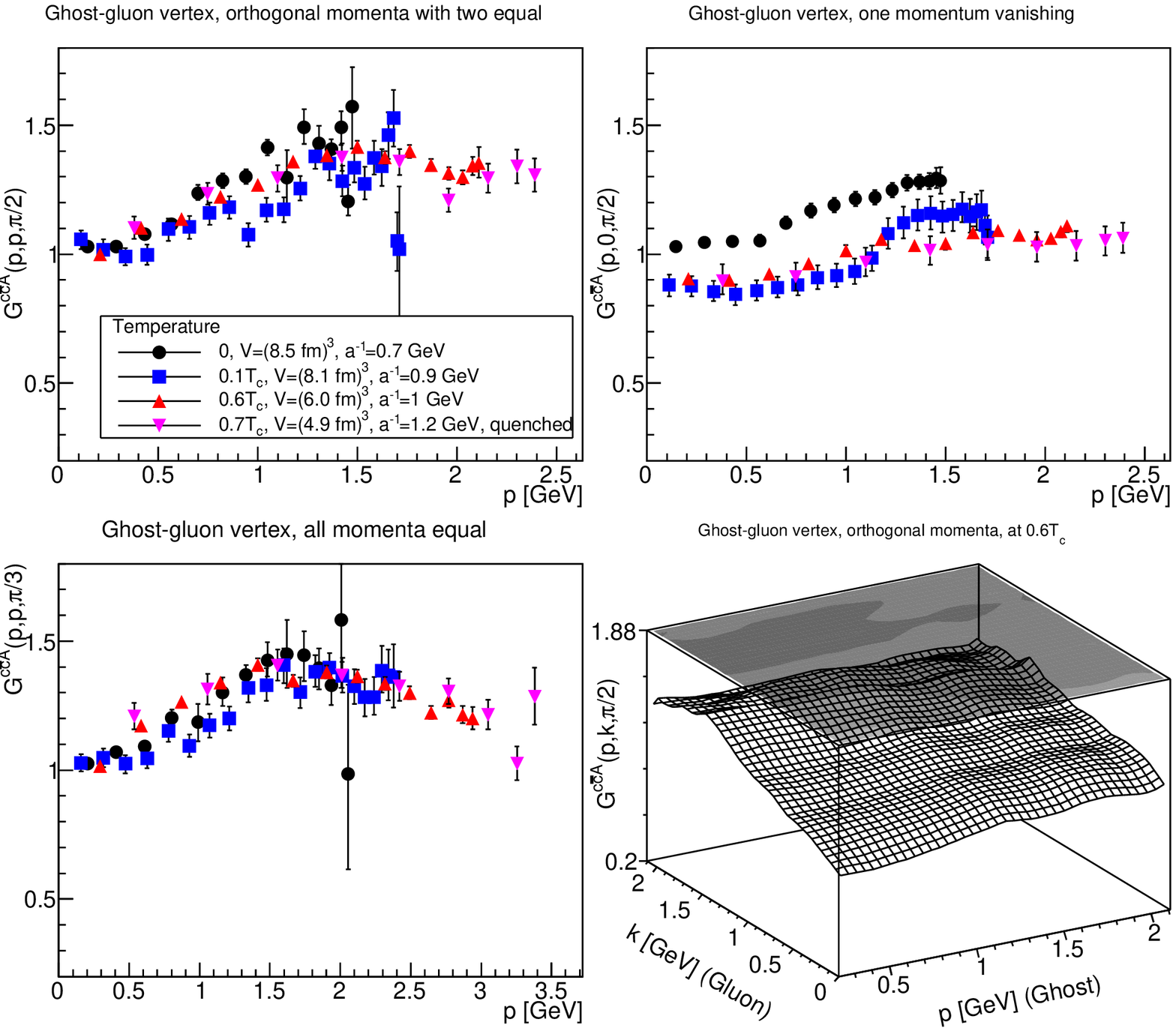}\\
\includegraphics[width=\textwidth]{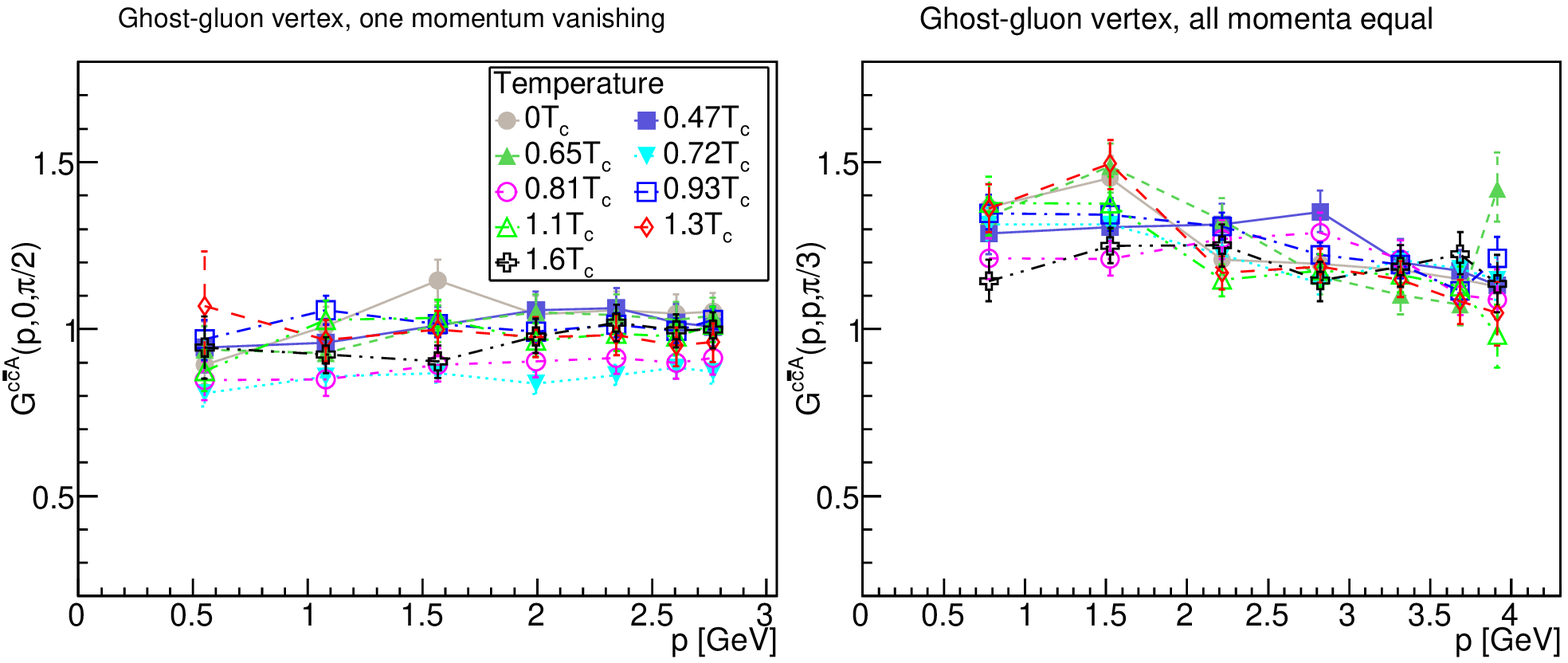}
\caption{\label{fig:ft:ggv}The ghost-gluon vertex dressing for different momentum configurations at finite temperature in comparison to quenched data from \cite{Fister:2014bpa}. The lower two panels contain the results at $\beta=2.1$. See text for details. Results have not been renormalized.}
\end{figure}

The results for the soft magnetic ghost-gluon vertex are shown in figure \ref{fig:ft:ggv}. The results show essentially no temperature-dependence, as in the quenched case \cite{Fister:2014bpa}. The only visible effect seems to be at vanishing gluon momentum, and then again in the same way as for the quenched case. However, this may actually be a finite-volume effect \cite{Maas:unpublished}, and should therefore not be overstated. This is especially seen in the $\beta=2.1$ case, where at fixed spatial volume no such effect occurs. Thus, also at finite temperature in the unquenched case this vertex is almost tree-level.

\begin{figure}
\includegraphics[width=\textwidth]{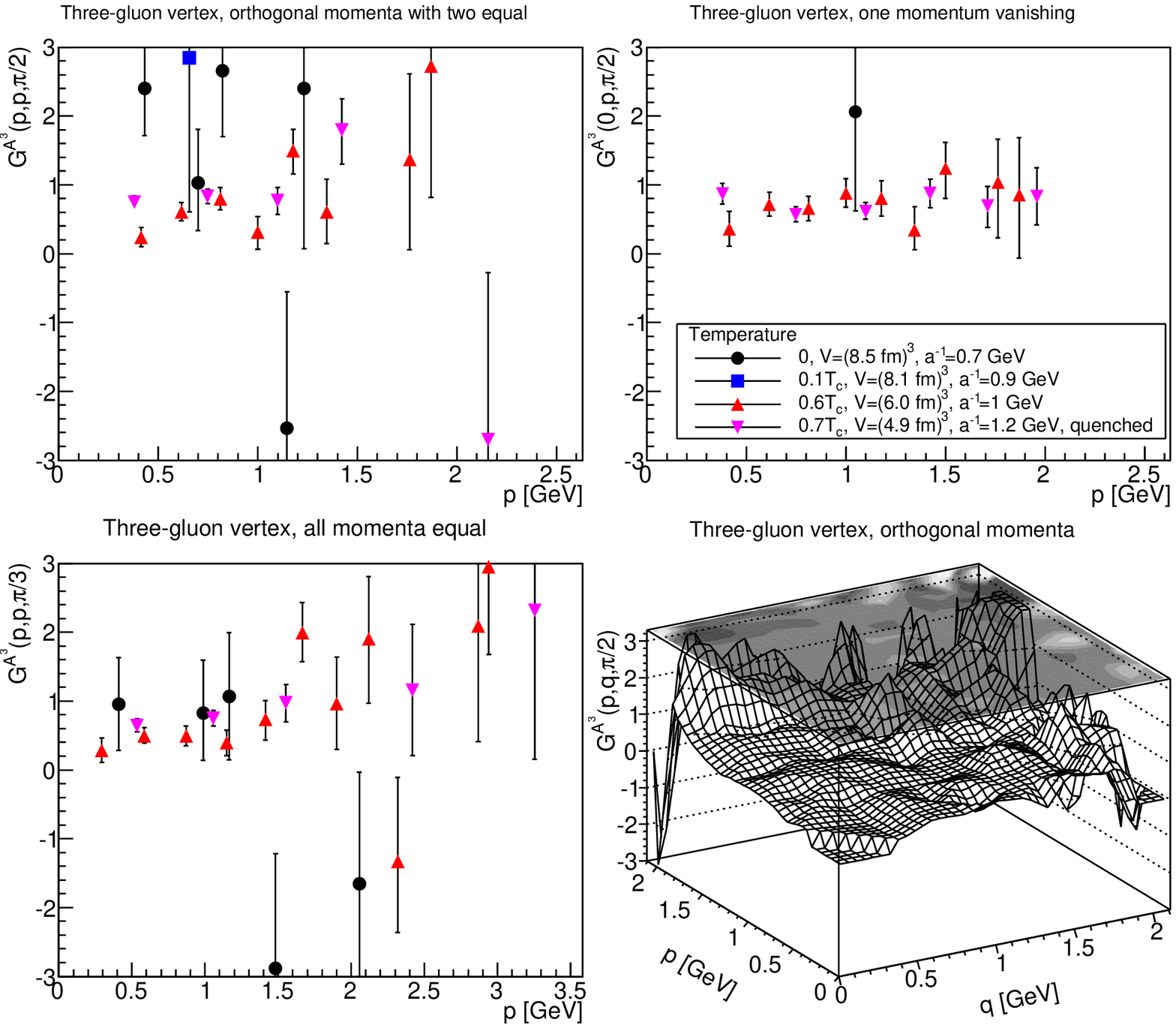}\\
\includegraphics[width=\textwidth]{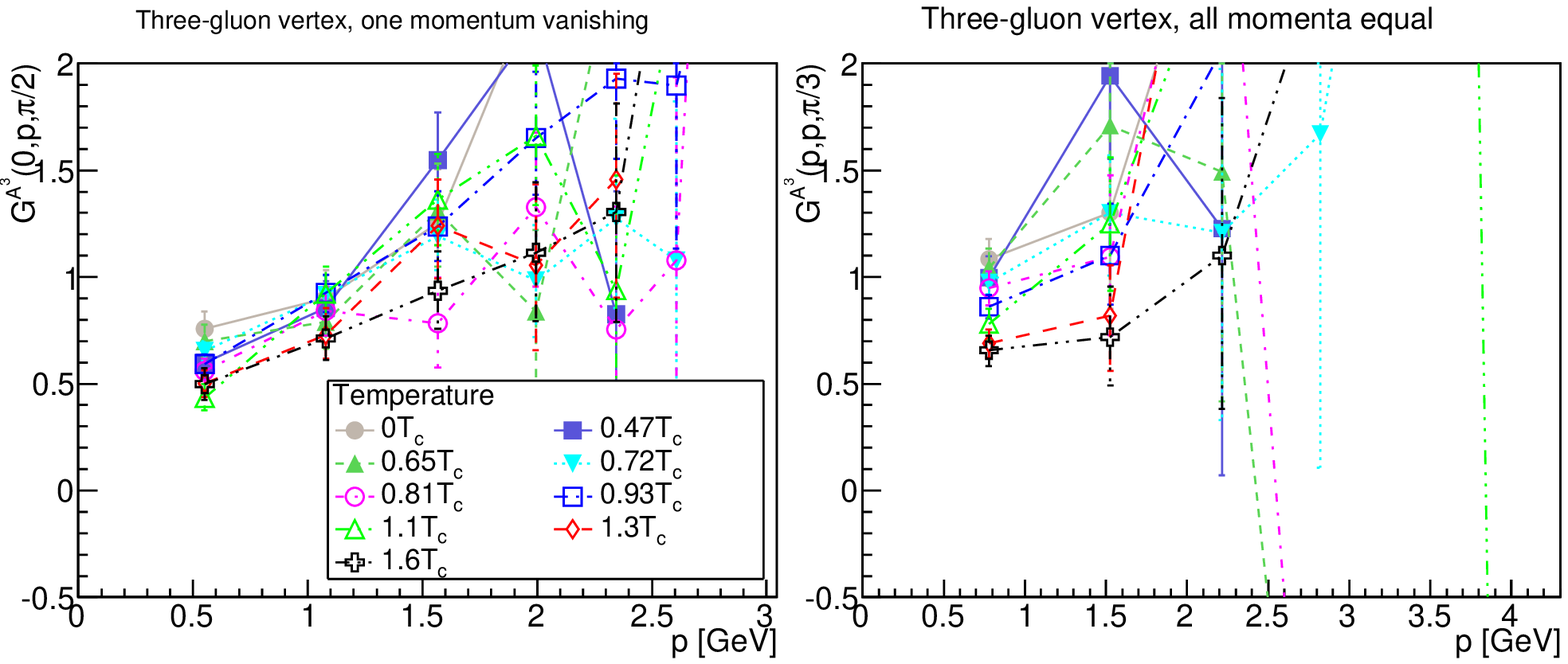}
\caption{\label{fig:ft:g3v}The three-gluon vertex dressing for different momentum configurations at finite temperature in comparison to quenched data from \cite{Fister:2014bpa}. The lower two panels contain the results at $\beta=2.1$. See text for details. Results have not been renormalized.}
\end{figure}

Because of the much larger statistical noise for the three-gluon
vertex \cite{Cucchieri:2006tf} its results, shown in figure \ref{fig:ft:g3v}, are much less conclusive. Essentially, no results
with reasonable statistical errors have been obtained at
$0.1T_c$. However, the results at $0.6T_c$ are reasonable, and again
rather close to the quenched case. In particular, they differ only weakly from zero temperature, and thus the three-gluon vertex is also in the unquenched case not substantially affected by low temperatures. In the case at $\beta=2.1$ a slight suppression above the phase transition is observed, which is in line with the quenched case \cite{Fister:2014bpa}. However, this lattice setting does not probe far enough into the infrared to be also sensitive to the substantial changes seen for this vertex in a narrow temperature interval around the phase transition, where it changes sign \cite{Fister:2014bpa}.

\section{Finite density results}\label{s:res}

As is discussed in appendix \ref{s:sys} the diquark source seems to have no statistically significant effect, while the volume has a slight effect. Thus, only volume effects will be discussed here. Note also that the gluon propagator has been investigated in detail already in \cite{Boz:2013rca} for the cases with $\beta\le 1.9$. With respect to these results we present them here for completeness, as they enter crucially both the running coupling and the three-gluon vertex. We also checked that the results in \cite{Boz:2013rca} coincide with the ones presented here, as both have been calculated using different numerical codes.

\begin{figure}
\includegraphics[width=\textwidth]{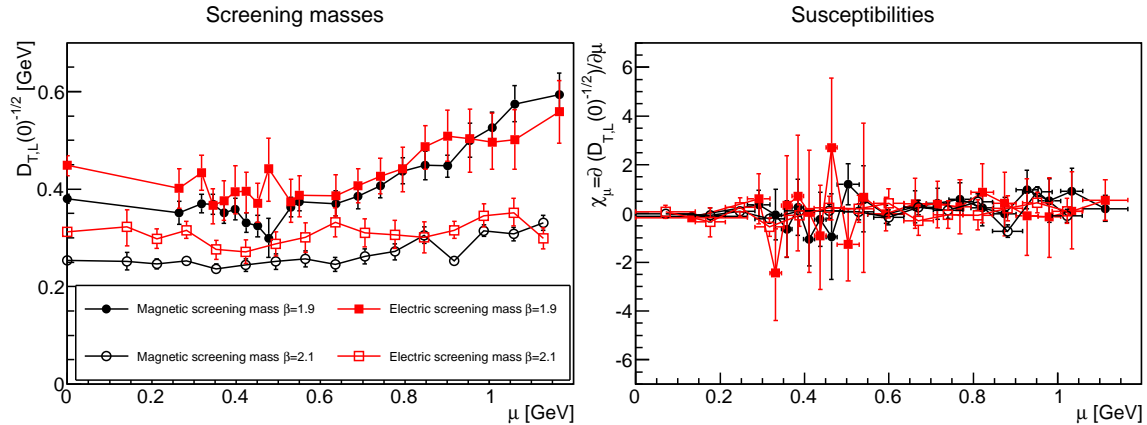}
\caption{\label{fig:sm}The dependence of the screening masses (left panel) and susceptibilities (right panel) with density at fixed spatial physical volume of (2.2 fm)$^3$ for $\beta=1.9$ and $\beta=2.1$ and
 temporal lattice extent 24 and 32, respectively. Results are not renormalized.}
\end{figure}

The first result is the development of the screening masses with
density, which is shown\footnote{Note that electric and magnetic
  screening masses do not coincide at zero chemical potential as the
  lattice is asymmetric and elongated in time direction. If a
  symmetric lattice is used they coincide, as do the propagators for
  all momenta.} in figure \ref{fig:sm}. In accordance with previous
investigations at $\beta\le 1.9$ \cite{Boz:2013rca,Cotter:2012mb}, no
pronounced change is seen, except for a slow increase after a
transition at $\mu\approx 750$ MeV, especially in the magnetic
sector. In contrast, at $\beta=2.1$ no such increase is
seen\footnote{In fact, at the two-$\sigma$ level there is still a
  systematic increase visible for the magnetic screening mass only,
  see figure \ref{fig:mj} in appendix \ref{a:sysj}. But there is no
  statistically reliable effect anymore.}. However, also no abrupt
change is seen at the silver-blaze point at about $\mu\approx 375$
MeV, which is a phase transition. The latter is in marked contrast to
the finite-temperature transition, see section \ref{s:tres} and
\cite{Maas:2011ez}, where in particular in the high-temperature phase
magnetic and electric screening mass differ substantially, and at
least the electric one strongly depends on the temperature. Thus, the
absence of a signal at higher densities should not be taken as an
indication that no phase change or transition takes place. However,
the absence of such a transition would be consistent with the
observations in  \cite{Astrakhantsev:2018uzd,Bornyakov:2017txe}. Thus,
the gluon propagator seems to show no density dependence. But because
it also does not react to the phase transition at the silver-blaze
point this cannot be taken as an indication of the absence of a phase
transition itself. It is in itself remarkable that a discontinuity in
the free energy is not also inducing a discontinuity in the
gauge-fixed correlation functions. This implies that they cannot be
used to determine the phase structure of a theory reliably on their own. The alternative is, of course, that the critical region becomes so narrow at $\beta=2.1$ that our spacing in $\mu$ is not sensitive to the transition. However, the absence of a trend at large chemical potentials in comparison to the $\beta=1.9$ case makes this interpretation unlikely.

\afterpage{\clearpage}

\begin{figure}
\includegraphics[width=0.5\textwidth]{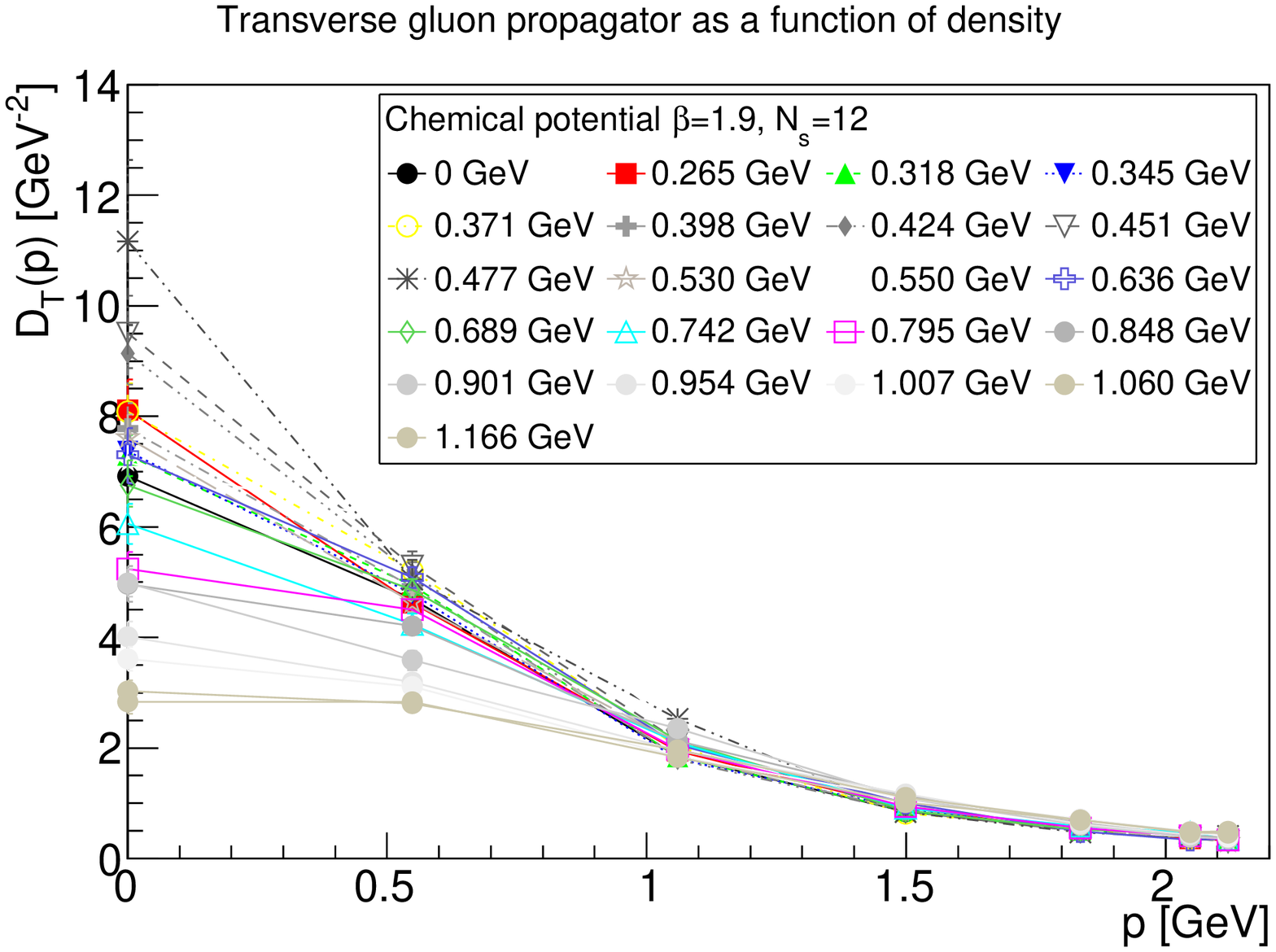}\includegraphics[width=0.5\textwidth]{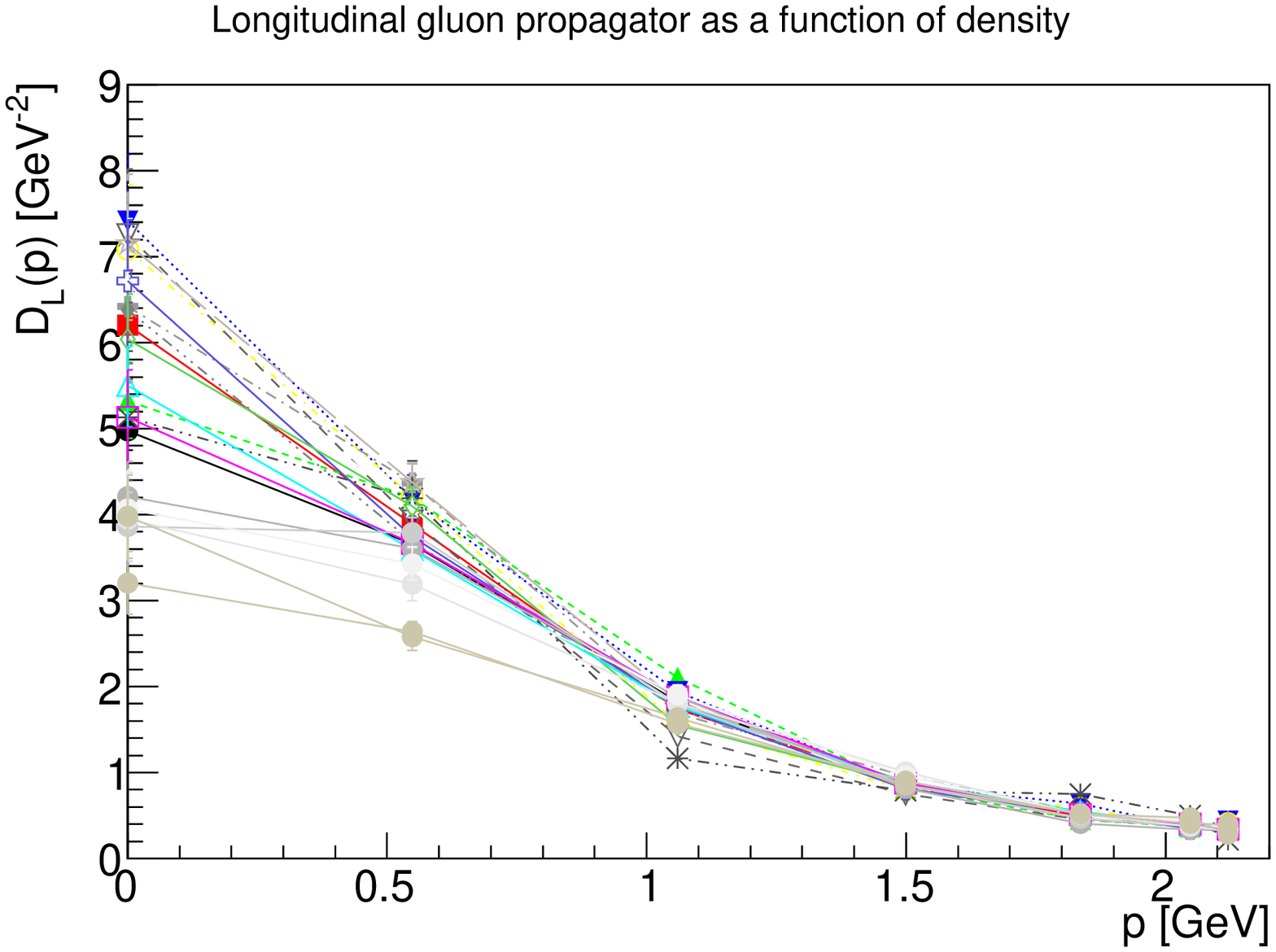}\\
\includegraphics[width=0.5\textwidth]{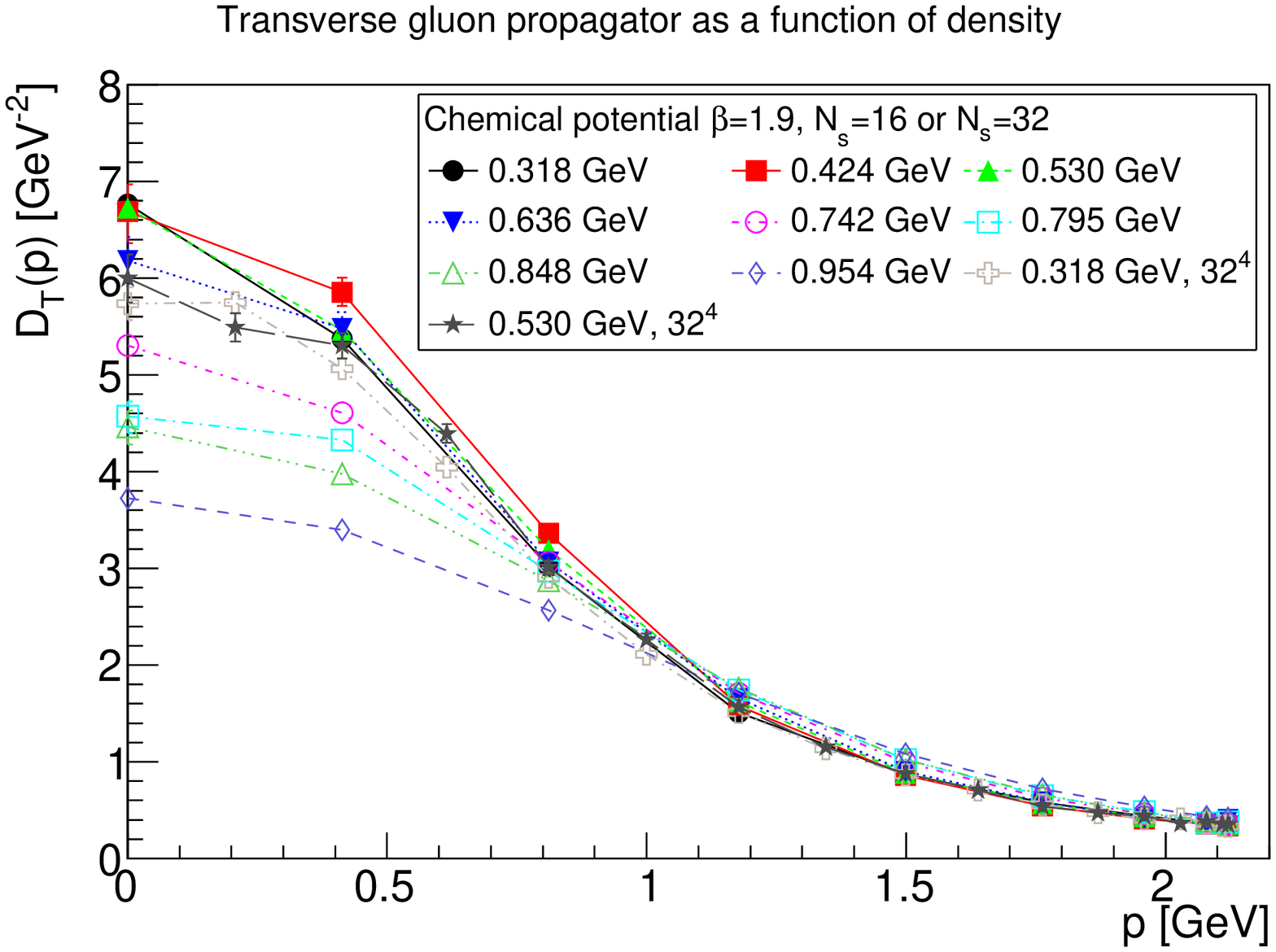}\includegraphics[width=0.5\textwidth]{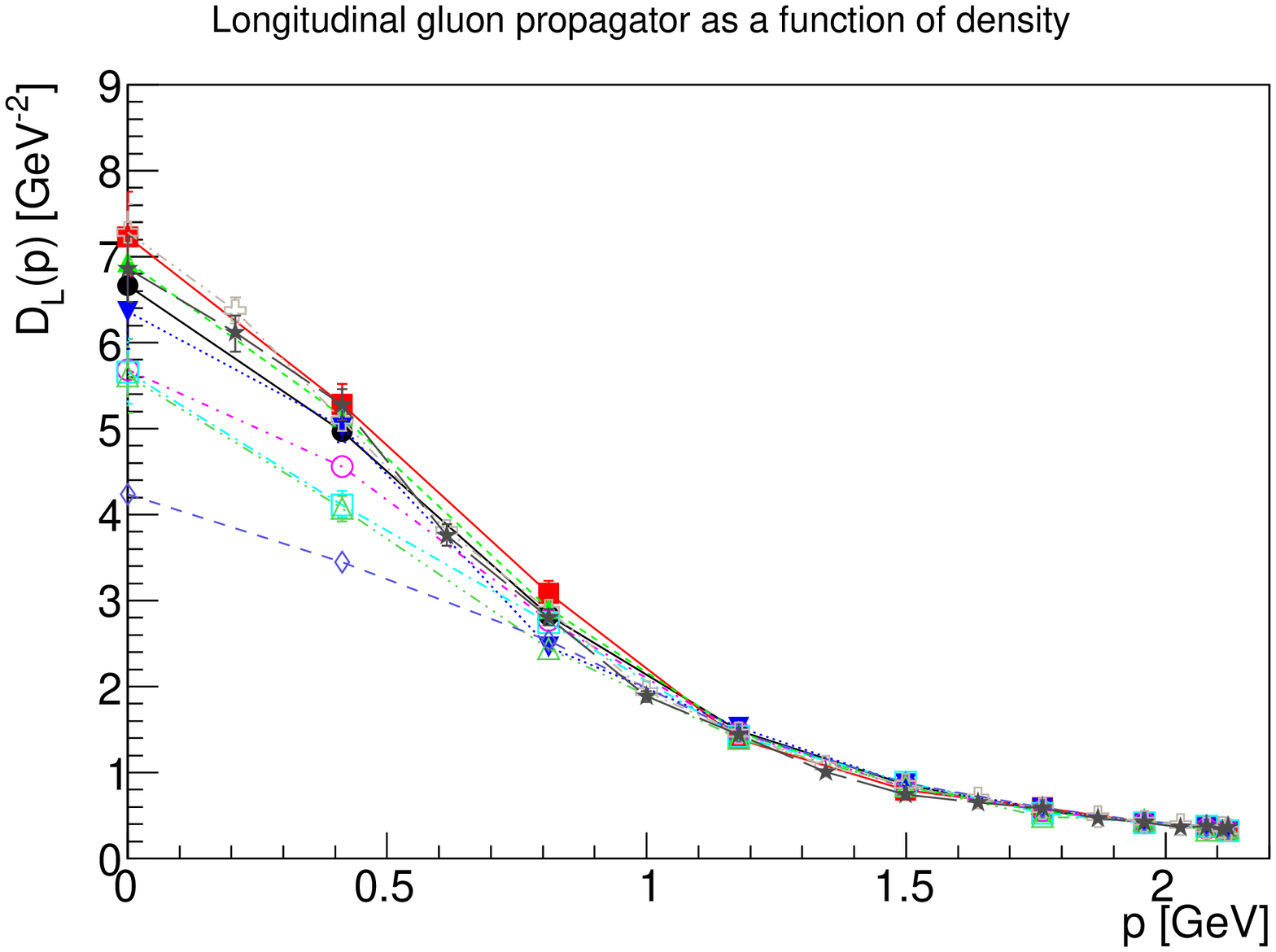}\\
\includegraphics[width=0.5\textwidth]{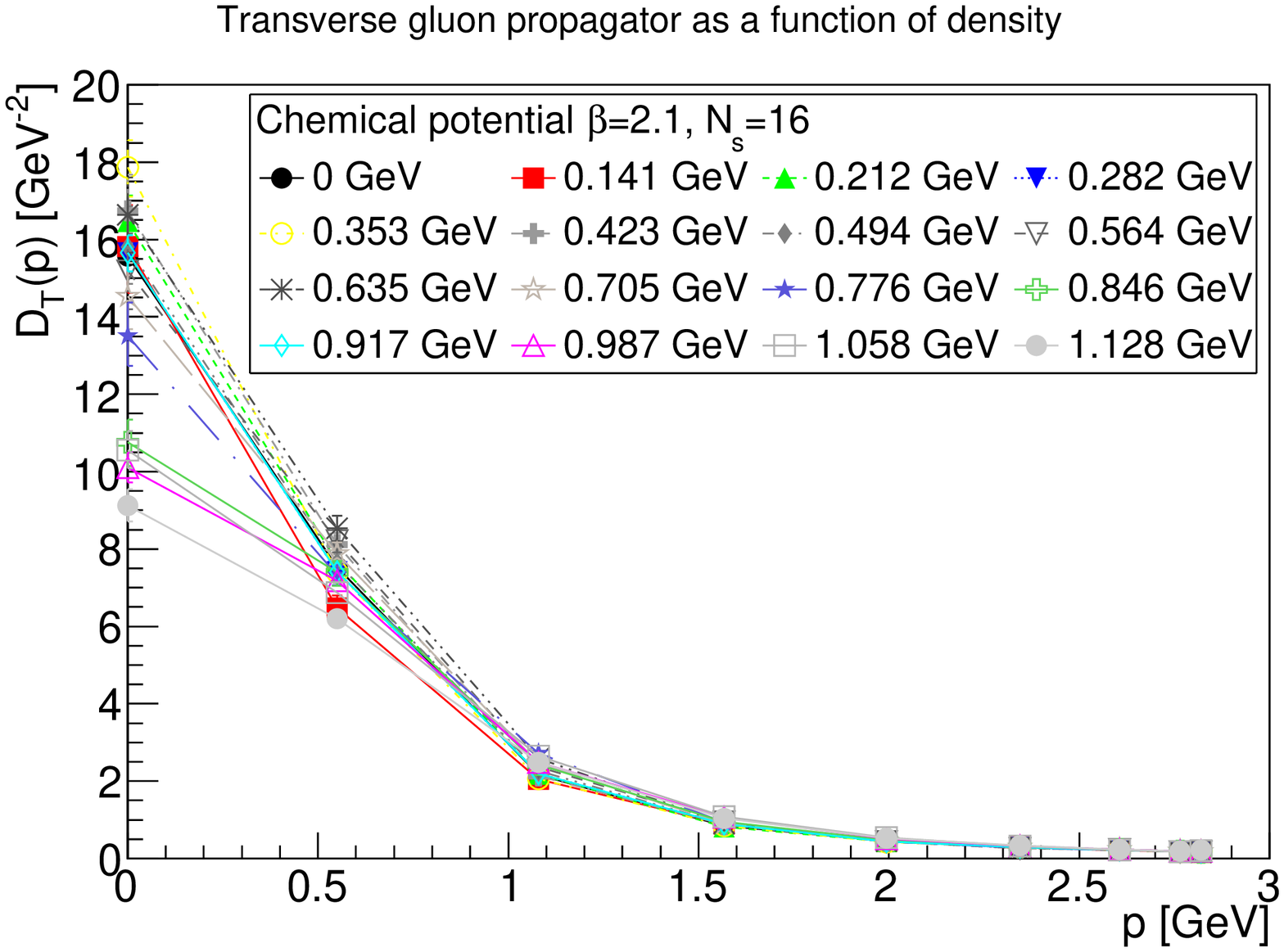}\includegraphics[width=0.5\textwidth]{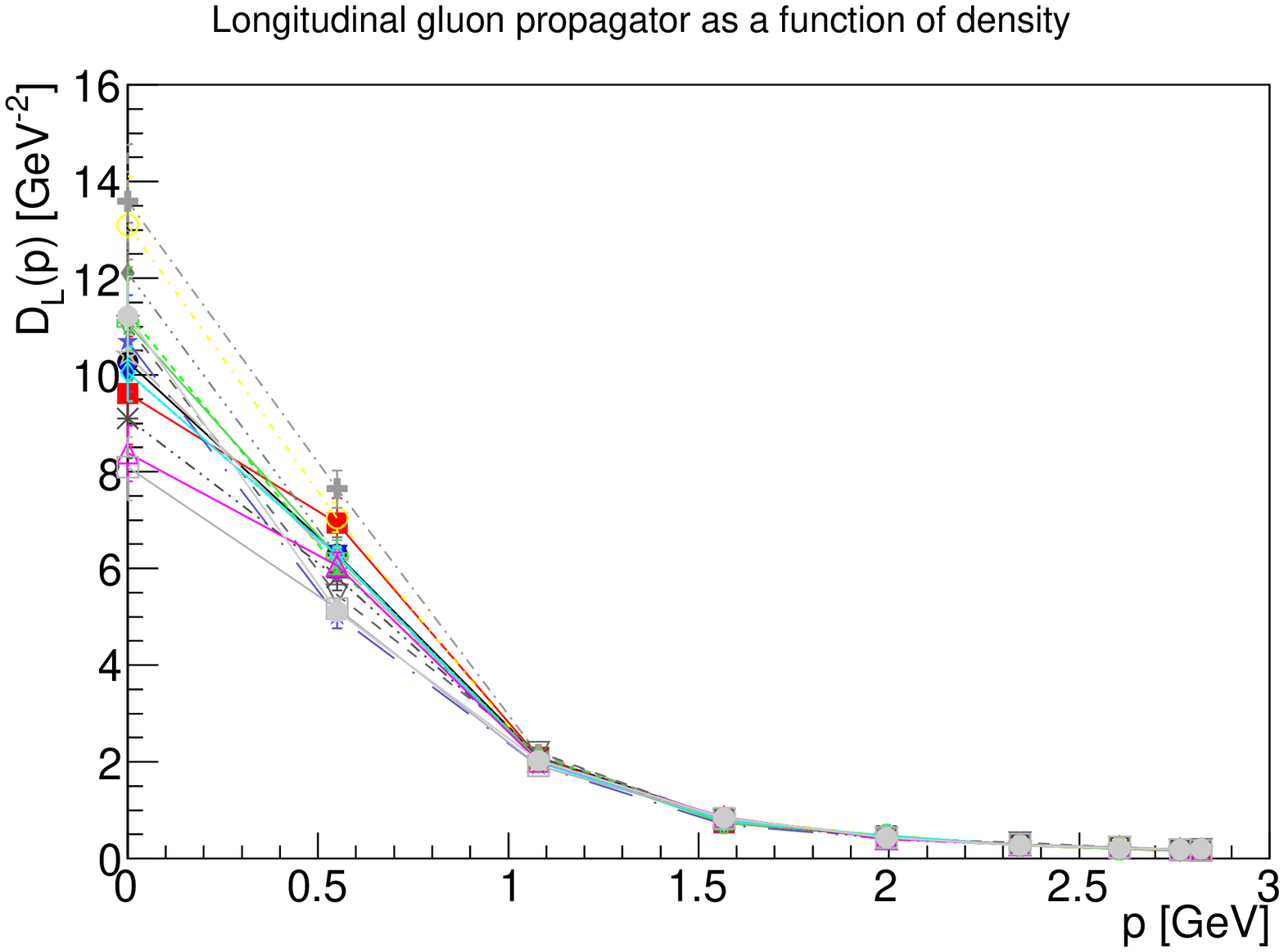}
\caption{\label{fig:gpmu}The dependence of the transverse (left panels) and longitudinal (right panels) gluon propagator on the chemical potential at $\beta=1.9$ and fixed volume $24\times 12^3$ (top panels) and $24\times 16^3$ or $32^4$ (middle panels) and at $\beta=2.1$ and $32\times 16^3$ (bottom panels).}
\end{figure}

\afterpage{\clearpage}

The same pattern is repeated in the full momentum dependence shown in figure \ref{fig:gpmu}. The only exception is a  dip in the magnetic screening mass around $\mu\approx 450$ MeV at $\beta=1.9$, which creates an infrared enhancement for the magnetic propagator. Note that for larger volumes, see appendix \ref{a:sysv}, no trend towards such a dip is observed, and neither is this the case at $\beta=2.1$. Hence, this is likely a statistical fluctuation and/or a lattice artifact. At finite momenta there is no discernible trend with chemical potential visible, except for the infrared suppression due to the increase in screening mass on the coarser lattice at $\beta=1.9$. This is also emphasized on the larger volume at fixed $\beta$ and the finer lattices at fixed spatial volume, where the evolution is smoother and essentially independent of chemical potential. Again, there is no visible difference in the transverse and longitudinal sector.

Also due to the limited statistics, it is only possible to state regarding the Schwinger function that it does not substantially change before the zero crossing, which appears to remain at all densities, and at about the same time scale of 1 fm.

\begin{figure}[ht!]
\includegraphics[width=0.5\textwidth]{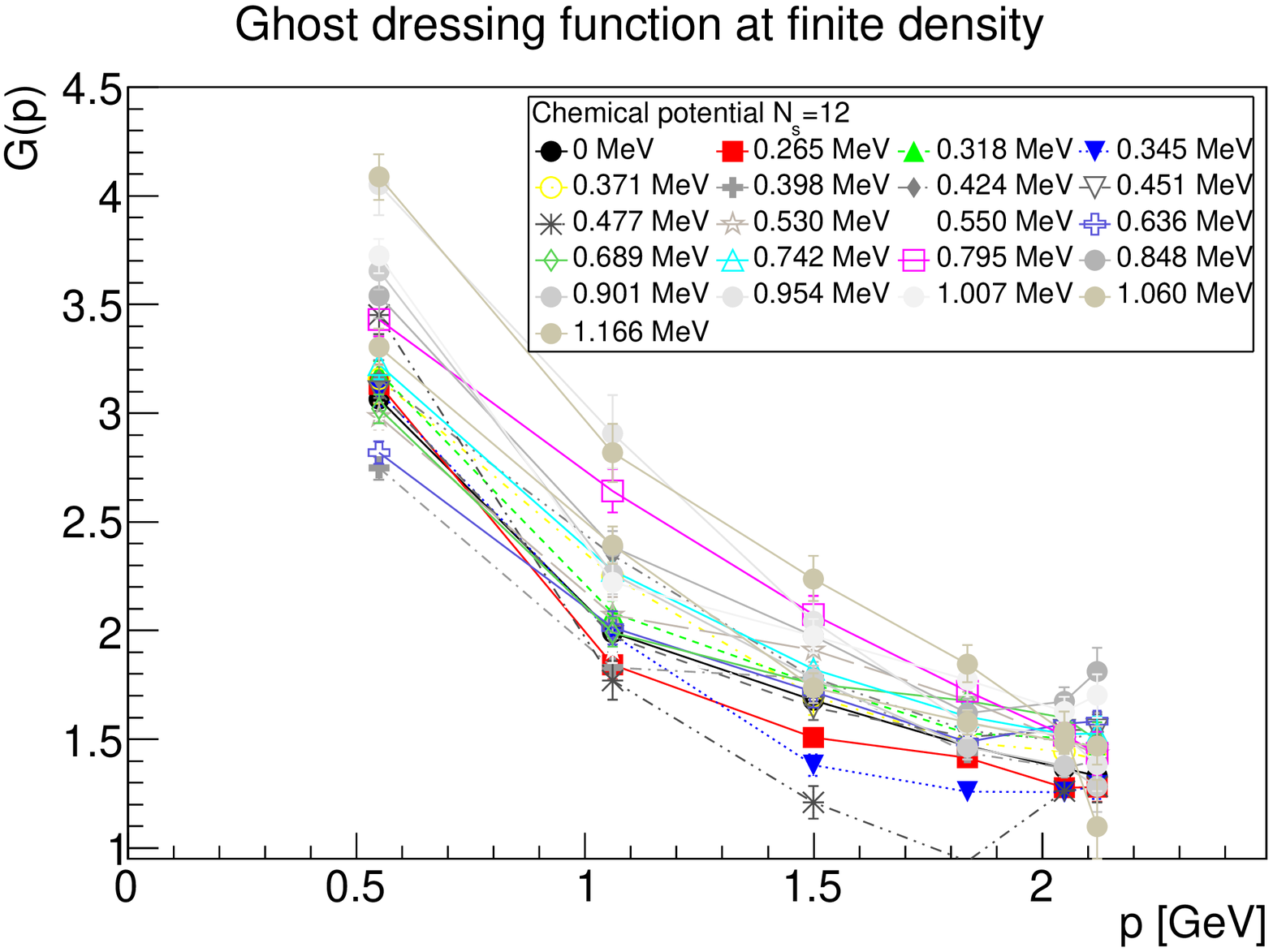}\includegraphics[width=0.5\textwidth]{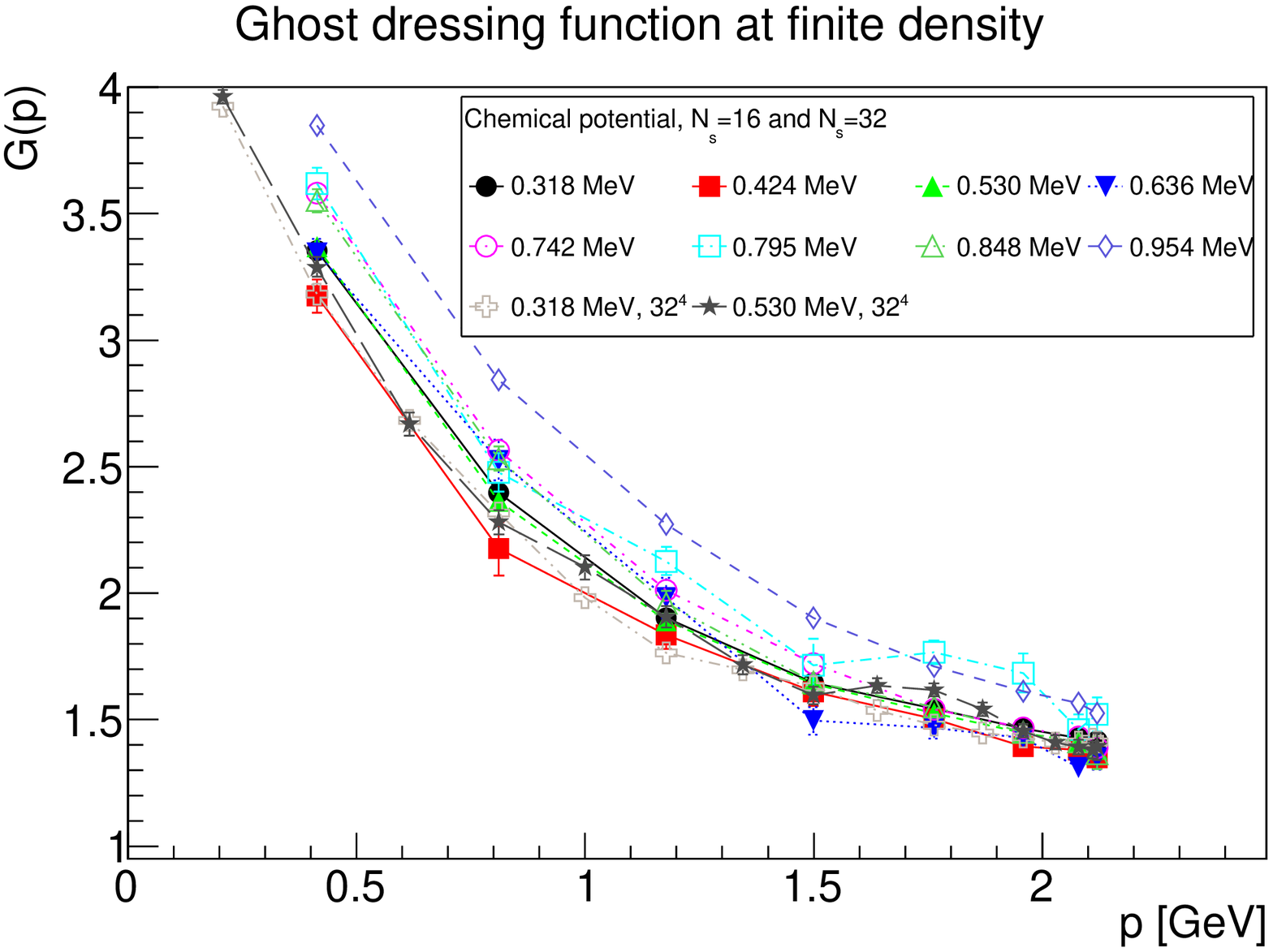}\\
\begin{center}
 \includegraphics[width=0.5\textwidth]{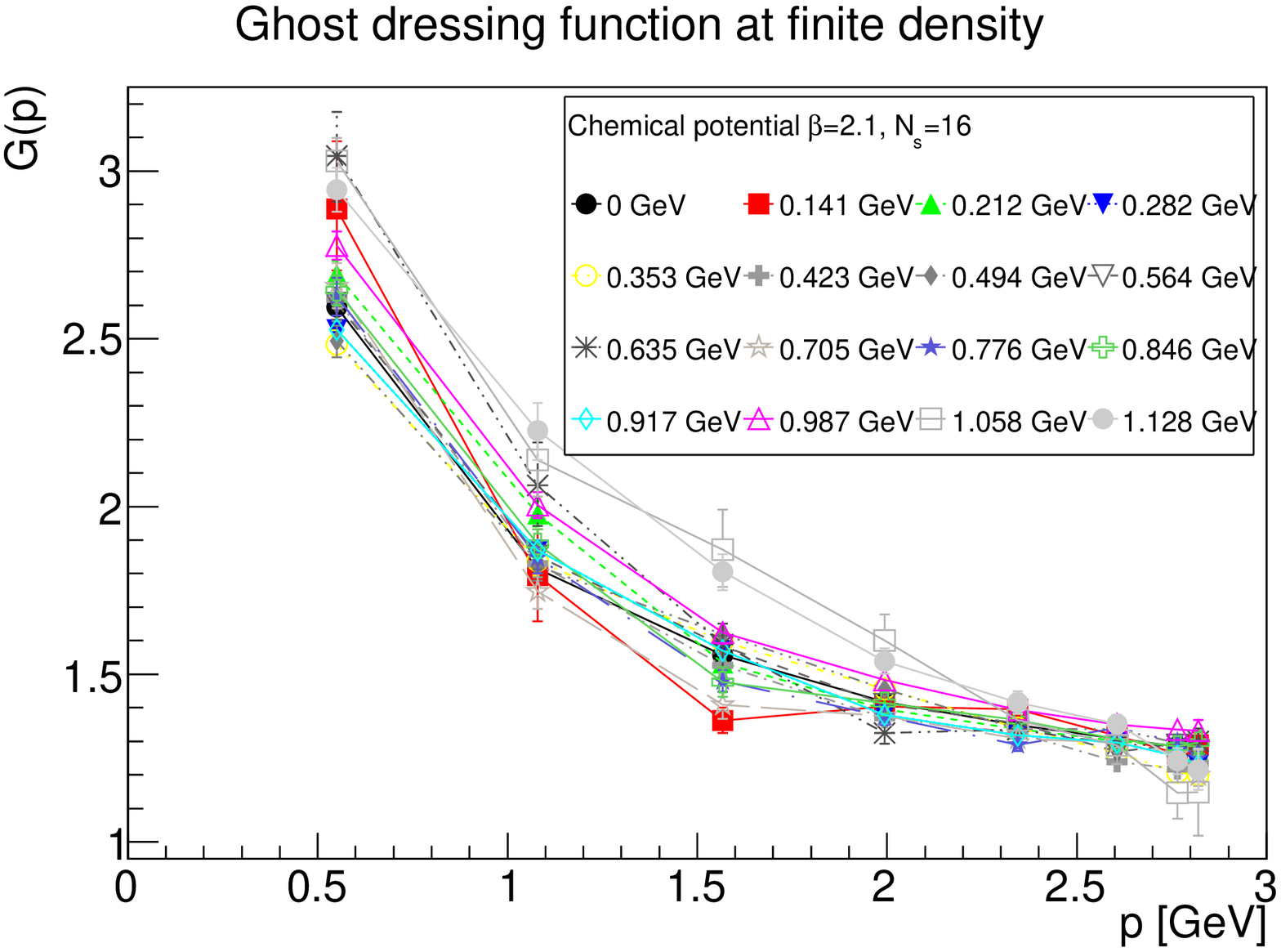}
\end{center}
\caption{\label{fig:ghpmu}The dependence of the ghost dressing function on the chemical potential at fixed volume $24\times 12^3$ at $\beta=1.9$ (top-left panel), at $24\times 16^3$ or $32^4$ at $\beta=1.9$ (top-right panel), and at $32\times 16^3$ at $\beta=2.1$ (bottom panel).}
\end{figure}

\afterpage{\clearpage}

The results for the ghost dressing function are shown in figure
\ref{fig:ghpmu}. Overall, the dressing function is almost unaffected
by the chemical potential. However there seems to be a slight trend at $\beta=1.9$ at low momenta that the dressing function becomes somewhat steeper at larger chemical potentials, but the effect is smaller when increasing the volume from $N_s=12$ to $N_s=16$, and is also not visible on the finer lattices at $\beta=2.1$ except for $\mu>1$ GeV. Hence, this may move to even larger chemical potentials on even finer lattices, if this is a lattice artifact. This would fit with the effect for the gluon propagator, where the effect also vanishes, or at best is moved towards much larger chemical potentials when making the lattice finer. Thus, if there is any effect, substantially more systematics will be needed to establish it.

\begin{figure}
\includegraphics[width=0.5\textwidth]{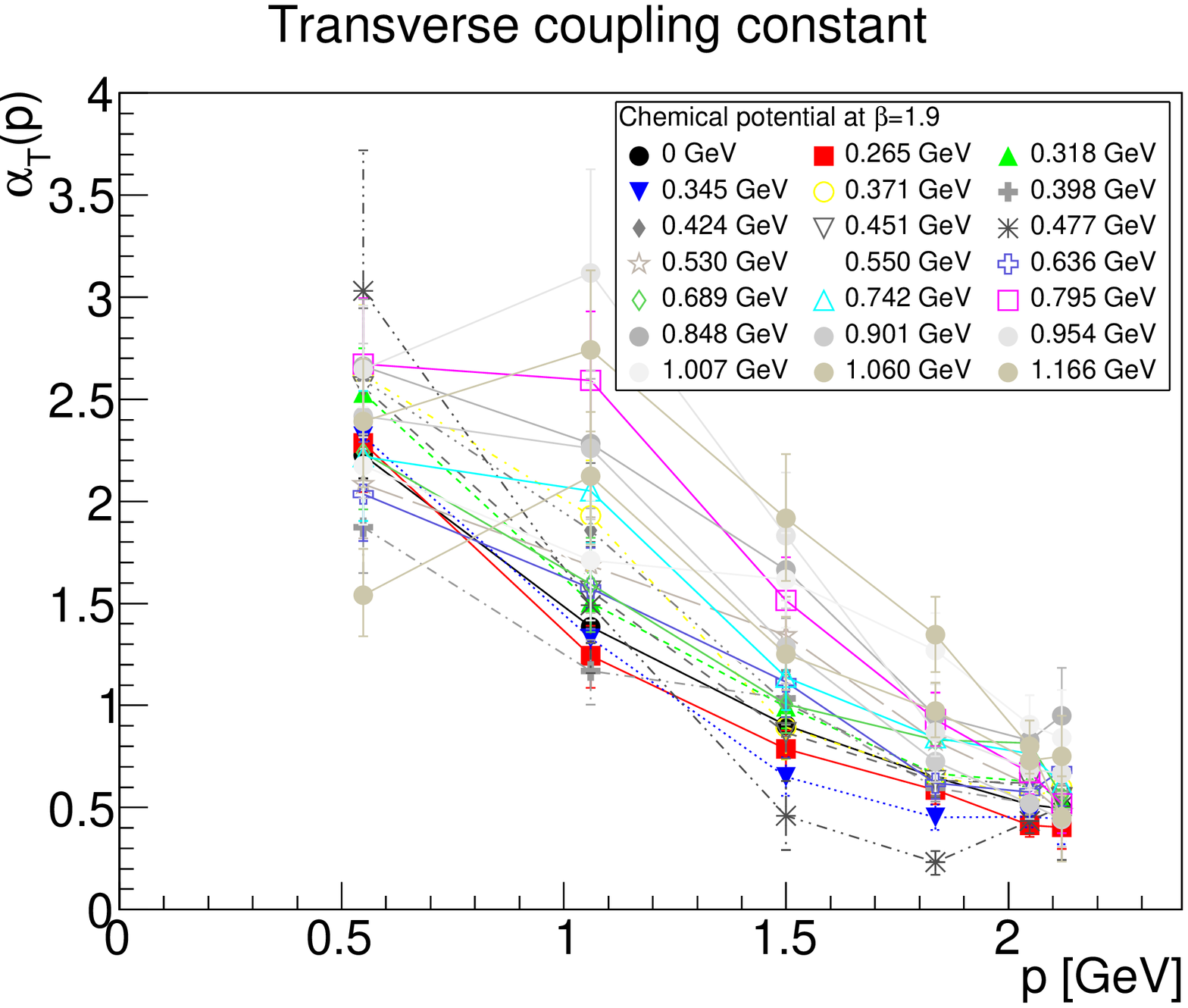}\includegraphics[width=0.5\textwidth]{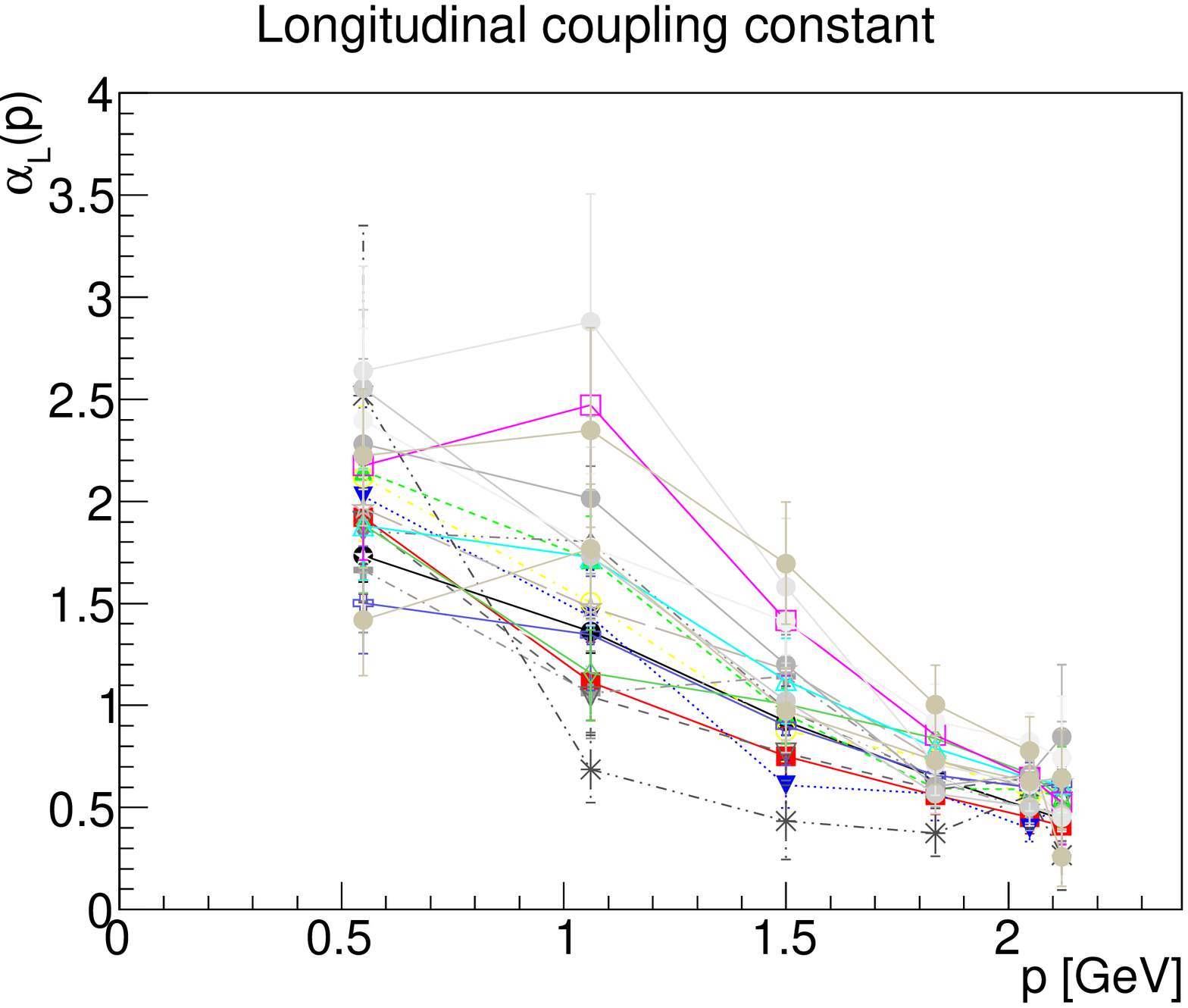}\\
\includegraphics[width=0.5\textwidth]{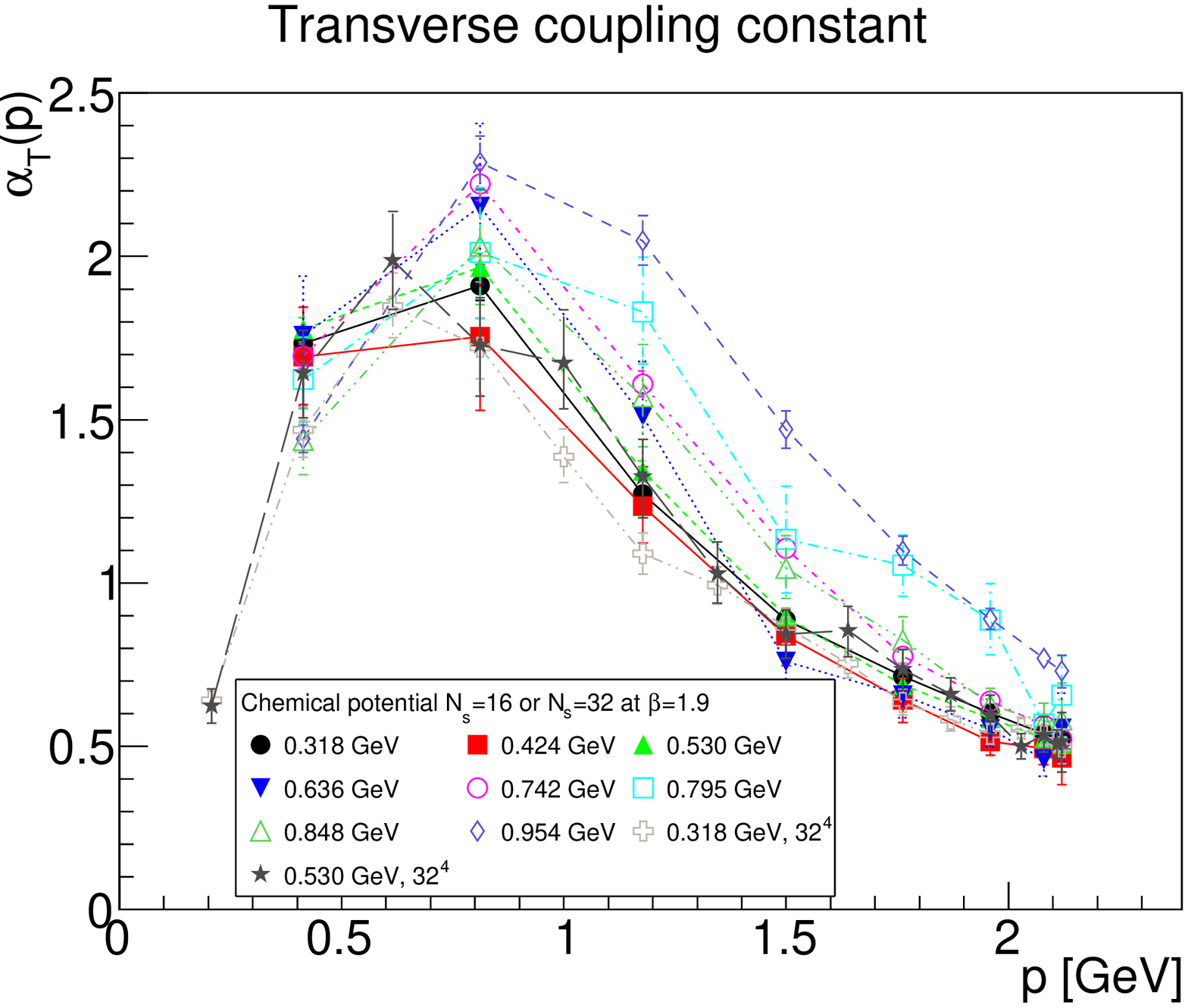}\includegraphics[width=0.5\textwidth]{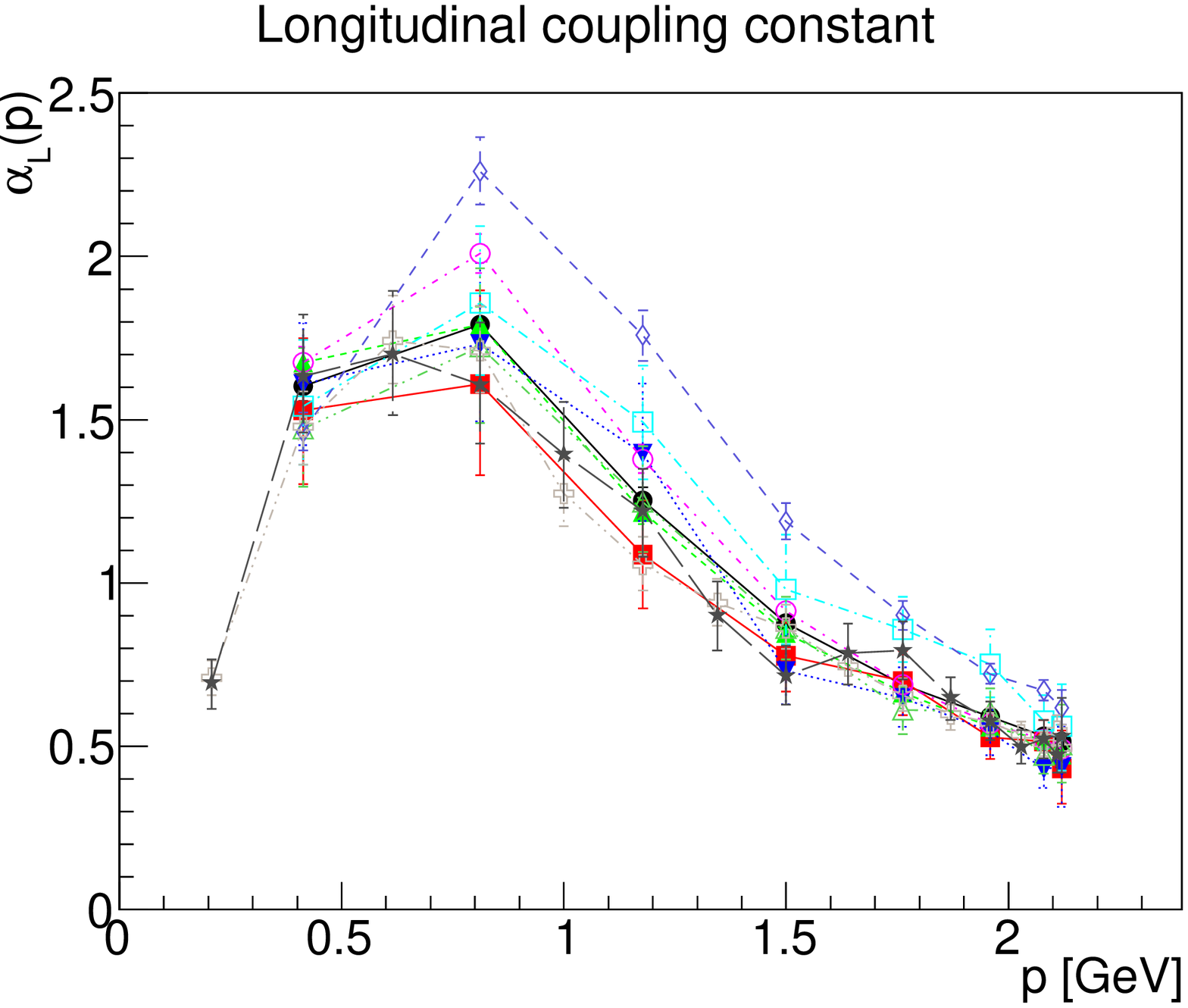}\\
\includegraphics[width=0.5\textwidth]{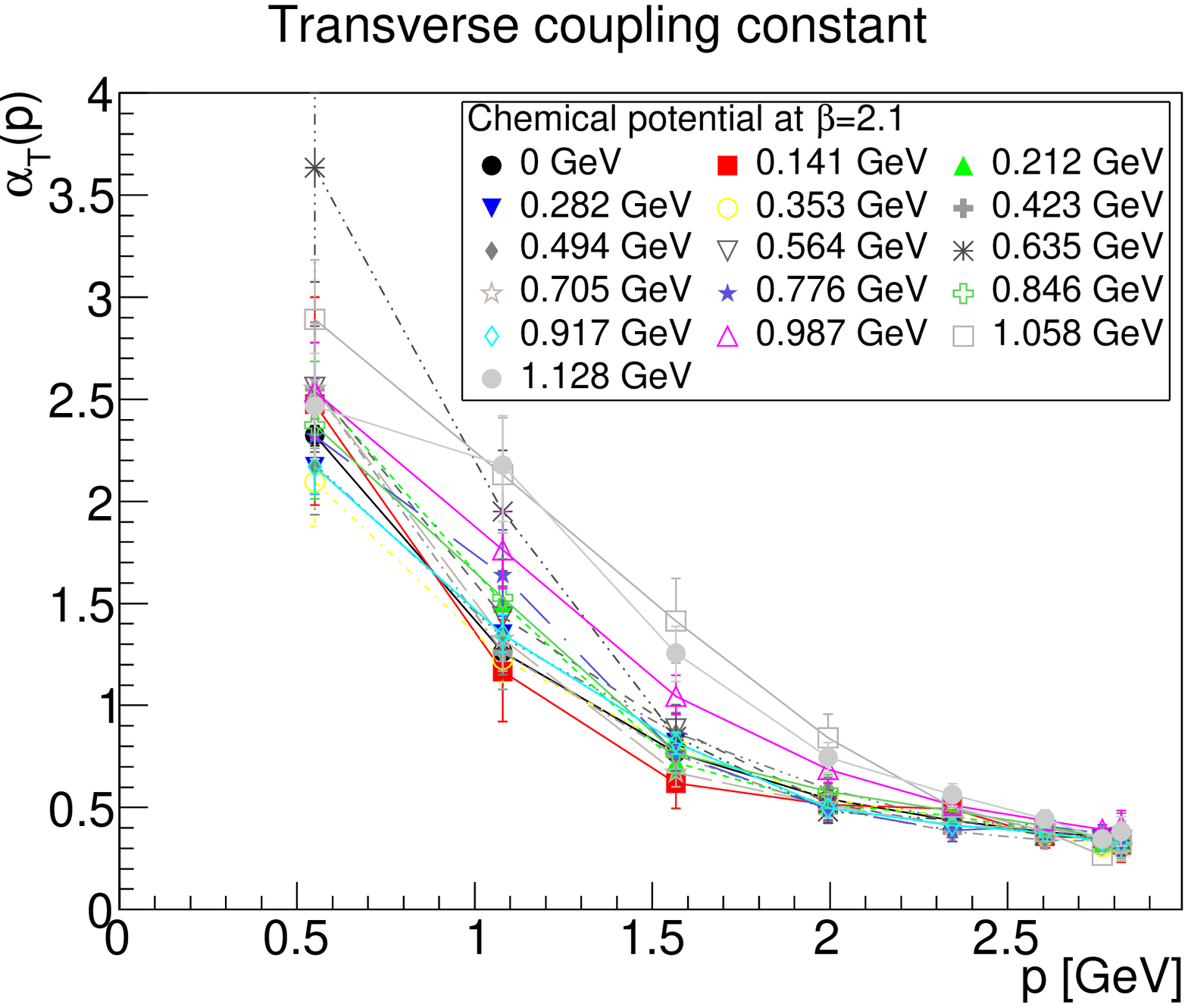}\includegraphics[width=0.5\textwidth]{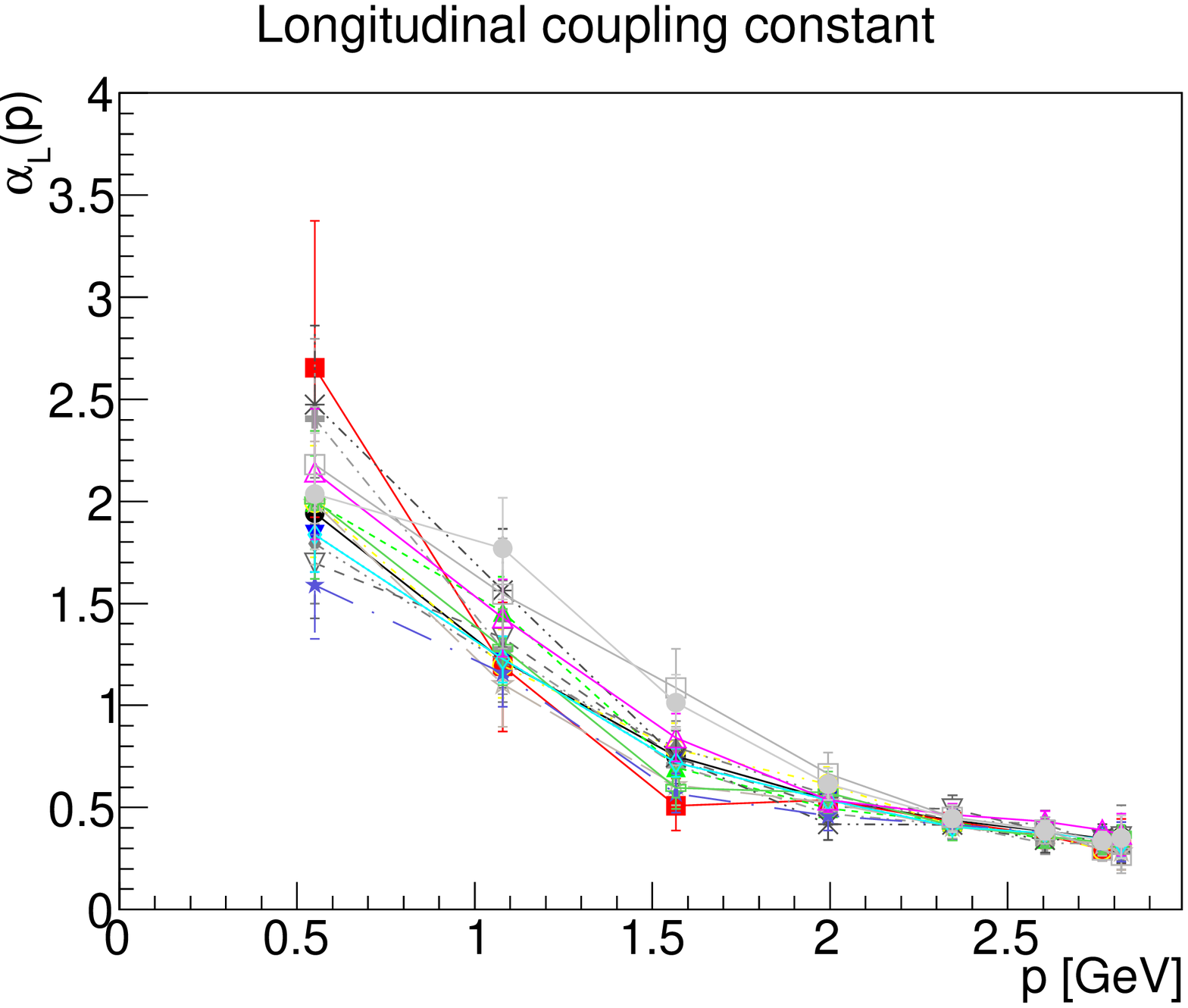}
\caption{\label{fig:alphamu}The dependence of the transverse (left panels) and longitudinal (right panels) running coupling on the chemical potential at fixed volume $24\times 12^3$ at $\beta=1.9$ (top panels), $24\times 16^3$  and $32^4$ at $\beta=1.9$ (middle panels), and $32\times 16^3$ at $\beta=2.1$ (bottom panels).}
\end{figure}

\afterpage{\clearpage}

These features of the gluon propagator and the ghost propagator are reflected in the running couplings shown in figure \ref{fig:alphamu}. Quite visible are strong finite-volume effects in comparison between $L\approx 2.2$ fm and $L>2.2$ fm, especially the appearance of a maximum. Apart from this, there is almost no density-dependence, except a slight infrared suppression which results from the enhancement of the gluonic screening masses seen in figure \ref{fig:sm} at $\beta=1.9$. This effect is again gone at $\beta=2.1$ at the same volume.

This result is probably the most remarkable result of the present study. While the Polyakov loop and the Wilson potential, both relevant to the properties of quarks, do show a pronounced density dependence\footnote{\label{fn5}Preliminary results at $\beta=2.1$ show that also this is likely a lattice artifact for the Polyakov loop, while results for the Wilson potential are yet inconclusive because of limited statistics \cite{Boz:2018phd,Skullerud:unpublished}. This is in line with the previously referred to absence of a transition at low chemical potentials.} \cite{Boz:2013rca,Cotter:2012mb} at $\beta=1.9$, the running coupling derived from the ghost-gluon vertex, encoding pure gauge dynamics, does not show any such effects at any $\beta$. The gauge sector at finite density appears to be largely inert. This includes the transition to a condensate at the silver-blaze point. This will also be confirmed below for the interaction vertices themselves.

In particular, this also implies that approximations using a vacuum gauge sector and containing all density-dependence in the quark sector alone, like \cite{Nickel:2006kc,Nickel:2006vf,Marhauser:2006hy,Nickel:2008ef,Contant:2017onc,Fischer:2018sdj}, are probably much better than should naively be expected. This would simplify calculations in other non-perturbative methods, like functional methods, substantially. Of course, whether this carries over to full QCD is an assumption so far. A test with other models, like G$_2$-QCD \cite{Maas:2012ts}, which shows a much more complicated pattern at finite density \cite{Wellegehausen:2013cya}, will be an important cross check. See \cite{Contant:2017onc} for first steps in this direction.

\begin{figure}
\includegraphics[width=\textwidth]{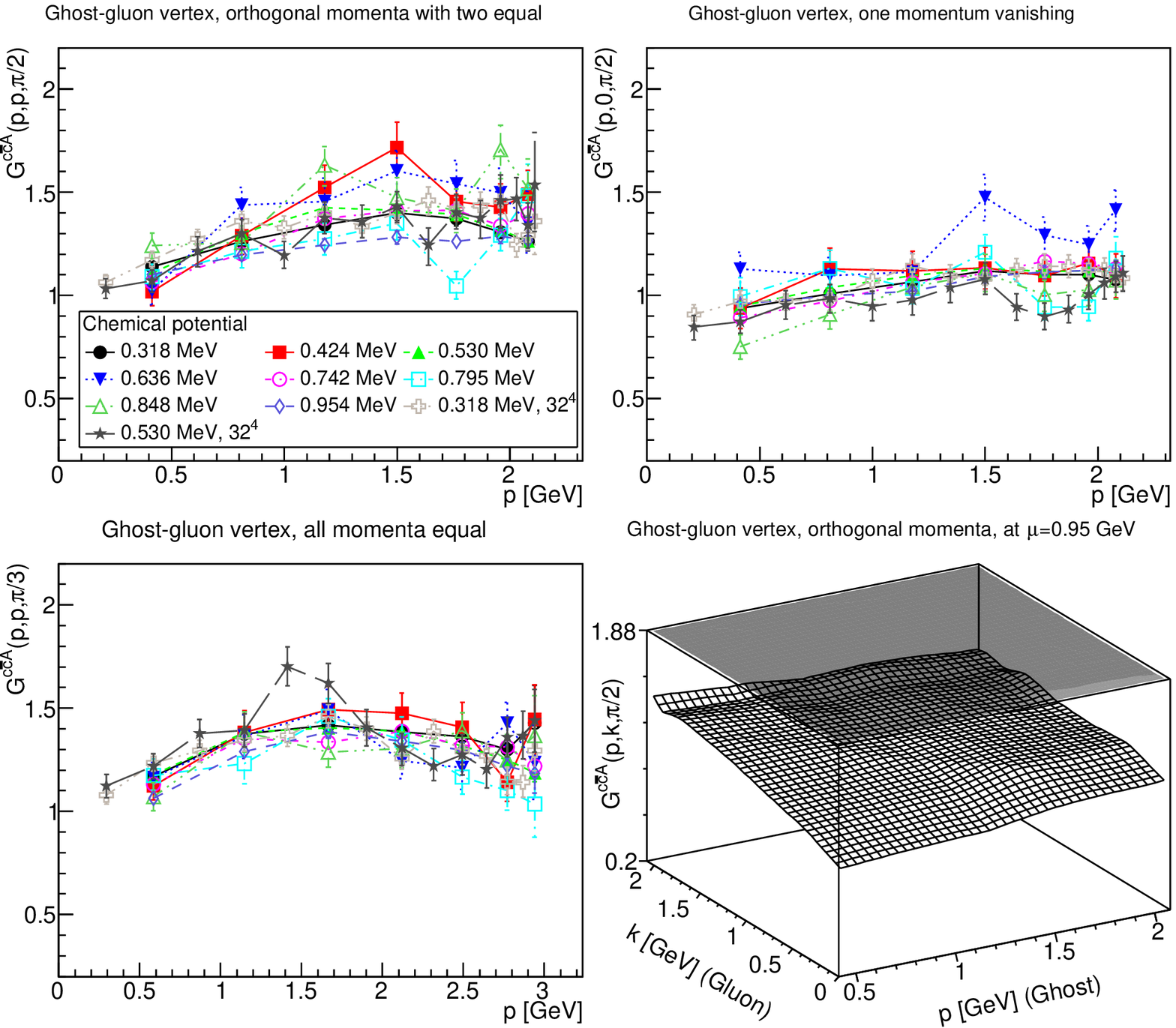}\\
\includegraphics[width=\textwidth]{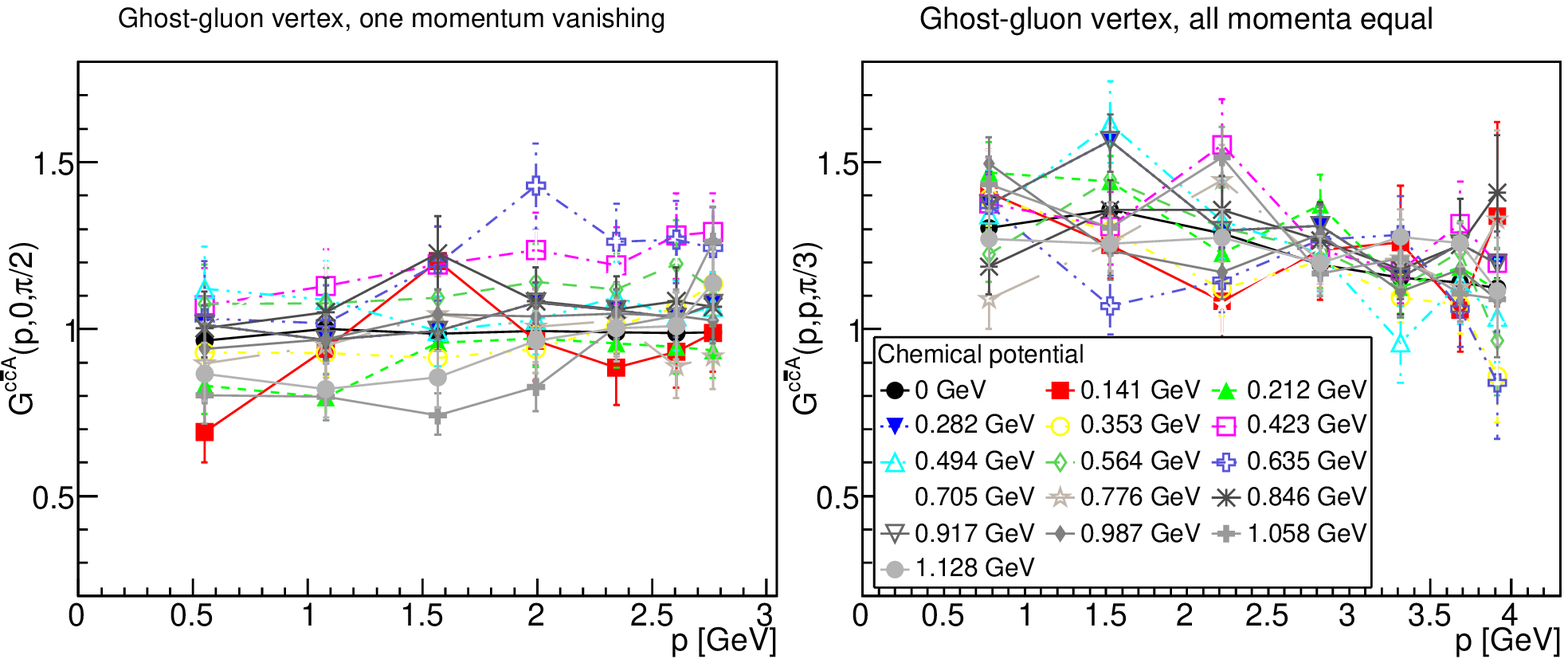}
\caption{\label{fig:ggvmu}The ghost-gluon vertex dressing for different momentum configurations at finite density in different momentum configurations for spatial size $N_s=16$ and $N_s=32$ and $\beta=1.9$ (top and middle panels). The middle-right panel is at the largest chemical potential of 954 MeV. The bottom panels are at $\beta=2.1$.}
\end{figure}

The results for the ghost-gluon vertex for the larger volumes of $\beta=1.9$ and for $\beta=2.1$ are shown in figure \ref{fig:ggvmu}. On the smaller volume at $\beta=1.9$, which is denser in the chemical potential, the fluctuations are substantially larger. These obscure that there is no statistically significant dependence on the density, as is visible for the larger volume and for the finer lattice. This is in line with the observations on the running coupling above which, after all, is derived from this vertex.

\begin{figure}
\includegraphics[width=\textwidth]{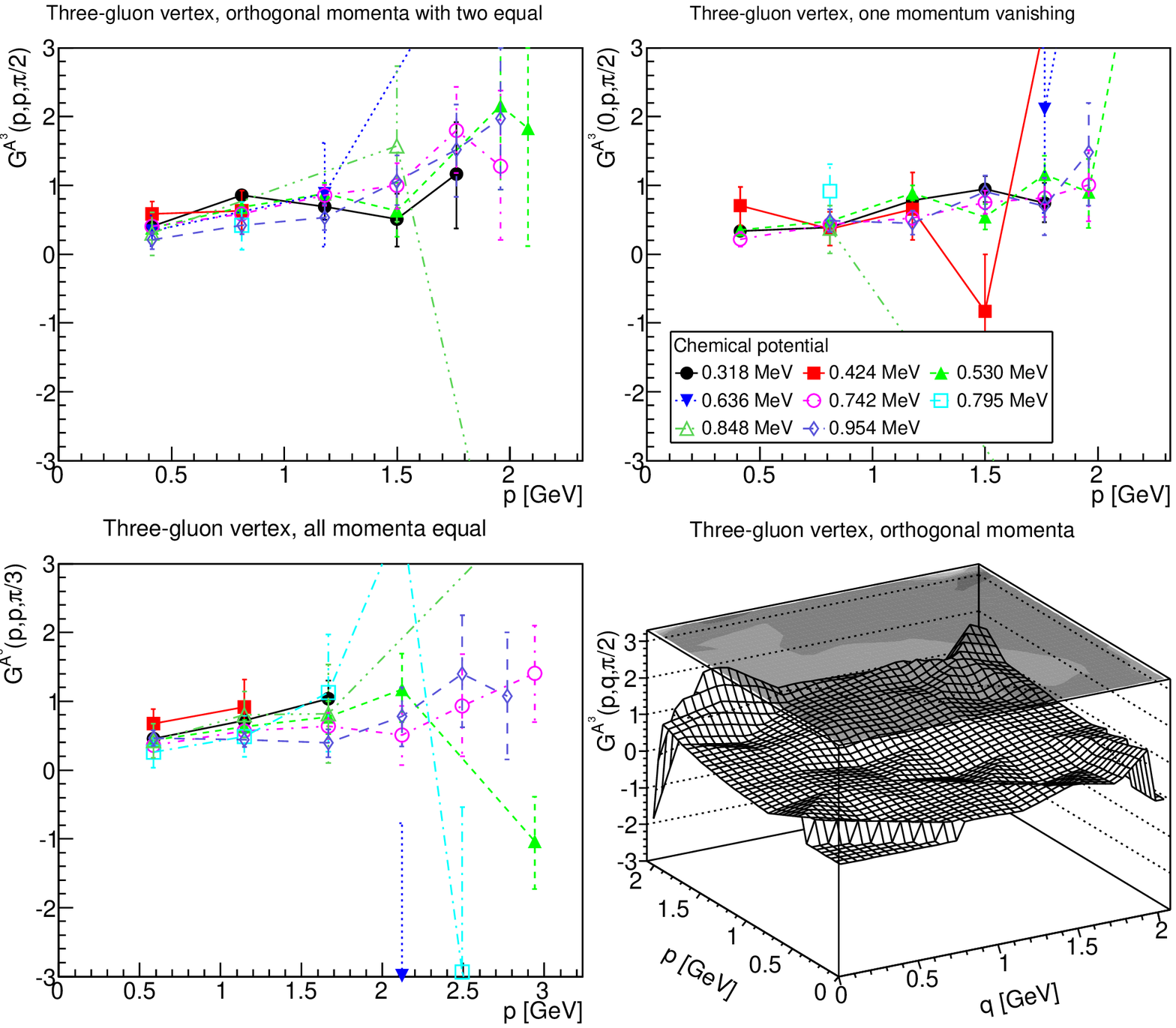}\\
\includegraphics[width=\textwidth]{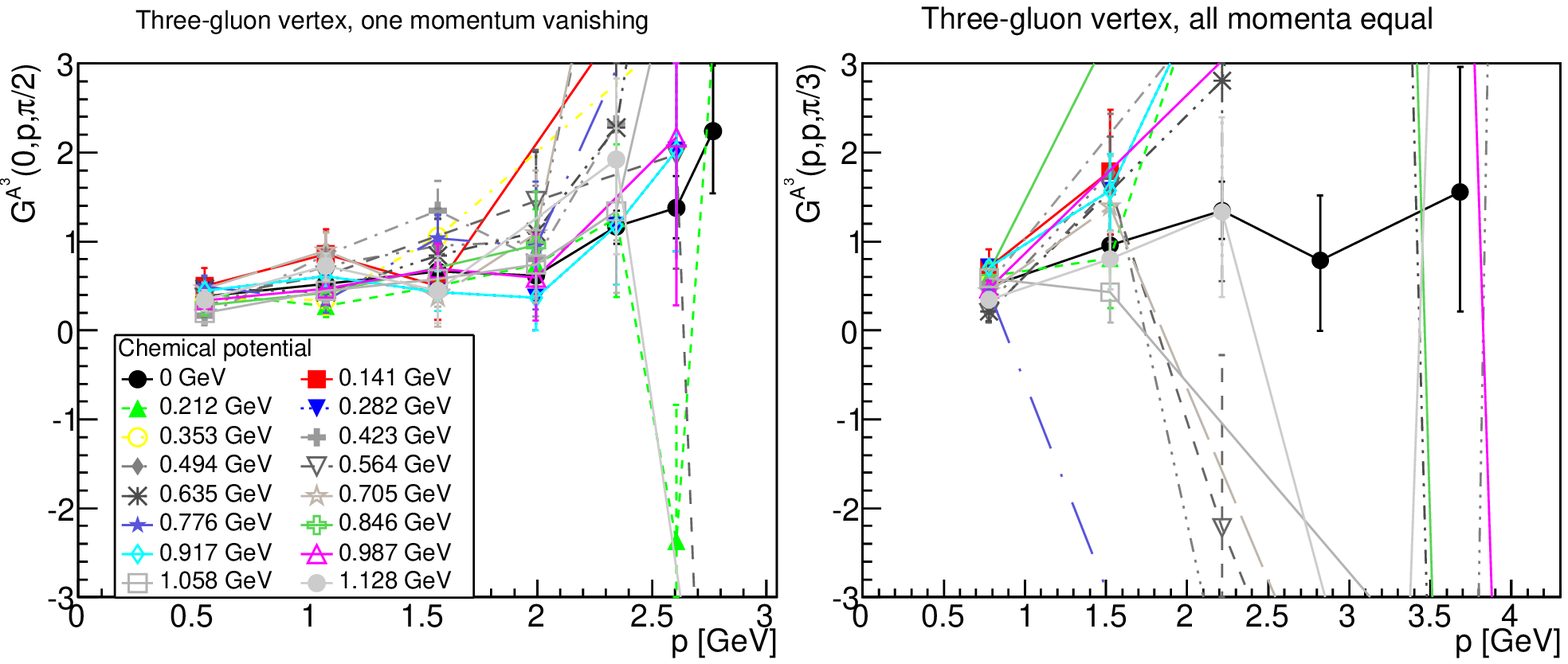}
\caption{\label{fig:g3vmu}The three-gluon vertex dressing for different momentum configurations at finite density in different momentum configurations for spatial size $N_s=16$ and $\beta=1.9$ (top and middle panels). The middle-right panel is at the largest chemical potential of 954 MeV. The bottom panels are at $\beta=2.1$. Points with relative errors larger than 100\% have been suppressed.}
\end{figure}

The results for the soft, magnetic three-gluon vertex are finally
shown in figure \ref{fig:g3vmu}. Within errors, no change with
chemical potential is seen. This is
in marked contrast to the situation at finite temperature
\cite{Fister:2014bpa}, where the same quantity shows a substantial
dependence on the temperature around the phase transition. Note that, in contrast to section \ref{s:tres}, at least at $\beta=1.9$ the volumes are large enough to reach into the relevant momentum regime. This inertness was not a foregone conclusion: after all, also at finite temperature the magnetic gluon propagator shows (almost) no dependence on the temperature, while the magnetic vertex does. Thus, this could not have been inferred from the behavior of the gluon propagators.

\section{Finite density and temperature}\label{s:pd}

\begin{figure}
\includegraphics[width=0.5\textwidth]{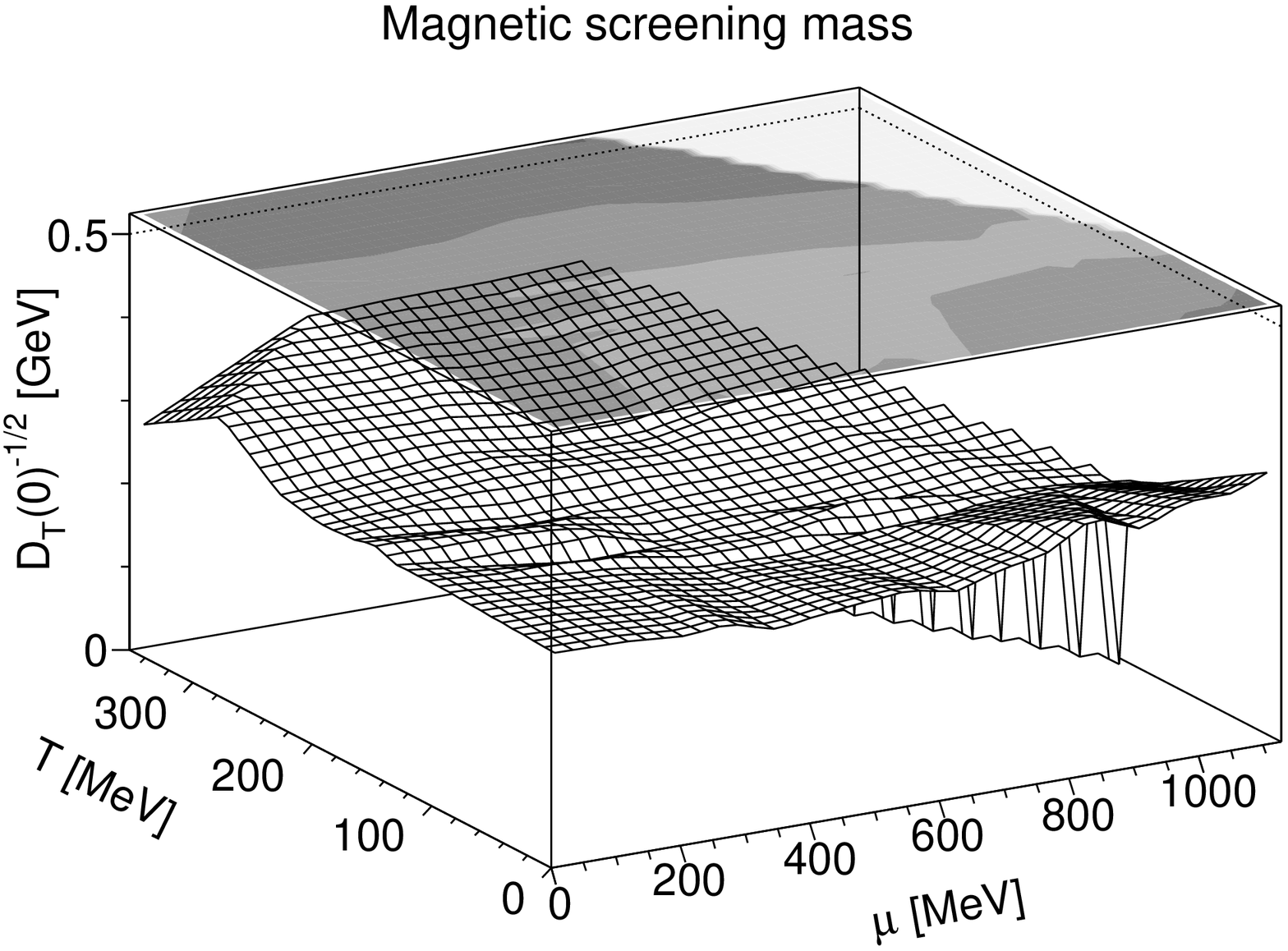}\includegraphics[width=0.5\textwidth]{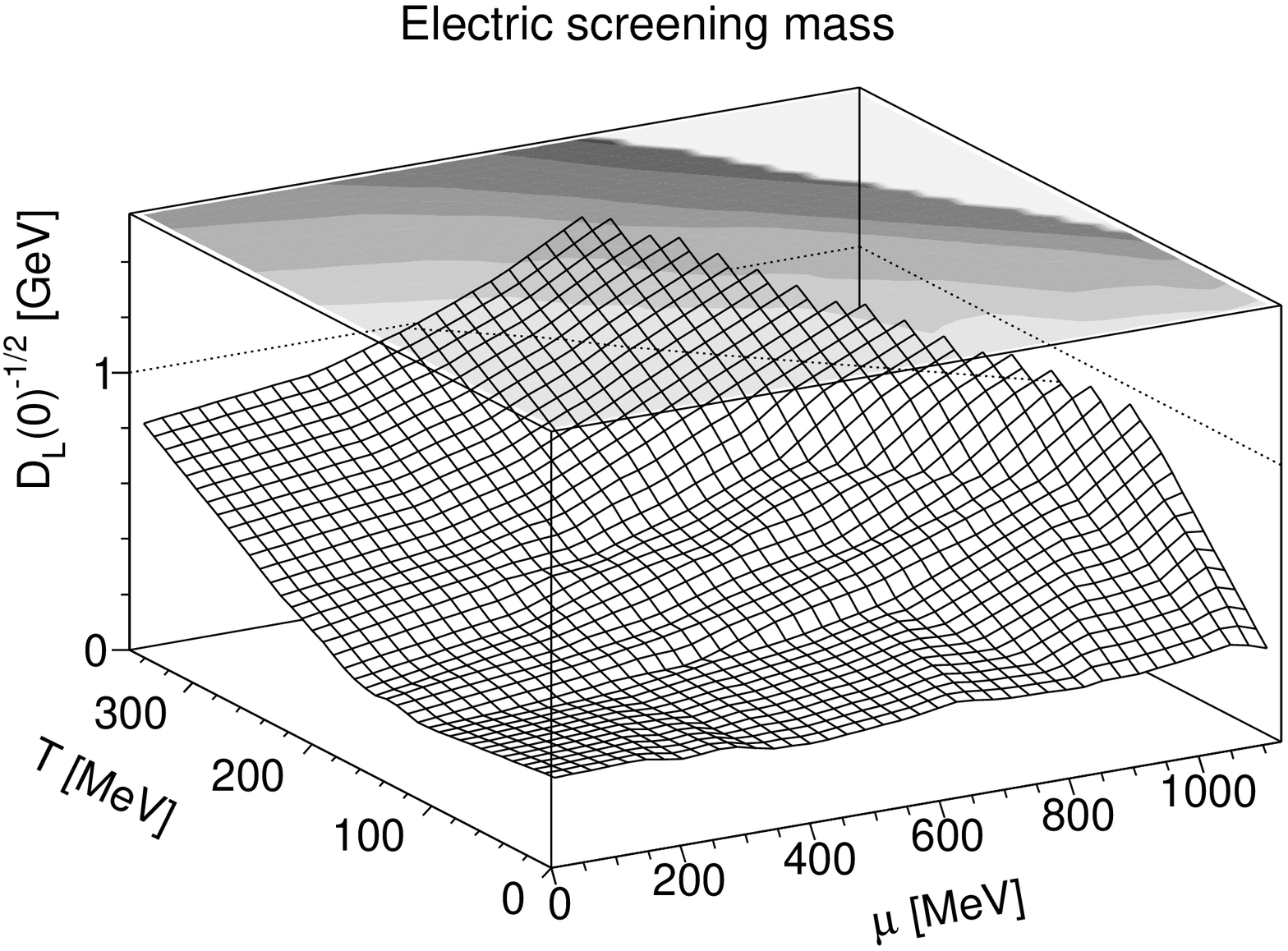}
\caption{\label{fig:mpd}The magnetic (left panel) and electric (right panel) screening mass in the phase diagram at $\beta=2.1$. Note the different scale on the left-hand side and on the right-hand side.}
\end{figure}

In total, the results show so far that unquenching leaves the qualitative behavior at zero density and finite temperature unchanged. At the same time, the gauge sector is essentially inert with respect to density at zero temperature. This leaves the interesting question whether the gauge correlation functions potentially reflect other structures in the phase diagram, e.\ g.\ a possible critical end-point \cite{Boz:2013rca}. As a first check, the electric and magnetic screening mass for the points shown in figure \ref{fig:pd} are shown in figure \ref{fig:mpd}.

The magnetic screening mass is essentially constant throughout the phase diagram, except for a slight increase at high temperature, which can already be inferred from figure \ref{fig:gpftr}. The electric screening mass, however, is only constant within a range which roughly traces out the 'hadronic' phase of two-color QCD \cite{Boz:2013rca}. Beyond that, it rises rapidly, i.\ e.\ chromoelectric correlators become strongly suppressed. This happens at lower temperature at larger chemical potentials. Thus, this seems to follow the conjectured curvature of the phase separation between the high-temperature phase and the hadronic low temperature phase. There is, however, no behavior which could be interpreted as any kind of critical endpoint.

\begin{figure}
\includegraphics[width=0.5\textwidth]{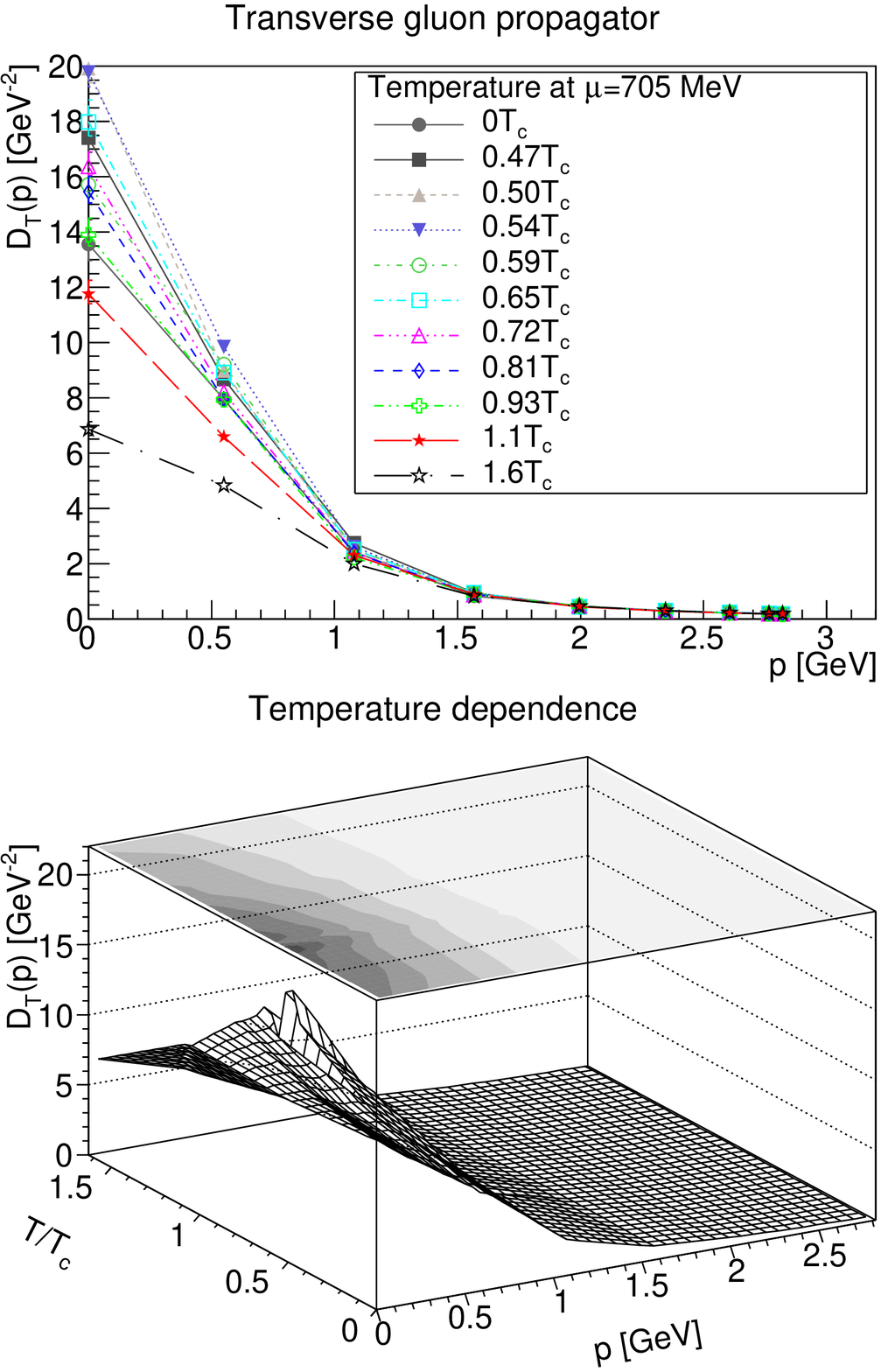}\includegraphics[width=0.5\textwidth]{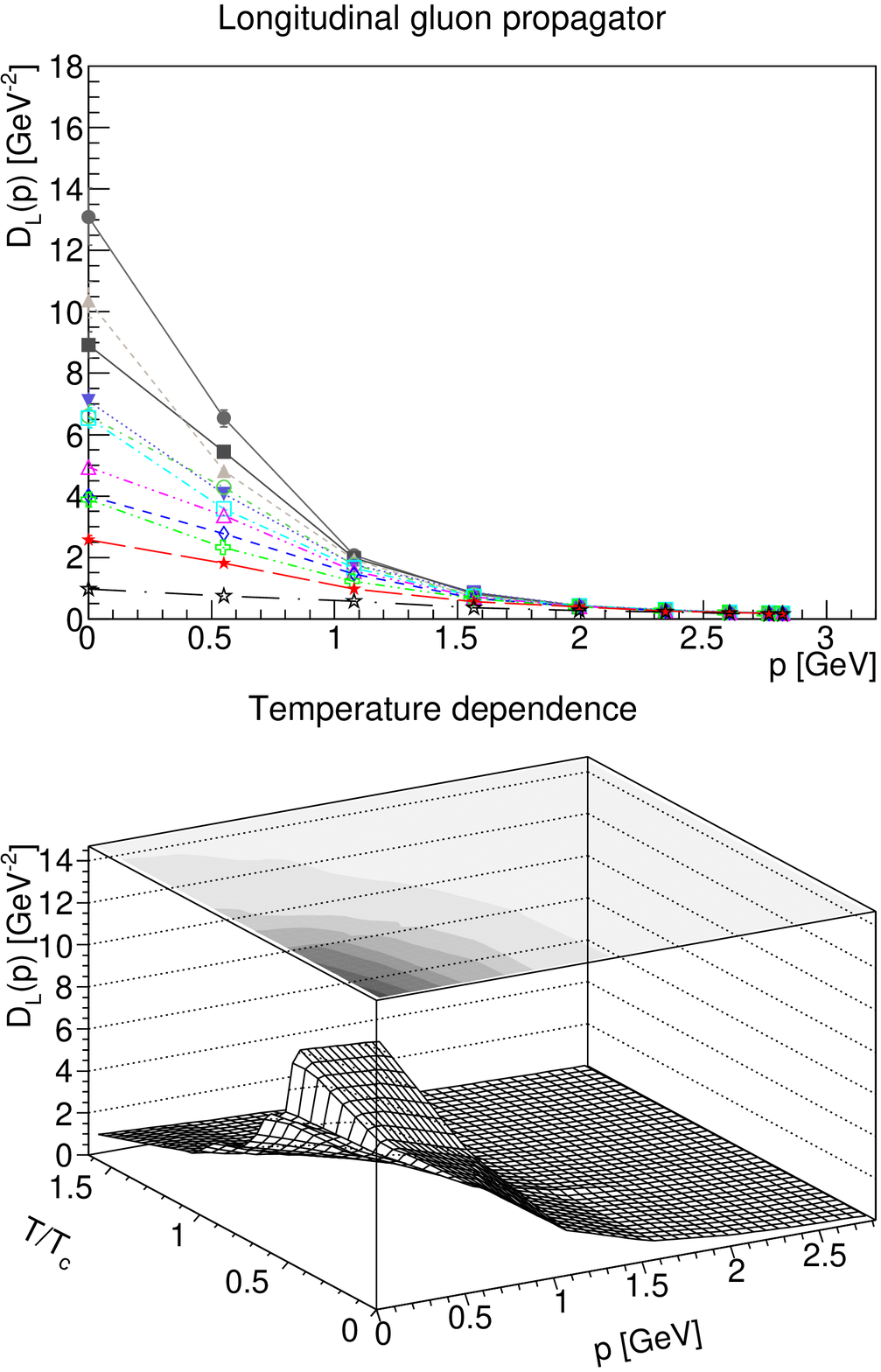}\\
\includegraphics[width=0.5\textwidth]{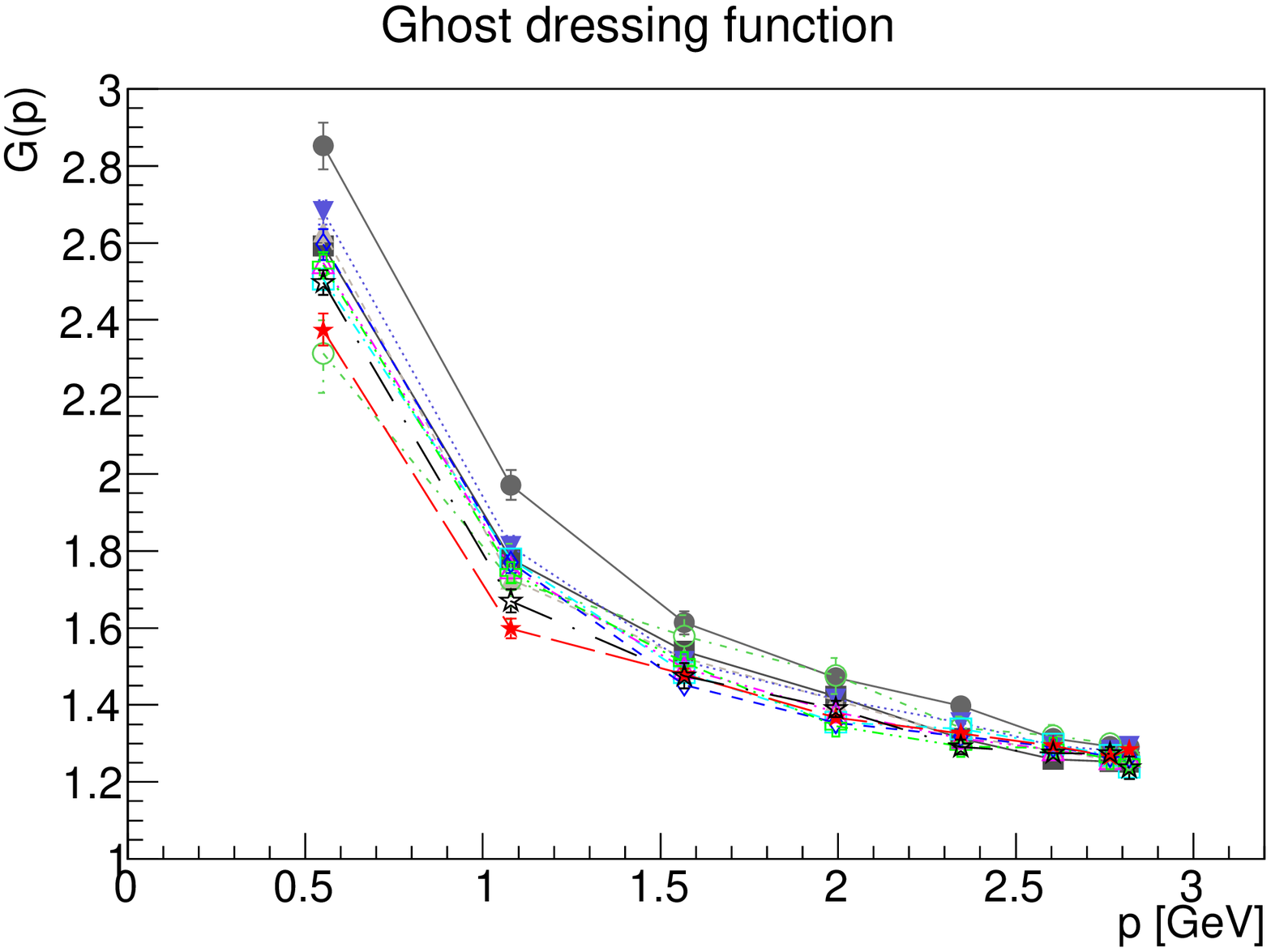}\includegraphics[width=0.5\textwidth]{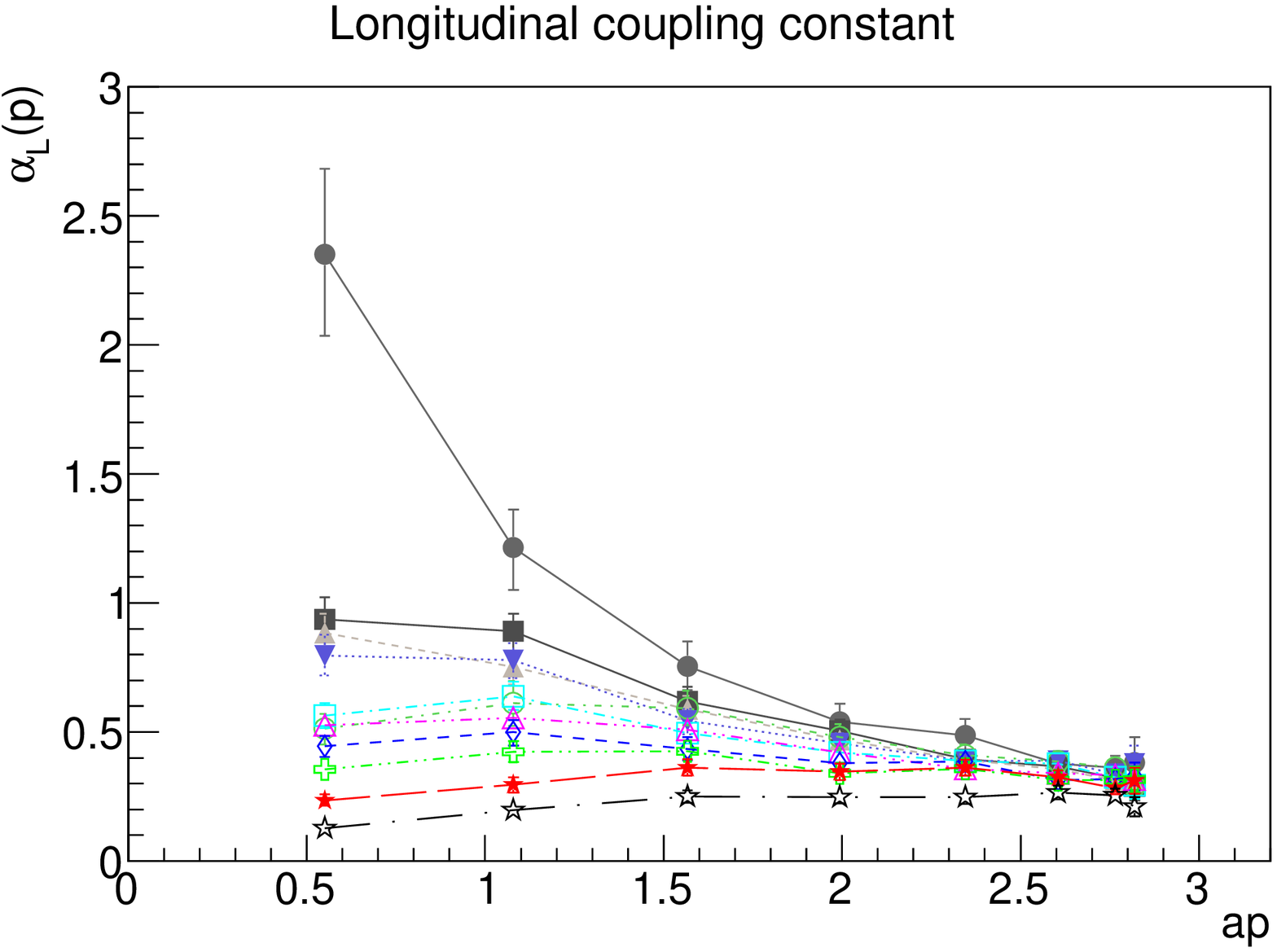}
\caption{\label{fig:ppd}The soft mode of the magnetic gluon propagator (top-left panels), the soft mode of the electric gluon propagator (top-right panels), the ghost dressing function (lower-left panel), and the running longitudinal coupling (lower-right panel) as a function of temperature at fixed chemical potential $\mu=705$ MeV from the $\beta=2.1$ data. Results have not been renormalized.}
\end{figure}

\begin{figure}
\includegraphics[width=\textwidth]{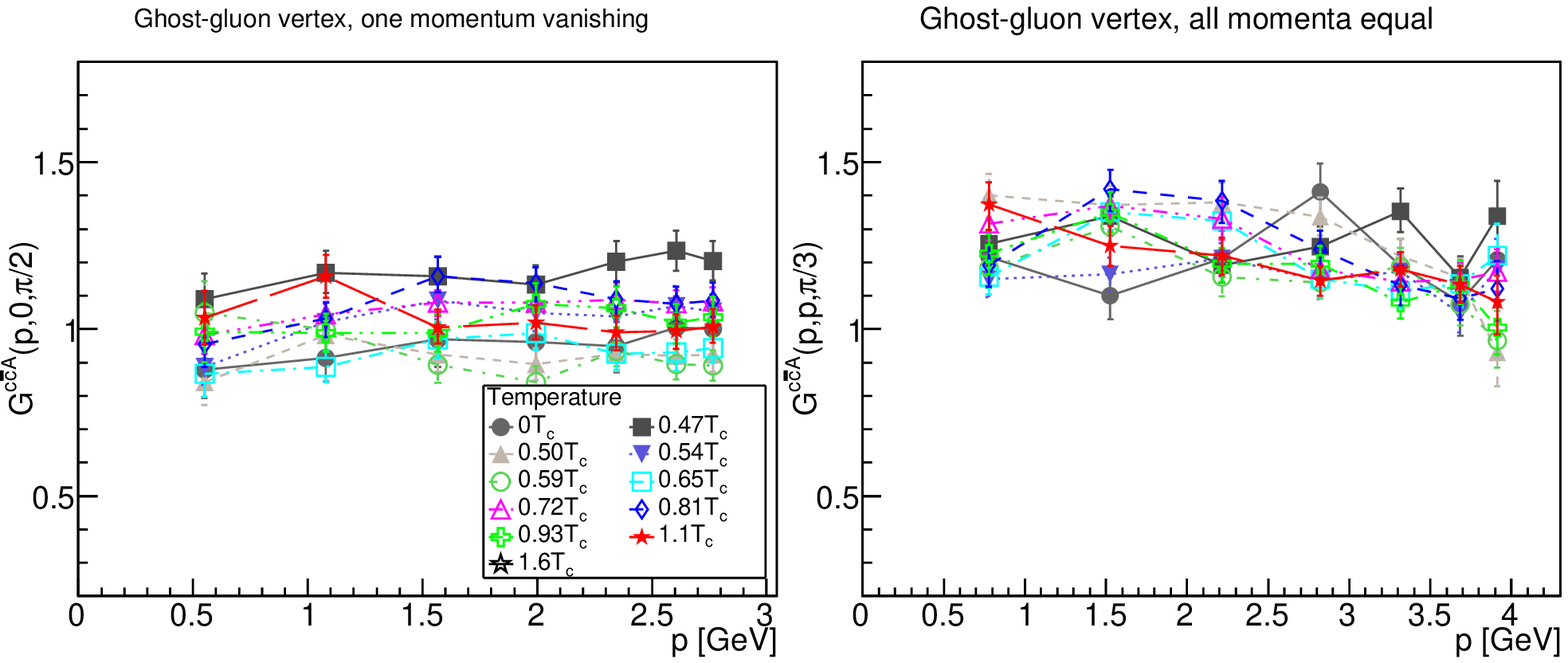}\\
\includegraphics[width=\textwidth]{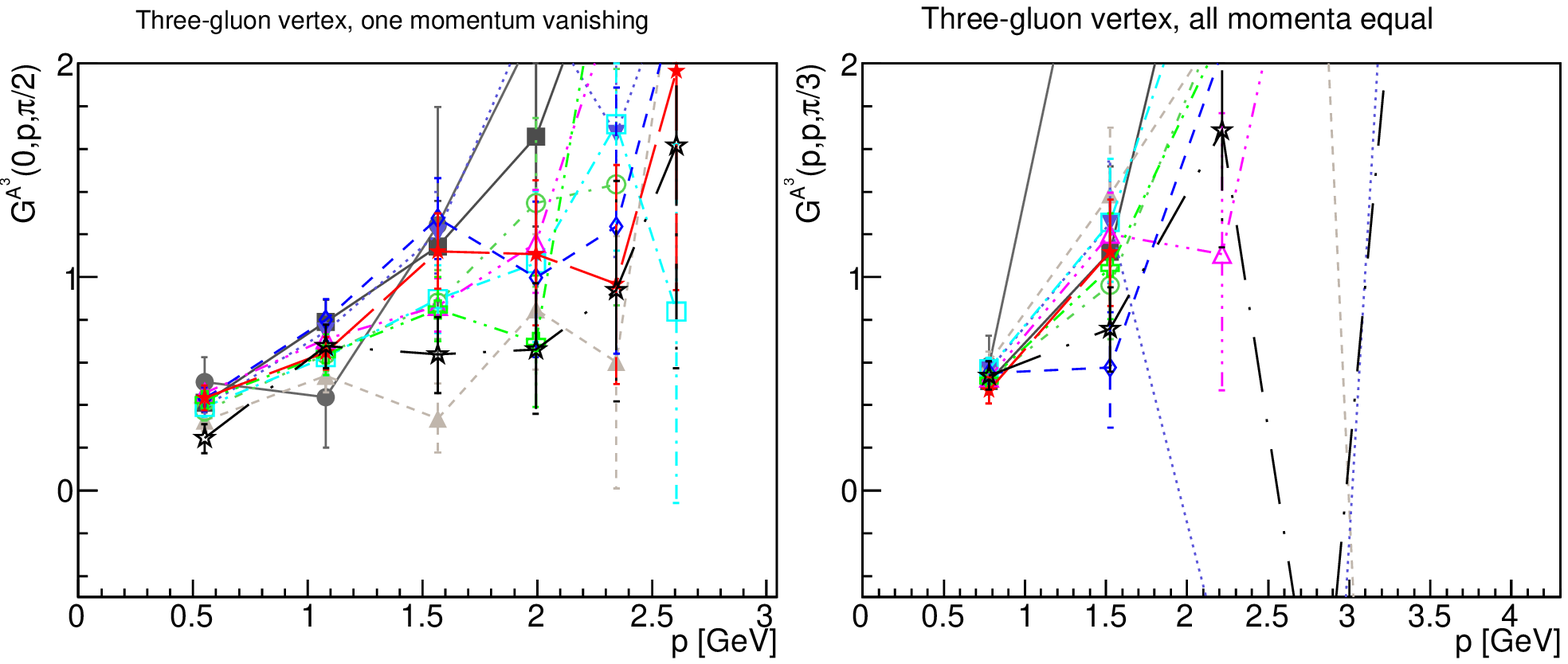}
\caption{\label{fig:vpd}The ghost-gluon vertex dressing (top panels) and three-gluon vertex dressing (lower panels) for different momentum configurations as a function of temperature at fixed chemical potential $\mu=705$ MeV from the $\beta=2.1$ data. Results have not been renormalized.}
\end{figure}

Thus, the behavior of the screening mass at finite chemical potential seems to be driven by the same mechanism as at zero chemical potential. This pattern repeats itself in all correlation functions, as is visible in figures \ref{fig:ppd} and \ref{fig:vpd} at fixed $\mu=705$ MeV and varying temperatures for propagators and vertices, respectively.

\begin{figure}
\includegraphics[width=0.5\textwidth]{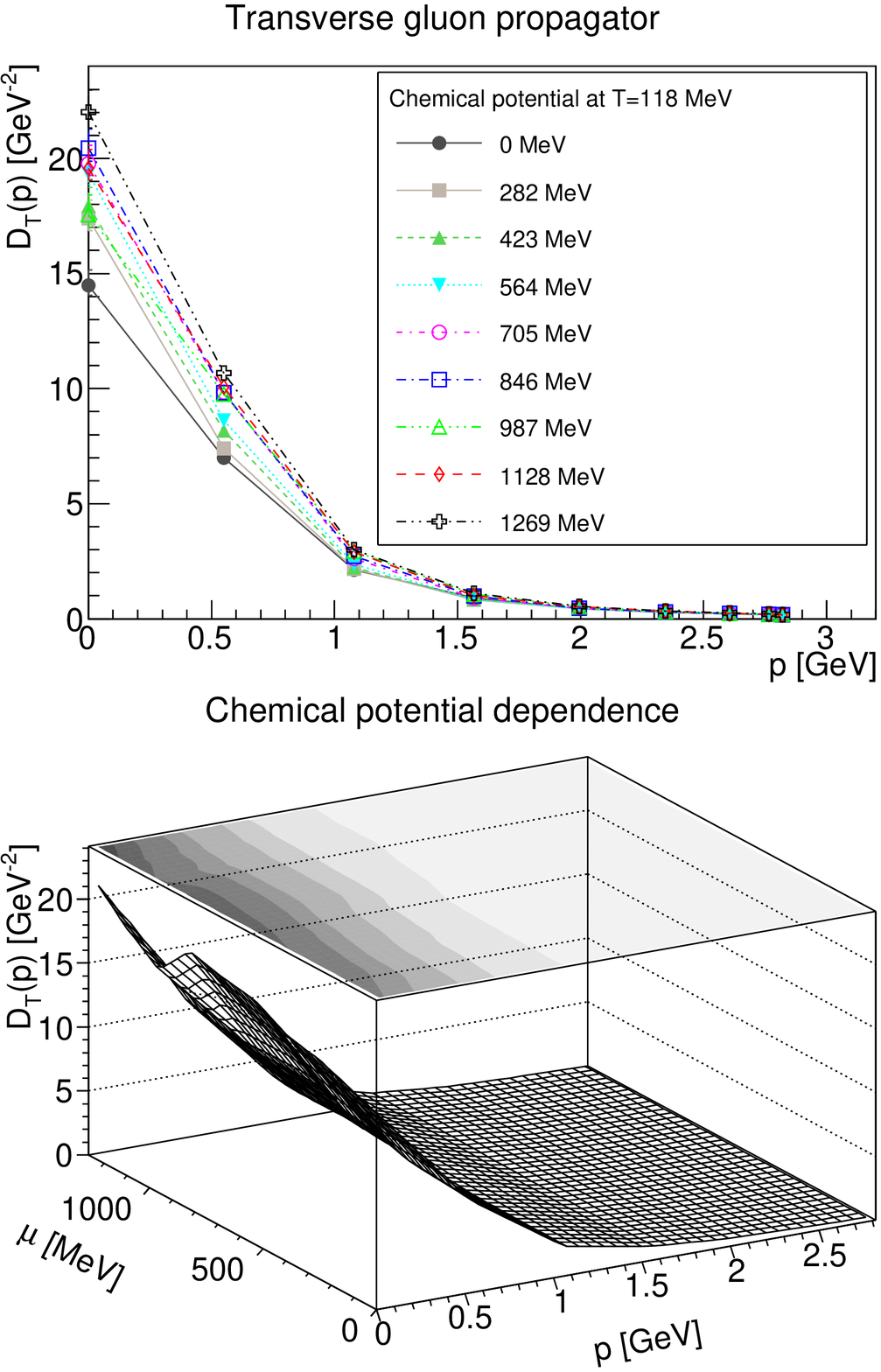}\includegraphics[width=0.5\textwidth]{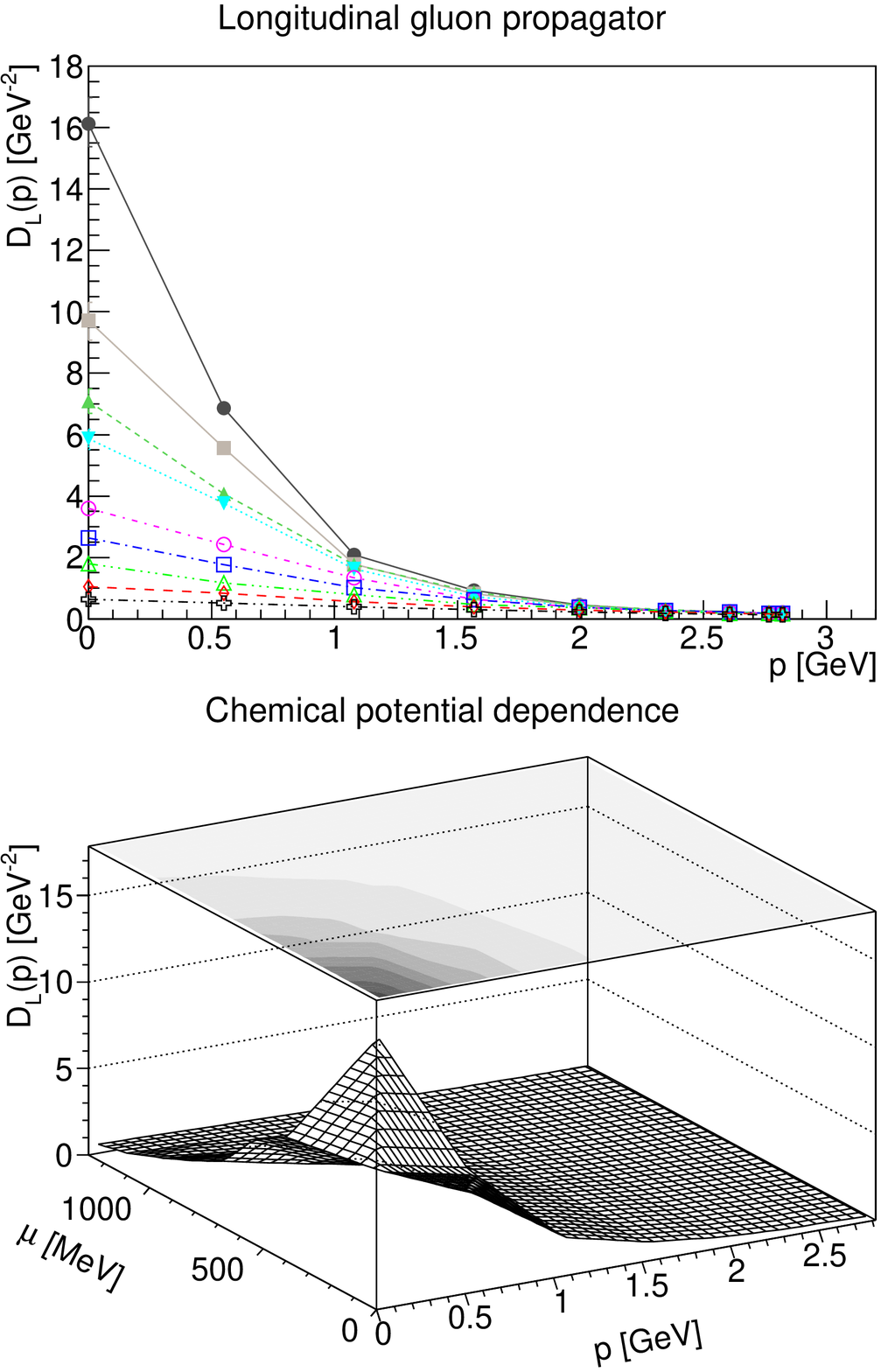}\\
\includegraphics[width=0.5\textwidth]{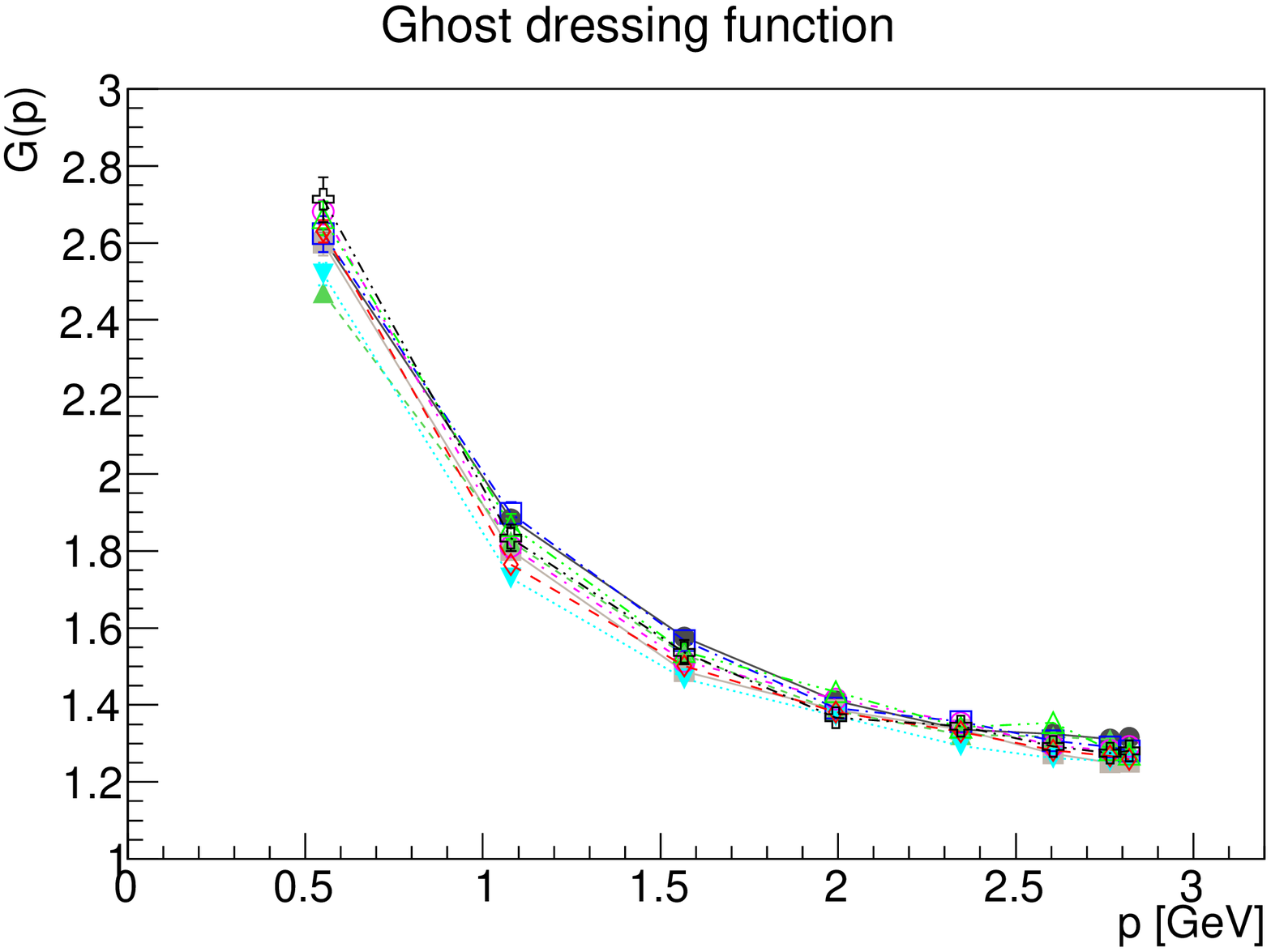}\includegraphics[width=0.5\textwidth]{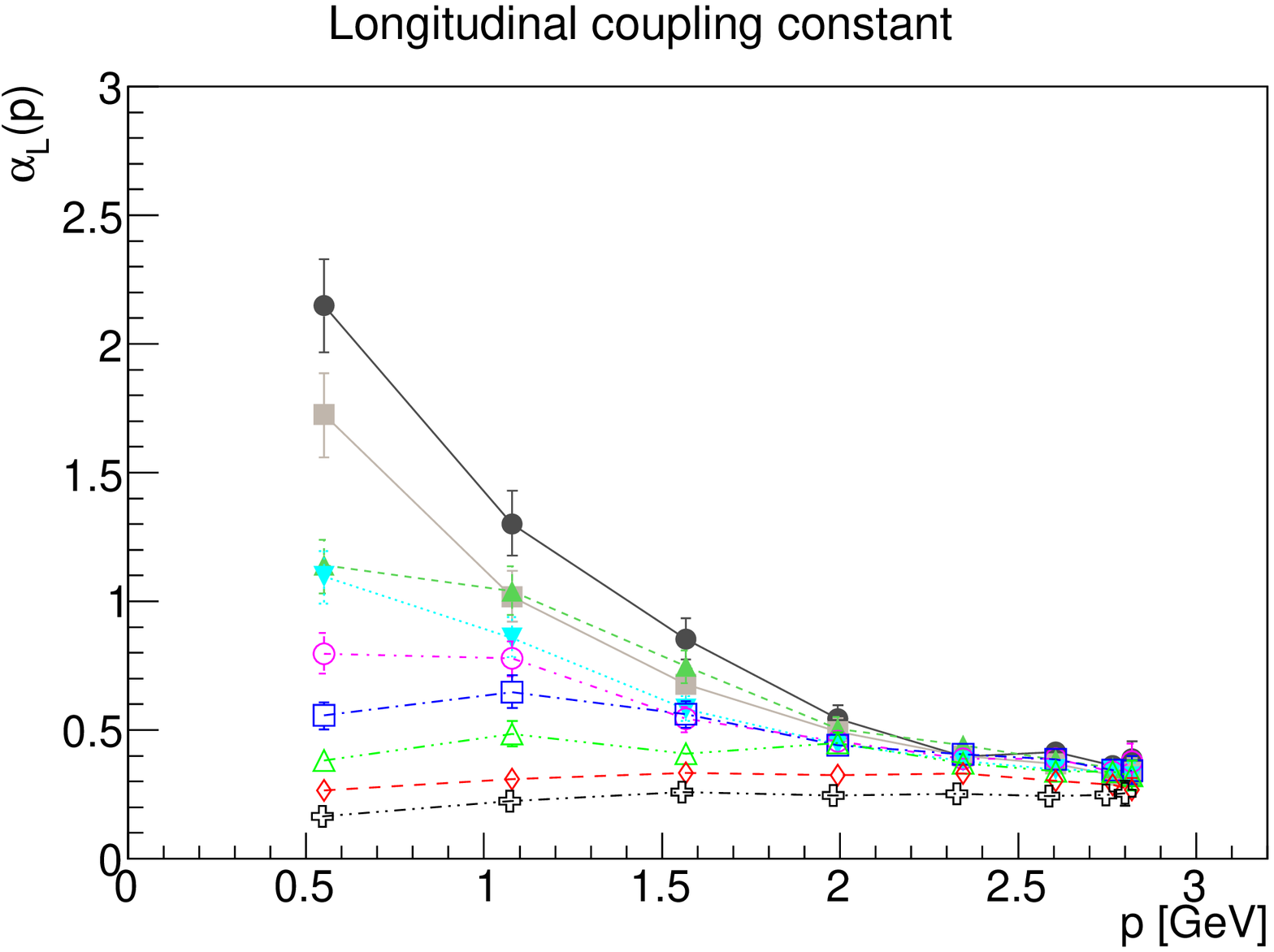}
\caption{\label{fig:p12}The soft mode of the magnetic gluon propagator (top-left panels), the soft mode of the electric gluon propagator (top-right panels), the ghost dressing function (lower-left panel), and the running longitudinal coupling (lower-right panel) as a function of chemical potential at fixed temperature $T=118$ MeV from the $\beta=2.1$ data. Results have not been renormalized.}
\end{figure}

\begin{figure}
\includegraphics[width=\textwidth]{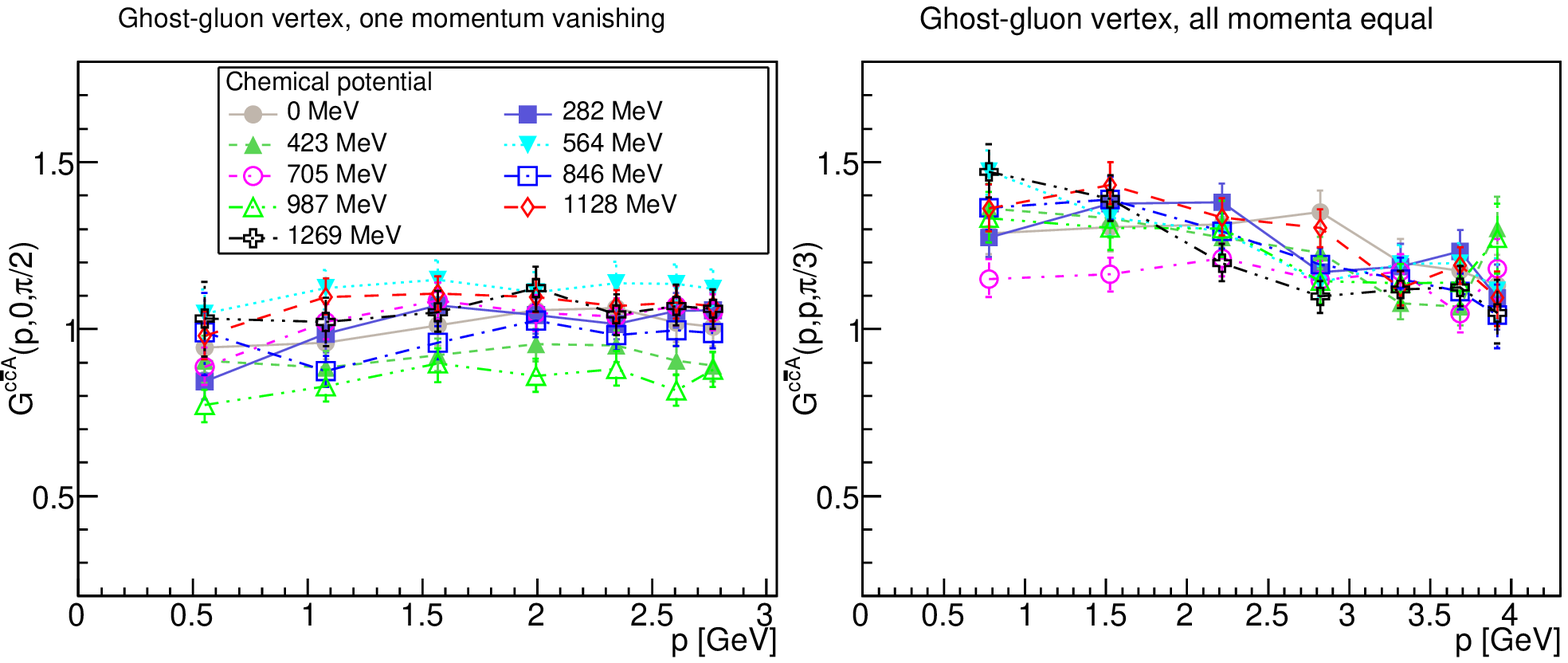}\\
\includegraphics[width=\textwidth]{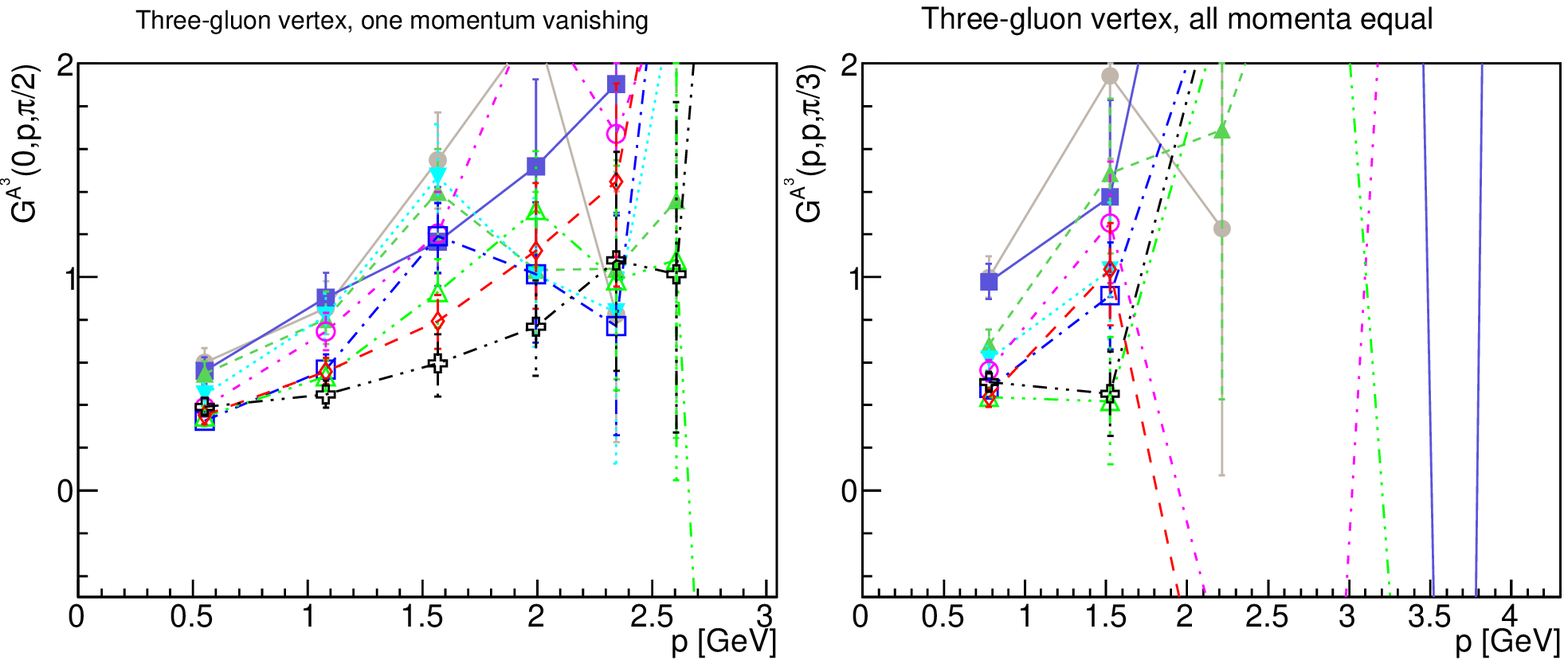}
\caption{\label{fig:v12}The ghost-gluon vertex dressing (top panels) and three-gluon vertex dressing (lower panels) for different momentum configurations as a function of chemical potential at fixed temperature $T=118$ MeV from the $\beta=2.1$ data. Results have not been renormalized.}
\end{figure}

Conversely, at fixed temperature eventually a point is reached in chemical potential where the correlation functions show the same behavior as when increasing the temperature. This is shown in figure \ref{fig:p12} and \ref{fig:v12} for the propagators and vertices, respectively, for fixed temperature $T=118$ MeV and varying chemical potential. Eventually, the electric propagator becomes suppressed. The other correlation functions show no pronounced dependence. But again, neither is the chemical potential mesh fine enough nor the spatial volume large enough to resolve effects like the ones seen around the finite-temperature phase transition.

As is visible in figure \ref{fig:mpd}, this behavior occurs at larger and larger chemical potentials the lower the temperature. Eventually, the results indicate that at zero (low) temperature, this occurs either at chemical potentials larger than the ones accessible here of about 1.1 GeV, or the effect ceases. Given the results in \cite{Astrakhantsev:2018uzd,Bornyakov:2017txe}, the former explanation seems more plausible. An eventual confirmation will require explicit tests.

\section{Conclusions}\label{s:con}

Summarizing, we have studied the behavior of the gauge sector in
two-color QCD both in the vacuum and at non-zero temperature and
chemical potential. At zero chemical potential the behavior is as expected from corresponding results for three-color QCD as well as Yang-Mills theory. At zero temperature and finite chemical potential no statistically and systematically significant change is seen as compared to the vacuum up to a chemical potential of about 1.1 GeV. Thus, the gauge sector is essentially inert. This is in marked contrast to quark sector observables, like the Wilson loop and the Polyakov loop, which show on coarse lattices\footnote{Compare footnote \ref{fn5}.} a dependence on the chemical potential \cite{Boz:2013rca,Cotter:2012mb}. This suggests that approximation schemes which assume such a behavior \cite{Nickel:2006kc,Nickel:2006vf,Marhauser:2006hy,Nickel:2008ef,Contant:2017onc,Fischer:2018sdj} are probably much better than expected. Inside the full phase diagram, the results indicate that this behavior persists everywhere in the low-temperature, low-density domain. Only outside the ``hadronic'' region established in \cite{Boz:2013rca,Cotter:2012mb} do the gauge correlation functions show a different behavior. This difference is essentially only a suppression of the electric interactions at low momenta. The magnetic and ghost interactions stay virtually vacuum-like throughout the phase diagram. In fact, it appears that the gauge sector effectively only depends on some fixed combination $aT^n+b\mu^m$, rather than on $T$ and $\mu$ separately. However, the electric screening mass seems to be, as at zero chemical potential, a useful tool to track the phase diagram. Unfortunately, given our coarse temperature and chemical-potential mesh, we cannot decide yet whether these correlation functions show particular behaviors close to the transition region, especially with respect to any critical endpoint.

These results are stringent benchmarks for any calculations of the gauge sector. Their usefulness for QCD phenomenology will rest on whether these results are generic, and thus applicable also to three-color QCD. There are two possible approaches to do so. One would
be to consider other theories which are accessible at finite density
in lattice calculations. In particular, G$_2$-QCD, which shows a substantially more involved phase structure at zero temperature \cite{Wellegehausen:2013cya}, would be a candidate. The other one would be to work with this assumption in other methods, and eventually push them to calculate observable quantities, e.\ g.\ neutron star properties. If they would provide reasonable accuracy in their description, this would provide an independent check of the inertness of the gauge sector, at least at intermediate densities.

\section*{Acknowledgments}

O.\ H.\ was supported by the FWF DK W1203-N16. We acknowledge the
networking support by the COST action CA15213 ``Theory of hot matter
and relativistic heavy-ion collisions''. T.S.B. and J.I.S. acknowledge
support from SFI grant 11-RFP.1-PHY3193.  This work used the DiRAC
Blue Gene Q Shared Petaflop system at the University of Edinburgh,
operated by the Edinburgh Parallel Computing Centre on behalf of the
STFC DiRAC HPC Facility (www.dirac.ac.uk). This equipment was funded
by BIS National E-infrastructure capital grant ST/K000411/1, STFC
capital grant ST/H008845/1, and STFC DiRAC Operations grants
ST/K005804/1 and ST/K005790/1. DiRAC is part of the National
E-Infrastructure.

\appendix

\section{Systematic errors}\label{s:sys}

There are several sources of systematic errors in our lattice calculations. Besides the usual vacuum source of finite volume and finite lattice spacing, thermodynamics introduces in addition the finite aspect ratio \cite{Maas:2011ez}. In addition, in the investigated system an explicit diquark source $j$ was introduced to induce diquark condensation \cite{Boz:2013rca,Cotter:2012mb}. The actual desired results would be obtained in the limit of $j\to 0$. 

As is seen in sections \ref{s:res} and \ref{s:pd}, essentially the only quantities substantially influenced by the thermodynamics are the screening masses. We will therefore concentrate here on the effects of the systematic errors sources on this quantity. However, we have, in detail, also investigated the impact on all other quantities, and did not find any cases in which stronger effects are present than the ones in the screening masses.

\subsection{Discretization}

The impact of discretization at fixed spatial volumes is already studied in the main text, especially in figure \ref{fig:sm}. At small chemical potentials the results for both $\beta=1.9$ and $\beta=2.1$ show essentially the same behavior. However, at chemical potentials above $a\mu\gtrapprox 0.7$ differences start to appear. This suggests that this marks the onset of discretization artifacts. Almost all our data for the finer lattices are not exceeding this range. Especially, all relevant effects arise already below this bound. Thus, we consider the fine results reasonably unaffected by discretization artifacts, but would assume that the results on the coarser lattices cannot be fully trusted above this, see also footnote \ref{fn5}. However, most quantities agree even above this threshold between both discretizations, suggesting that this is often a mild effect.

\subsection{Diquark source}\label{a:sysj}

\begin{figure}
\includegraphics[width=0.5\textwidth]{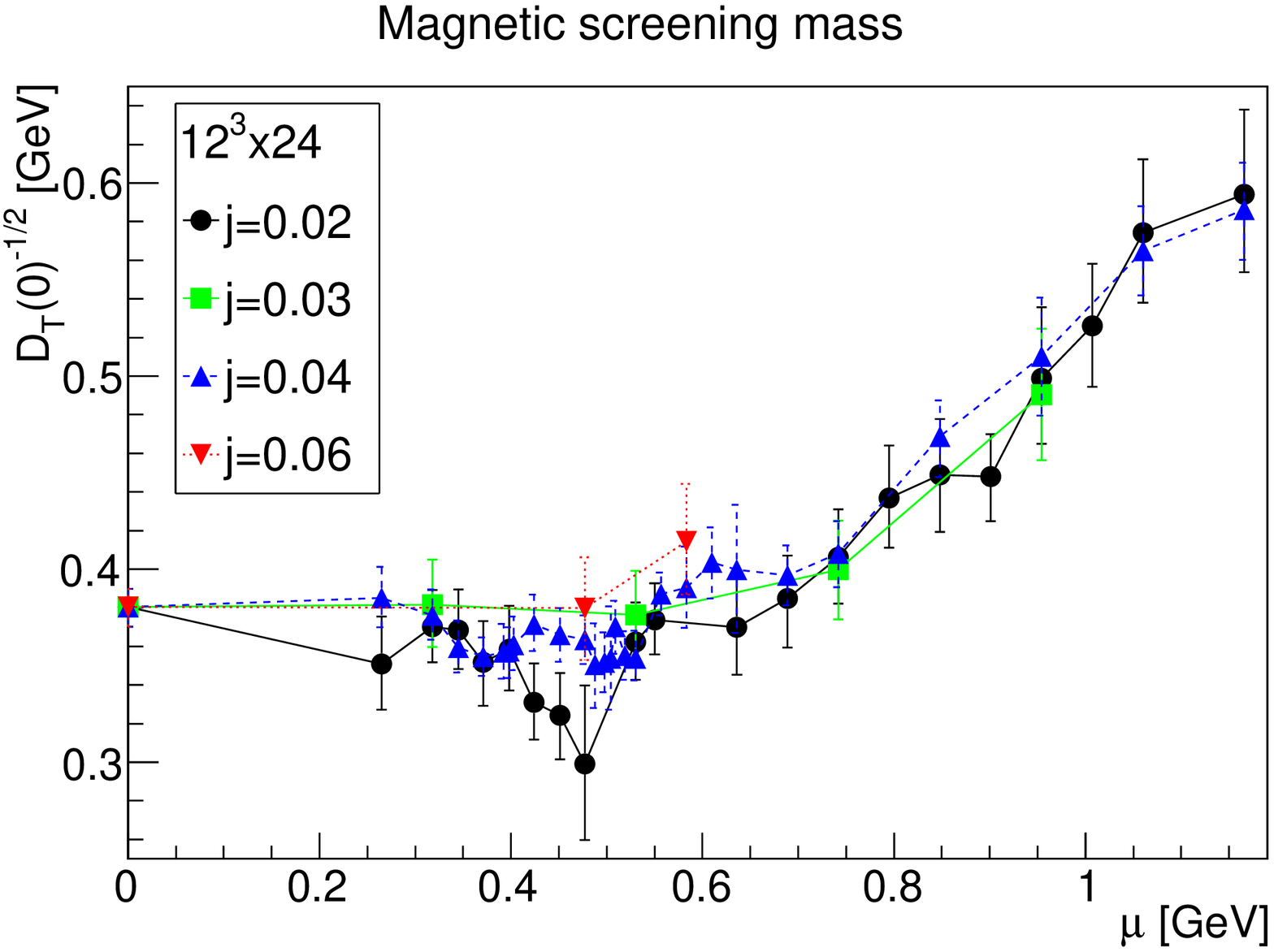}\includegraphics[width=0.5\textwidth]{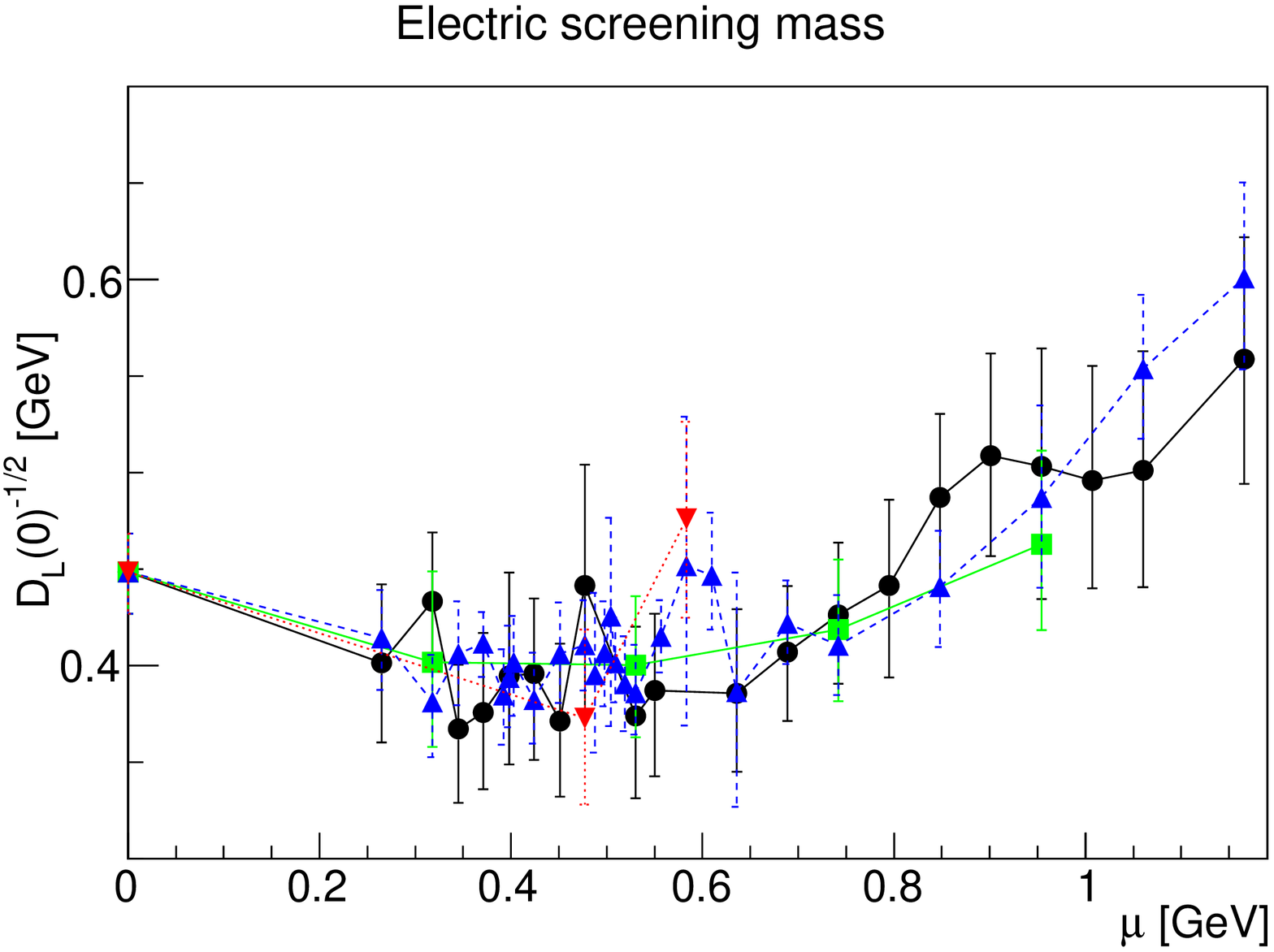}\\
\includegraphics[width=0.5\textwidth]{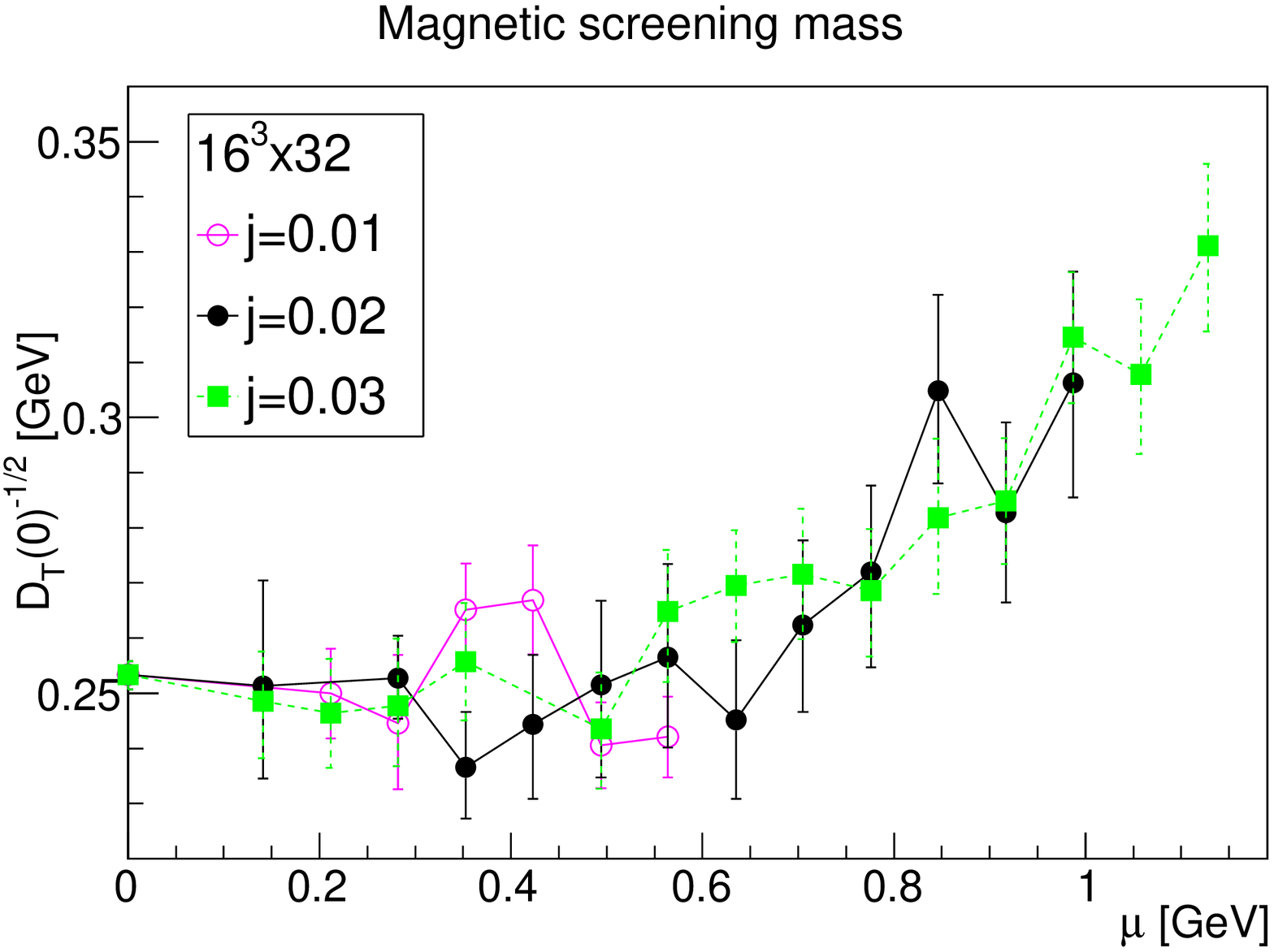}\includegraphics[width=0.5\textwidth]{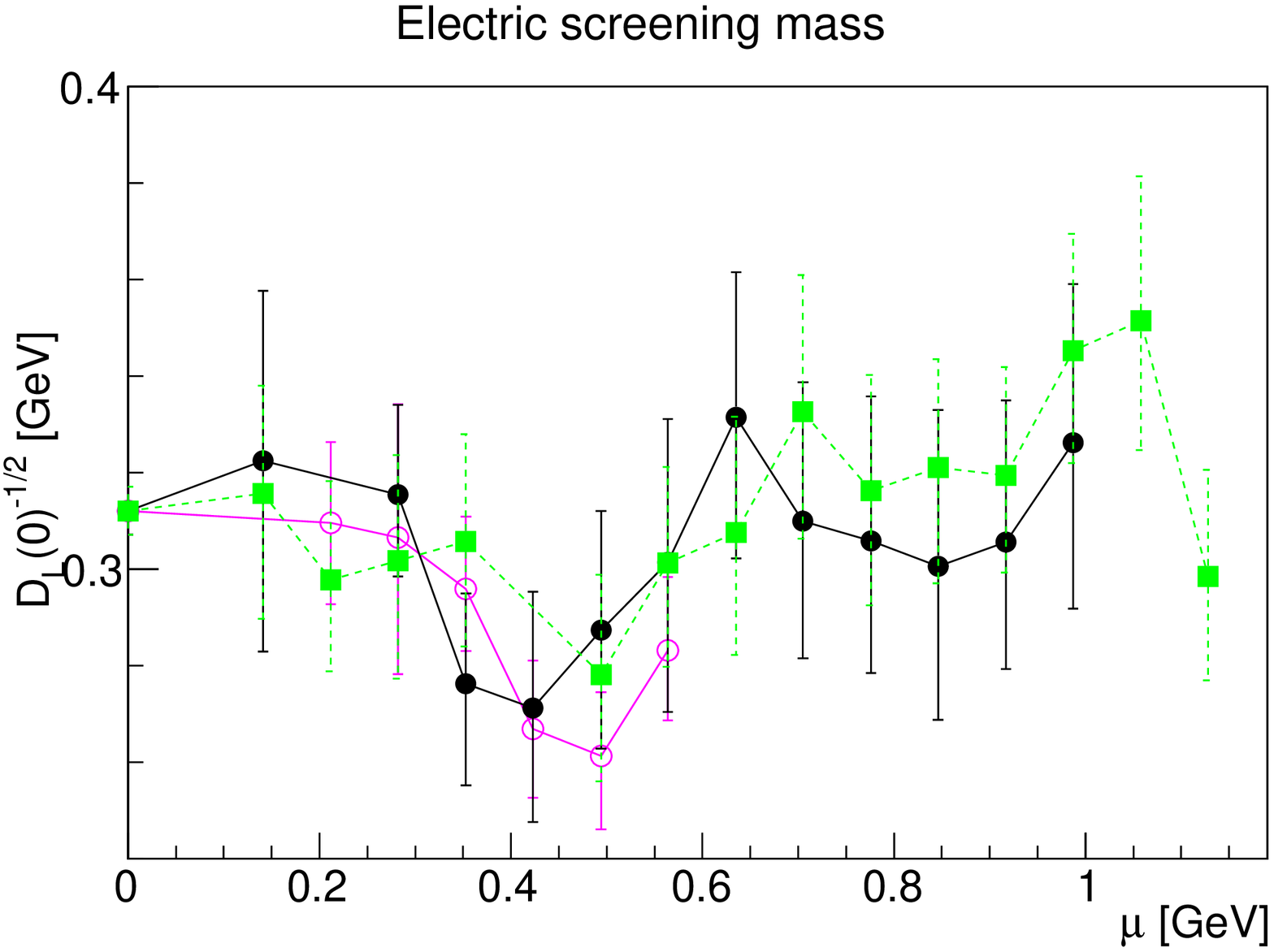}
\caption{\label{fig:mj}The dependence of the magnetic (left panels) and electric (right panels) screening masses on the diquark sources at fixed volume at $\beta=1.9$ and $12^3\times 24$ (top panels) and at $\beta=2.1$ and $16^3\times 32$ (bottom panels).}
\end{figure}

The dependence of the screening masses on the diquark source as a function of chemical potential is shown in figure \ref{fig:mj}. There is no statistically significant dependence on the diquark source visible at any chemical potential or for the different $\beta$ values, nor is there a difference between the magnetic and electric screening masses. Thus, within the available statistical accuracy there is no effect, and thus any extrapolation to zero diquark sources \cite{Boz:2013rca, Cotter:2012mb} is not meaningful for the investigated observables. Hence, the dependence on the diquark source is neglected throughout.

\subsection{Volume dependence}\label{a:sysv}

\begin{figure}
\includegraphics[width=\textwidth]{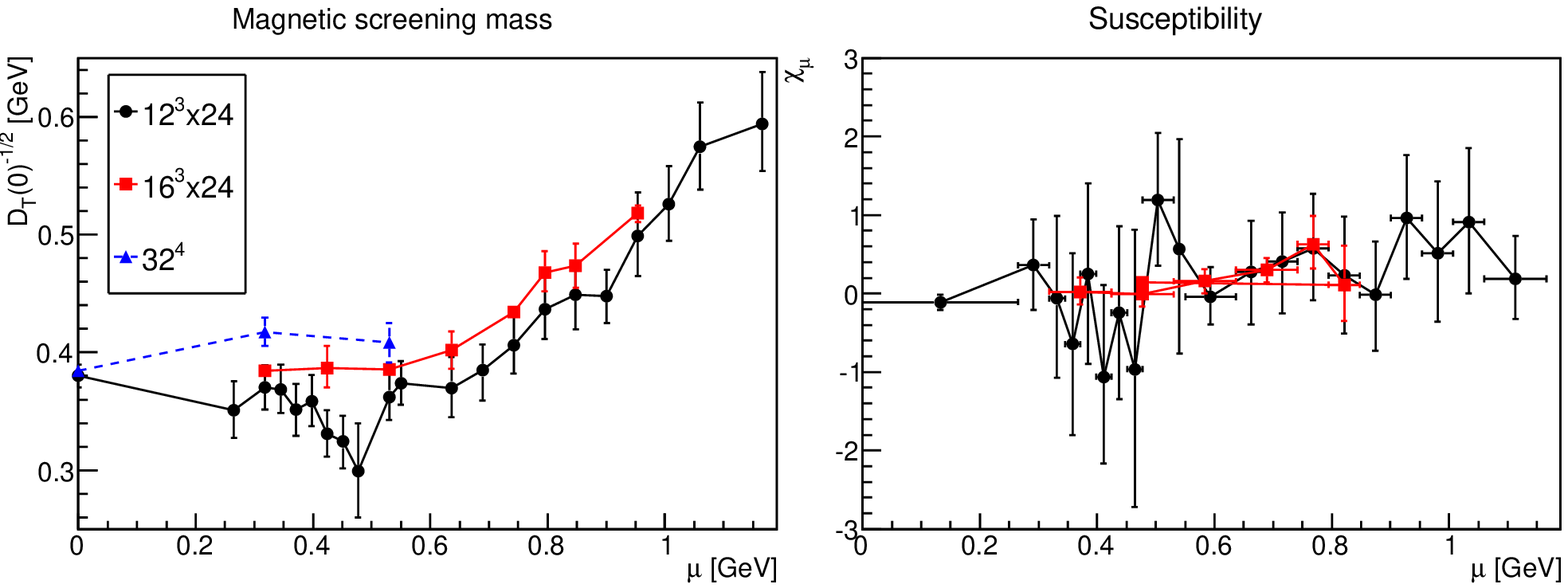}\\
\includegraphics[width=\textwidth]{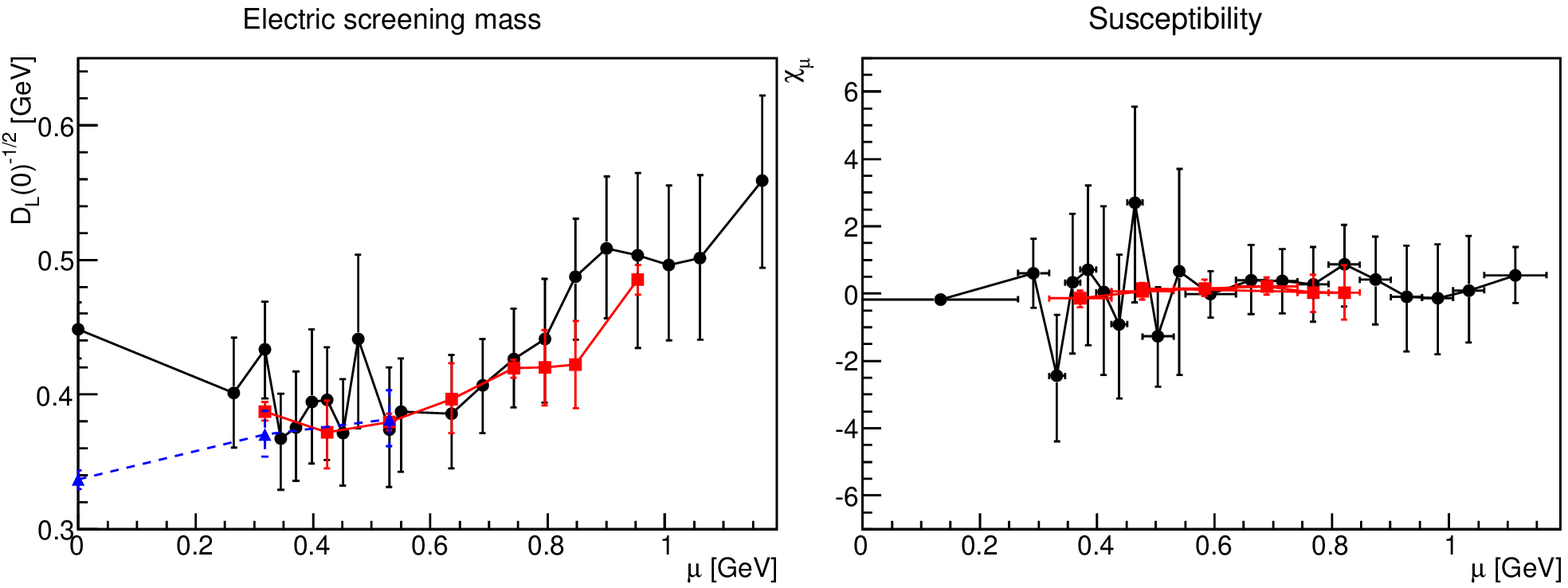}
\caption{\label{fig:mv}The dependence of the magnetic (top panels) and electric (bottom panels) screening masses (left panels) and susceptibility (right panels) on the physical volume and aspect ratio at $\beta=1.9$.}
\end{figure}

The situation is somewhat different when it comes to finite-volume effects, as shown in figure \ref{fig:mv}. In the magnetic
case a slight, but significant, dependence is seen, especially when it
comes to the, almost ten times larger, largest volume.  In the
electric case, no such effect is seen once at finite density. Still,
within errors no qualitative effect is seen even in the magnetic case,
and even the quantitative effect is only moderate. Still, this
provides a clear motivation for investigating the volume dependence more closely in the main text.

\subsection{Aspect ratio}\label{a:ar}

Formally, the aspect ratio between the lattice time extent and spatial extent
$(aN_t)/(aN_s)$ should tend to zero at finite temperature and 1 at
zero temperature, the latter for any chemical potential. At a finite
number of lattice points, this can only be approximated, which can
have substantial impact \cite{Maas:2011ez}. At finite temperature this
requires an investigation of the temperature dependence for different $\beta$ values, which is not possible here except for a very few temperatures below the phase transition, which were displayed in section \ref{s:tres}, and did not yield an effect within the other uncertainties. At finite densities already the results displayed in figure \ref{fig:mv} suggest an impact at $\beta=1.9$ for the magnetic case. Moreover, as long as $N_t$ is finite, a lattice system is not really at zero temperature, but there is a residual temperature. While we set this residual temperature to zero in the main text if $N_t\ge N_s$, this is therefore strictly speaking not true. This effect will mix at finite $N_t$ with the aspect ratio effects, and thus we cannot disentangle both of them. Thus, the following should be considered to be a combination of the systematic influence of both of them.

\begin{figure}
\includegraphics[width=0.5\textwidth]{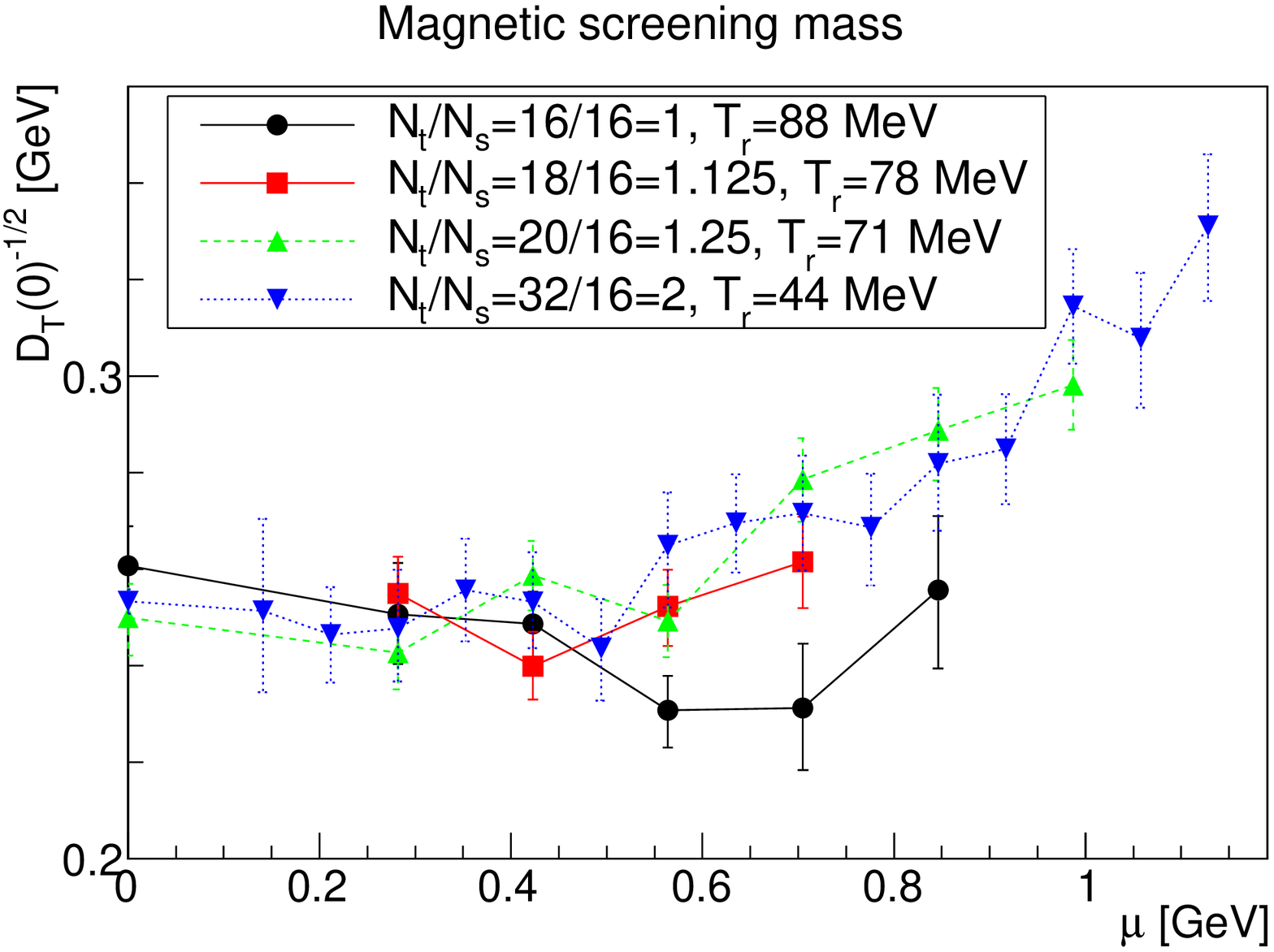}\includegraphics[width=0.5\textwidth]{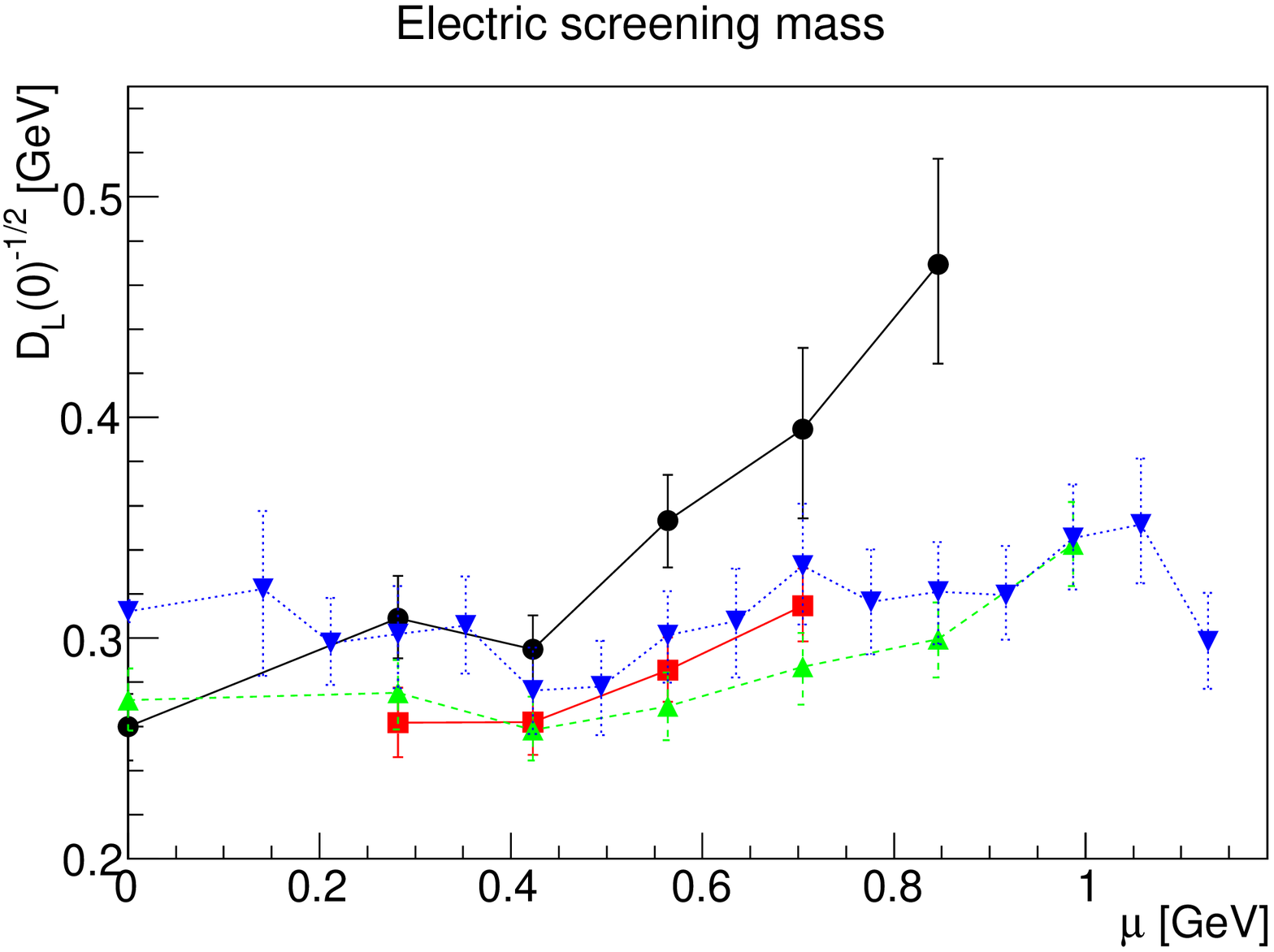}
\caption{\label{fig:mar}The dependence of the magnetic (left panel) and electric (right panel) screening masses on the aspect ratio/residual temperature $T_r$ at $\beta=2.1$ at fixed spatial volume.}
\end{figure}

At $\beta=2.1$ several different values of $N_t$ at fixed $N_s$, and thus different aspect ratios and residual temperatures, are available, and the results are shown in figure \ref{fig:mar}. All results at an aspect ratio larger than 1 agree within statistical errors, without any systematic trend. The only difference arises for an aspect ratio of 1. While the effect is still not statistically significant there is a systematic trend that the magnetic screening mass above $\mu\approx 0.5$ GeV is smaller than the one at an aspect ratio larger than one, and the reverse for the electric screening mass. However, the rise of the electric screening mass is at chemical potentials where the relatively high residual temperature may already indicate the effect seen in figure \ref{fig:mpd}.  Still, the effect is comparatively small when considering the statistical errors, such that in the main text only the aspect ratio of two is considered, for which the finest mesh in chemical potential is available.

\section{Configurations}\label{s:conf}

The list of configurations and lattice parameters used is given in table \ref{conf-sys}

\begin{longtable}{|c|c|c|c|c|c|c|c|c|c|}
\caption{\label{conf-sys}Employed lattice parameters and number of configurations. Note that all lattices with $N_s\le N_t$ will be considered to be at zero temperature.}\\
\hline
$N_s$	& $N_t$	& $\beta$ & $\kappa$ & $a^{-1}$ [GeV] & $L$ [fm] & $T$ [MeV] & $\mu$ [MeV] ($a\mu$) & $aj$ & Configuration \cr
\hline \endfirsthead
\hline
\multicolumn{10}{|l|}{Table \ref{conf-sys} continued}\\
\hline
$N_s$	& $N_t$	& $\beta$ & $\kappa$ & $a^{-1}$ [GeV] & $L$ [fm] & $T$ [MeV] & $\mu$ [MeV] ($a\mu$) & $aj$ & Configuration \cr
\hline \endhead
\hline
\multicolumn{10}{|r|}{Continued on next page}\\
\hline\endfoot
\endlastfoot
\hline
\hline
32	& 32	& 1.6	& 0.1820	& 0.741	& 8.51	& 0	& 0		& 0	& 2000	\cr
\hline
32	& 32	& 1.7	& 0.1780	& 0.857	& 7.36	& 0	& 0		& 0	& 1014	\cr
\hline
12	& 24	& 1.9	& 0.1680	& 1.06	& 2.23	& 0	& 0		& 0	& 313	\cr
32	& 32	& 1.9	& 0.1680	& 1.06	& 5.95	& 0	& 0		& 0	& 640	\cr
\hline
16	& 16	& 2.1	& 0.1577	& 1.41	& 2.21	& 0	& 0		& 0	& 200	\cr
16	& 20	& 	&		&	&	&	&		& 0	& 200	\cr
16	& 32	&  	&  	 	&  	& 	& 	& 		& 0	& 2060 \cr
\hline
\hline
16	& 32	& 2.1	& 0.1577	& 1.41	& 2.21	& 0	& 141 (0.100)	& 0.02	& 41	\cr
        &	&	&		&	&	&	&		& 0.03	& 102	\cr
\hline
16	& 32	& 2.1	& 0.1577	& 1.41	& 2.21	& 0	& 212 (0.150)	& 0.01	& 207	\cr
	&	&	&		&	&	&	&		& 0.03	& 104	\cr
\hline
12	& 24	& 1.9	& 0.1680	& 1.06	& 2.23	& 0	& 265 (0.250)	& 0.02	& 50	\cr%
	&	&	&		&	&	&	&		& 0.04	& 127	\cr%
\hline
16	& 16	&2.1	& 0.1577	& 1.41	& 2.21	& 0	& 282 (0.200)	& 0.02	& 126	\cr
	&	&	&		&	&	&	&		& 0.03	& 197	\cr
16	& 18	&	&		&	&	&	&		& 0.03	& 220	\cr
16	& 20	& 	&		&	&	&	&		& 0.03	& 200	\cr
16	& 32	& 	& 		& 	& 	& 	& 		& 0.01	& 66	\cr
	&	&	&		&	&	&	&		& 0.02	& 202	\cr
	&	&	&		&	&	&	&		& 0.03	& 100	\cr
\hline
12	& 24	& 1.9	& 0.1680	& 1.06	& 2.23	& 0	& 318 (0.300)	& 0.02	& 102	\cr%
	&	&	&		&	&	&	&		& 0.03	& 54	\cr%
	&	&	&		&	&	&	&		& 0.04	& 160	\cr%
\hline
16	& 24	& 1.9	& 0.1680	& 1.06	& 2.98	& 0	& 318 (0.300)	& 0.04	& 1960\cr
\hline
32	& 32	& 1.9	& 0.1680	& 1.06	& 5.95	& 0	& 318 (0.300)	& 0.04	& 299	\cr
\hline
12	& 24	& 1.9	& 0.1680	& 1.06	& 2.23	& 0	& 345 (0.325)	& 0.02	& 48	\cr%
	&	&	&		&	&	&	&		& 0.04	& 128	\cr%
\hline
16	& 32	& 2.1	& 0.1577	& 1.41	& 2.21	& 0	& 353 (0.250)	& 0.01	& 211	\cr
	&	&	&		&	&	&	&		& 0.02	& 96	\cr
	&	&	&		&	&	&	&		& 0.03	& 100	\cr
\hline
12	& 24	& 1.9	& 0.1680	& 1.06	& 2.23	& 0	& 371 (0.350)	& 0.02	& 49	\cr%
	&	&	&		&	&	&	&		& 0.04	& 284	\cr%
\hline
12	& 24	& 1.9	& 0.1680	& 1.06	& 2.23	& 0	& 392 (0.370)	& 0.04	& 126	\cr%
\hline
12	& 24	& 1.9	& 0.1680	& 1.06	& 2.23	& 0	& 398 (0.375)	& 0.02	& 52	\cr%
	&	&	&		&	&	&	&		& 0.04	& 153	\cr%
\hline
12	& 24	& 1.9	& 0.1680	& 1.06	& 2.23	& 0	& 403 (0.380)	& 0.04	& 126	\cr%
\hline
16	& 16	& 2.1	& 0.1577	& 1.41	& 2.21	& 0	& 423 (0.300)	& 0.02	& 204	\cr
	&	&	&		&	&	&	&		& 0.03	& 200	\cr
16	& 18	&	&		&	&	&	&		& 0.03	& 204	\cr
16	& 20	& 	&		&	&	&	&		& 0.03	& 210	\cr
16	& 32	& 	& 		& 	& 	& 	&		& 0.01	& 204	\cr
	&	&	&		&	&	&	&		& 0.02	& 60	\cr
	&	&	&		&	&	&	&		& 0.03	& 102	\cr
\hline
12	& 24	& 1.9	& 0.1680	& 1.06	& 2.23	& 0	& 424 (0.400)	& 0.02	& 42	\cr%
	&	&	&		&	&	&	&		& 0.04	& 138	\cr%
\hline
16	& 24	& 1.9	& 0.1680	& 1.06	& 2.98	& 0	& 424 (0.400)	& 0.04	& 100	\cr
\hline
12	& 24	& 1.9	& 0.1680	& 1.06	& 2.23	& 0	& 451 (0.425)	& 0.02	& 52	\cr%
	&	&	&		&	&	&	&		& 0.04	& 136	\cr%
\hline
12	& 24	& 1.9	& 0.1680	& 1.06	& 2.23	& 0	& 477 (0.450)	& 0.02	& 68	\cr
	&	&	&		&	&	&	&		& 0.04	& 181	\cr%
	&	&	&		&	&	&	&		& 0.06	& 34	\cr%
\hline
12	& 24	& 1.9	& 0.1680	& 1.06	& 2.23	& 0	& 488 (0.460)	& 0.04	& 60	\cr%
\hline
16	& 32	& 2.1	& 0.1577	& 1.41	& 2.21	& 0	& 494 (0.350)	& 0.01	& 200	\cr
	&	&	&		&	&	&	&		& 0.02	& 60	\cr
	&	&	&		&	&	&	&		& 0.03	& 100	\cr
\hline
12	& 24	& 1.9	& 0.1680	& 1.06	& 2.23	& 0	& 498 (0.470)	& 0.04	& 158	\cr%
\hline
12	& 24	& 1.9	& 0.1680	& 1.06	& 2.23	& 0	& 504 (0.475)	& 0.04	& 50	\cr%
\hline
12	& 24	& 1.9	& 0.1680	& 1.06	& 2.23	& 0	& 509 (0.480)	& 0.04	& 164	\cr%
\hline
12	& 24	& 1.9	& 0.1680	& 1.06	& 2.23	& 0	& 519 (0.490)	& 0.04	& 165	\cr%
\hline
12	& 24	& 1.9	& 0.1680	& 1.06	& 2.23	& 0	& 530 (0.500)	& 0.02	& 49	\cr%
	&	&	&		&	&	&	&		& 0.03	& 54	\cr%
	&	&	&		&	&	&	&		& 0.04	& 158	\cr%
\hline
16	& 24	& 1.9	& 0.1680	& 1.06	& 2.98	& 0	& 530 (0.500)	& 0.04	& 2000	\cr
\hline
32	& 32	& 1.9	& 0.1680	& 1.06	& 5.95	&	& 530 (0.500)	& 0.04	& 126	\cr
\hline
12	& 24	& 1.9	& 0.1680	& 1.06	& 2.23	& 0	& 557 (0.525)	& 0.04	& 283	\cr%
\hline
16	& 16	& 2.1	& 0.1577	& 1.41	& 2.21	& 0	& 564 (0.400)	& 0.02	& 200	\cr
	&	&	&		&	&	&	&		& 0.03	& 216	\cr
16	& 18	&	&		&	&	&	&		& 0.03	& 212	\cr
16	& 20	& 	&		&	&	&	&		& 0.03	& 204	\cr
16	& 32	& 	&		& 	& 	& 	&		& 0.01	& 200	\cr
	&	&	&		&	&	&	&		& 0.02	& 63	\cr
	&	&	&		&	&	&	&		& 0.03	& 100	\cr
\hline
12	& 24	& 1.9	& 0.1680	& 1.06	& 2.23	& 0	& 583 (0.550)	& 0.02	& 52	\cr
	&	&	&		&	&	&	&		& 0.04	& 52	\cr%
	&	&	&		&	&	&	&		& 0.06	& 36	\cr%
\hline
12	& 24	& 1.9	& 0.1680	& 1.06	& 2.23	& 0	& 610 (0.575)	& 0.04	& 166	\cr%
\hline
16	& 32	& 2.1	& 0.1577	& 1.41	& 2.21	& 0	& 635 (0.450)	& 0.02	& 61	\cr
	&	&	&		&	&	&	&		& 0.03	& 100	\cr
\hline
12	& 24	& 1.9	& 0.1680	& 1.06	& 2.23	& 0	& 636 (0.600)	& 0.02	& 50	\cr%
	&	&	&		&	&	&	&		& 0.04	& 31	\cr%
\hline
16	& 24	& 1.9	& 0.1680	& 1.06	& 2.98	& 0	& 636 (0.600)	& 0.04	& 100	\cr
\hline
12	& 24	& 1.9	& 0.1680	& 1.06	& 2.23	& 0	& 689 (0.650)	& 0.02	& 52	\cr%
	&	&	&		&	&	&	&		& 0.04	& 149	\cr%
\hline
16	& 16	&2.1	& 0.1577	& 1.41	& 2.21	& 0	& 705 (0.500)	& 0.02	& 64	\cr
	&	&	&		&	&	&	&		& 0.03	& 200	\cr
16	& 18	&	&		&	&	&	&		& 0.03	& 210	\cr
16	& 20	& 	&		&	&	&	&		& 0.03	& 200	\cr
16	& 32	& 	&		& 	& 	& 	&		& 0.02	& 60	\cr
	&	&	&		&	&	&	& 		& 0.03	& 100	\cr
\hline
12	& 24	& 1.9	& 0.1680	& 1.06	& 2.23	& 0	& 742 (0.700)	& 0.02	& 50	\cr%
	&	&	&		&	&	&	&		& 0.03	& 50	\cr%
	&	&	&		&	&	&	&		& 0.04	& 116	\cr%
\hline
16	& 24	& 1.9	& 0.1680	& 1.06	& 2.98	& 0	& 742 (0.700)	& 0.04	& 2000	\cr
\hline
16	& 32	& 2.1	& 0.1577	& 1.41	& 2.21	& 0	& 776 (0.550)	& 0.02	& 60	\cr
	&	&	&		&	&	&	&		& 0.03	& 100	\cr
\hline
12	& 24	& 1.9	& 0.1680	& 1.06	& 2.23	& 0	& 795 (0.750)	& 0.02	& 50	\cr%
\hline
16	& 24	& 1.9	& 0.1680	& 1.06	& 2.98	& 0	& 795 (0.750)	& 0.04	& 120	\cr
\hline
16	& 16	&2.1	& 0.1577	& 1.41	& 2.21	& 0	& 846 (0.600)	& 0.02	& 60	\cr
	&	&	&		&	&	&	&		& 0.03	& 200	\cr
16	& 20	&	&		&	&	&	&	 	& 0.03	& 203	\cr
16	& 32	& 	&		& 	& 	& 	&		& 0.02	& 60	\cr
	&	&	&		&	&	&	&		& 0.03	& 100	\cr
\hline
12	& 24	& 1.9	& 0.1680	& 1.06	& 2.23	& 0	& 848 (0.800)	& 0.02	& 50	\cr%
	&	&	&		&	&	&	&		& 0.04	& 142	\cr%
\hline
16	& 24	& 1.9	& 0.1680	& 1.06	& 2.98	& 0	& 848 (0.800)	& 0.04	& 120	\cr
\hline
12	& 24	& 1.9	& 0.1680	& 1.06	& 2.23	& 0	& 901 (0.850)	& 0.02	& 50	\cr%
\hline
16	& 32	& 2.1	& 0.1577	& 1.41	& 2.21	& 0	& 917 (0.650)	& 0.02	& 60	\cr
	&	&	&		&	&	&	& 		& 0.03	& 100	\cr
\hline
12	& 24	& 1.9	& 0.1680	& 1.06	& 2.23	& 0	& 954 (0.900)	& 0.02	& 48	\cr%
	&	&	&		&	&	&	&		& 0.03	& 51	\cr%
	&	&	&		&	&	&	&		& 0.04	& 67	\cr%
\hline
16	& 24	& 1.9	& 0.1680	& 1.06	& 2.98	& 0	& 954 (0.900)	& 0.04	& 1100	\cr
\hline
16	& 20	& 2.1	& 0.1577	& 1.41	& 2.21	& 0	& 987 (0.700)	& 0.03	& 200	\cr
16	& 32	&	&		&	&	& 	&		& 0.02	& 60	\cr
	&	&	&		&	&	&	&		& 0.03	& 104	\cr
\hline
12	& 24	& 1.9	& 0.1680	& 1.06	& 2.23	& 0	& 1007 (0.950)	& 0.02	& 50	\cr%
\hline
16	& 32	& 2.1	& 0.1577	& 1.41	& 2.21	& 0	& 1058 (0.750)	& 0.03	& 102	\cr
\hline
12	& 24	& 1.9	& 0.1680	& 1.06	& 2.23	& 0	& 1060 (1.000)	& 0.02	& 50	\cr%
	&	&	&		&	&	&	&		& 0.04	& 126	\cr%
\hline
16	& 32	& 2.1	& 0.1577	& 1.41	& 2.21	& 0	& 1128 (0.800)	& 0.03	& 100	\cr
\hline
12	& 24	& 1.9	& 0.1680	& 1.06	& 2.23	& 0	& 1166 (1.100)	& 0.02	& 50	\cr
	&	&	&		&	&	&	&		& 0.04	& 88	\cr%
\hline
\hline
48	& 32	& 1.7	& 0.1780	& 0.857	& 8.10	& 27	& 0		& 0	& 302	\cr
\hline
32	& 8	& 1.9	& 0.1680	& 1.06	& 5.95	& 133	& 0		& 0	& 2000	\cr
\hline
16	& 12	& 2.1	& 0.1577	& 1.41	& 2.21	& 118	& 0		& 0	& 200	\cr
\hline
16	& 10	& 2.1	& 0.1577	& 1.41	& 2.21	& 141	& 0		& 0	& 200	\cr
\hline
16	& 9	& 2.1	& 0.1577	& 1.41	& 2.21	& 157	& 0		& 0	& 400	\cr
\hline
16	& 8	& 2.1	& 0.1577	& 1.41	& 2.21	& 176	& 0		& 0	& 200	\cr
\hline
16	& 7	& 2.1	& 0.1577	& 1.41	& 2.21	& 201	& 0		& 0	& 400	\cr
\hline
16	& 6	& 2.1	& 0.1577	& 1.41	& 2.21	& 235	& 0		& 0	& 200	\cr
\hline
16	& 5	& 2.1	& 0.1577	& 1.41	& 2.21	& 282	& 0		& 0	& 200	\cr
\hline
16	& 4	& 2.1	& 0.1577	& 1.41	& 2.21	& 353	& 0		& 0	& 200	\cr
\hline
\hline
16	& 14	& 2.1	& 0.1577	& 1.41	& 2.21	& 101	& 282 (0.200)	& 0.03	& 216	\cr
\hline
16	& 14	& 2.1	& 0.1577	& 1.41	& 2.21	& 101	& 423 (0.300)	& 0.02	& 206	\cr
	&	&	&		&	&	&	&		& 0.03	& 220	\cr
\hline
16	& 14	& 2.1	& 0.1577	& 1.41	& 2.21	& 101	& 564 (0.400)	& 0.02	& 210	\cr
	&	&	&		&	&	&	&		& 0.03	& 200	\cr
\hline
16	& 14	& 2.1	& 0.1577	& 1.41	& 2.21	& 101	& 705 (0.500)	& 0.02	& 100	\cr
	&	&	&		&	&	&	&		& 0.03	& 209	\cr
\hline
16	& 13	& 2.1	& 0.1577	& 1.41	& 2.21	& 108	& 282 (0.200)	& 0.03	& 200	\cr
\hline
16	& 13	& 2.1	& 0.1577	& 1.41	& 2.21	& 108	& 423 (0.300)	& 0.02	& 310	\cr
	&	&	&		&	&	&	&		& 0.03	& 200	\cr
\hline
16	& 13	& 2.1	& 0.1577	& 1.41	& 2.21	& 108	& 564 (0.400)	& 0.02	& 214	\cr
	&	&	&		&	&	&	&		& 0.03	& 218	\cr
\hline
16	& 13	& 2.1	& 0.1577	& 1.41	& 2.21	& 108	& 705 (0.500)	& 0.02	& 100	\cr
	&	&	&		&	&	&	&		& 0.03	& 200	\cr
\hline
16	& 12	& 2.1	& 0.1577	& 1.41	& 2.21	& 118	& 282 (0.200)	& 0.02	& 214	\cr
	&	&	&		&	&	&	&		& 0.03	& 200	\cr
\hline
16	& 12	& 2.1	& 0.1577	& 1.41	& 2.21	& 118	& 423 (0.300)	& 0.02	& 204	\cr
	&	&	&		&	&	&	&		& 0.03	& 216	\cr
\hline
16	& 12	& 2.1	& 0.1577	& 1.41	& 2.21	& 118	& 564 (0.400)	& 0.02	& 216	\cr
	&	&	&		&	&	&	&		& 0.03	& 216	\cr
\hline
16	& 12	& 2.1	& 0.1577	& 1.41	& 2.21	& 118	& 705 (0.500)	& 0.02	& 208	\cr
	&	&	&		&	&	&	&		& 0.03	& 200	\cr
\hline
16	& 12	& 2.1	& 0.1577	& 1.41	& 2.21	& 118	& 846 (0.600)	& 0.02	& 213	\cr
	&	&	&		&	&	&	&		& 0.03	& 224	\cr
\hline
16	& 12	& 2.1	& 0.1577	& 1.41	& 2.21	& 118	& 987 (0.700)	& 0.03	& 192	\cr
\hline
16	& 12	& 2.1	& 0.1577	& 1.41	& 2.21	& 118	& 1128 (0.800)	& 0.03	& 235	\cr
\hline
16	& 12	& 2.1	& 0.1577	& 1.41	& 2.21	& 118	& 1269 (0.900)	& 0.03	& 200	\cr
\hline
16	& 11	& 2.1	& 0.1577	& 1.41	& 2.21	& 128	& 282 (0.200)	& 0.03	& 204	\cr
\hline
16	& 11	& 2.1	& 0.1577	& 1.41	& 2.21	& 128	& 423 (0.300)	& 0.03	& 200	\cr
\hline
16	& 11	& 2.1	& 0.1577	& 1.41	& 2.21	& 128	& 564 (0.400)	& 0.03	& 220	\cr
\hline
16	& 11	& 2.1	& 0.1577	& 1.41	& 2.21	& 128	& 705 (0.500)	& 0.03	& 200	\cr
\hline
16	& 10	& 2.1	& 0.1577	& 1.41	& 2.21	& 141	& 282 (0.200)	& 0.03	& 200	\cr
\hline
16	& 10	& 2.1	& 0.1577	& 1.41	& 2.21	& 141	& 423 (0.300)	& 0.02	& 240	\cr
	&	&	&		&	&	&	&		& 0.03	& 240	\cr
\hline
16	& 10	& 2.1	& 0.1577	& 1.41	& 2.21	& 141	& 564 (0.400)	& 0.02	& 300	\cr
	&	&	&		&	&	&	&		& 0.03	& 227	\cr
\hline
16	& 10	& 2.1	& 0.1577	& 1.41	& 2.21	& 141	& 705 (0.500)	& 0.02	& 210	\cr
	&	&	&		&	&	&	&		& 0.03	& 210	\cr
\hline
16	& 9	& 2.1	& 0.1577	& 1.41	& 2.21	& 157	& 282 (0.200)	& 0.03	& 204	\cr
\hline
16	& 9	& 2.1	& 0.1577	& 1.41	& 2.21	& 157	& 423 (0.300)	& 0.03	& 500	\cr
\hline
16	& 9	& 2.1	& 0.1577	& 1.41	& 2.21	& 157	& 564 (0.400)	& 0.03	& 500	\cr
\hline
16	& 9	& 2.1	& 0.1577	& 1.41	& 2.21	& 157	& 705 (0.500)	& 0.03	& 400	\cr
\hline
16	& 8	& 2.1	& 0.1577	& 1.41	& 2.21	& 176	& 282 (0.200)	& 0.03	& 200	\cr
\hline
16	& 8	& 2.1	& 0.1577	& 1.41	& 2.21	& 176	& 423 (0.300)	& 0.02	& 200	\cr
	&	&	&		&	&	&	&		& 0.03	& 200	\cr
\hline
16	& 8	& 2.21	& 0.1577	& 1.41	& 2.21	& 176	& 564 (0.400)	& 0.02	& 200	\cr
	&	&	&		&	&	&	&		& 0.03	& 200	\cr
\hline
16	& 8	& 2.21	& 0.1577	& 1.41	& 2.21	& 176	& 705 (0.500)	& 0.03	& 200	\cr
\hline
16	& 7	& 2.1	& 0.1577	& 1.41	& 2.21	& 201	& 282 (0.200)	& 0.03	& 200	\cr
\hline
16	& 7	& 2.1	& 0.1577	& 1.41	& 2.21	& 201	& 423 (0.300)	& 0.03	& 200	\cr
\hline
16	& 7	& 2.1	& 0.1577	& 1.41	& 2.21	& 201	& 564 (0.400)	& 0.03	& 200	\cr
\hline
16	& 7	& 2.1	& 0.1577	& 1.41	& 2.21	& 201	& 705 (0.500)	& 0.03	& 200	\cr
\hline
16	& 6	& 2.1	& 0.1577	& 1.41	& 2.21	& 235	& 282 (0.200)	& 0.03	& 200	\cr
\hline
16	& 6	& 2.1	& 0.1577	& 1.41	& 2.21	& 235	& 423 (0.300)	& 0.02	& 200	\cr
	&	&	&		&	&	&	&		& 0.03	& 200	\cr
\hline
16	& 6	& 2.1	& 0.1577	& 1.41	& 2.21	& 235	& 564 (0.400)	& 0.02	& 200	\cr
	&	&	&		&	&	&	&		& 0.03	& 220	\cr
\hline
16	& 6	& 2.1	& 0.1577	& 1.41	& 2.21	& 235	& 705 (0.500)	& 0.02	& 200	\cr
	&	&	&		&	&	&	&		& 0.03	& 200	\cr 
\hline
16	& 4	& 2.1	& 0.1577	& 1.41	& 2.21	& 353	& 282 (0.200)	& 0.03	& 200	\cr
\hline
16	& 4	& 2.1	& 0.1577	& 1.41	& 2.21	& 353	& 423 (0.300)	& 0.02	& 200	\cr
	&	&	&		&	&	&	&		& 0.03	& 216	\cr
\hline
16	& 4	& 2.1	& 0.1577	& 1.41	& 2.21	& 353	& 564 (0.400)	& 0.02	& 200	\cr
	&	&	&		&	&	&	&		& 0.03	& 200	\cr
\hline
16	& 4	& 2.1	& 0.1577	& 1.41	& 2.21	& 353	& 705 (0.500)	& 0.02	& 200	\cr
	& 	&	&		&	&	&	&		& 0.03	& 200	\cr
\hline
\end{longtable}

\bibliographystyle{JHEP}
\bibliography{bib}

\end{document}